\documentclass{article}

\usepackage[margin=1in]{geometry}
\usepackage{lmodern}
\usepackage[T1]{fontenc}
\usepackage[utf8]{inputenc}
\usepackage{amsmath}
\allowdisplaybreaks
\usepackage{amssymb,amsthm,mathtools, bbm}
\usepackage{mathrsfs}
\usepackage[svgnames,table]{xcolor}
\usepackage{tikz}
\usepackage{booktabs}
\usepackage{multirow}
\usepackage{thmtools} 
\usepackage{thm-restate}
\usepackage{physics}
\usepackage{hyperref}
\usepackage{hhline}
\usepackage[capitalise,nameinlink]{cleveref}
\definecolor{links}{RGB}{11, 85, 255}
\definecolor{cites}{RGB}{0, 200, 0}
\definecolor{urls}{RGB}{255, 116, 0}
\hypersetup{colorlinks={true},linkcolor={links},citecolor=[named]{cites},urlcolor={urls}}
\usepackage[numbers,compress]{natbib}
\usepackage{doi}
\usepackage{tikz} 
\usepackage{pgfplots}
\pgfplotsset{compat=1.14}
\usepgfplotslibrary{fillbetween}
\usepackage{hyphenat}
\usepackage[font=footnotesize]{caption}
\usepackage{subcaption}
\usepackage{appendix}
\usepackage{nicefrac}
\usepackage{tcolorbox}
\usepackage{enumitem} 
\usepackage{todonotes}
\usepackage{xspace}

\usepackage{algorithm}
\usepackage[noend]{algpseudocode}

\definecolor{johannesgreen}{HTML}{006B3C}

\definecolor{revisioncolor}{HTML}{007A87}

\newcommand{\sedit}[1]{\textcolor{blue}{#1}}

\newcommand{\COPT}{\textsc{ConOpt}}

\newcommand{\opt}{\mu^\star}

\newcommand{\range}{%
    \hyperref[alg:mhr-bootstrap]{\textsc{Initialize mhr}}\xspace}
\newcommand{\optimize}{%
    \hyperref[alg:mhr-optimization]{\textsc{Optimize}}\xspace}
\newcommand{\mhr}{%
    \hyperref[alg:mhr]{\textsc{Bilateral mhr}}\xspace}

\newcommand{\init} 
{\hyperref[alg:regular-initialization-sketch]{\textsc{Initialize regular}}\xspace}
\newcommand{\refine}
{\hyperref[alg:regular-refinement]{\textsc{Successive Opt}}\xspace}

\newcommand{\oneround}{%
    \hyperref[alg:adv-inner]{\textsc{OneRound}}\xspace}

\newcommand{\KL}[2]{D_{\text{KL}}(#1 \parallel #2)}
\newcommand{\TV}[1]{\left | \left | #1\right| \right |_{\textsc{TV}}}

\newcommand{\Ej}[1]{\mathbb{E}^j\left[#1\right]}
\newcommand{\Eo}[1]{\mathbb{E}^0\left[#1\right]}
\newcommand{\Eup}[2]{\mathbb{E}^{#1}\left[#2\right]}
\newcommand{\bern}{\textnormal{Bern}}
\newcommand{\clean}{\mathcal E}

\newcommand{\cA}{\mathcal{A}}
\renewcommand{\AA}{\mathscr{A}}
\newcommand{\cD}{\mathcal{D}}

\newcommand{\cF}{\mathcal{F}}
\newcommand{\cG}{\mathcal{G}}
\newcommand{\cH}{\mathcal{H}}
\newcommand{\cI}{\mathcal{I}}
\newcommand{\cK}{\mathcal{K}}
\newcommand{\cL}{\mathcal{L}}

\newcommand{\cQ}{\mathcal{Q}}

\newcommand{\cS}{\mathcal{S}}
\newcommand{\cT}{\mathcal{T}}

\newcommand{\cX}{\mathcal{X}}

\newcommand{\cY}{\mathcal Y}

\newcommand{\Prb}{\mathbb{P}}
\newcommand{\ind}[1]{\mathbf{1}_{\{#1\}}}
\newcommand{\cC}{\mathcal{C}}

\newcommand{\R}{\mathbb{R}}
\DeclareMathOperator*{\argmax}{arg\,max}
\DeclareMathOperator*{\argmin}{arg\,min}

\renewcommand{\P}[1]{{\mathbb{P}}\left(#1\right)}
\newcommand{\Psub}[2]{{\mathbb{P}}_{#1}\left(#2\right)}

\newcommand{\E}[1]{{\mathbb{E}}\left[#1\right]}
\newcommand{\Esub}[2]{{\mathbb{E}}_{#1}\left[#2\right]}

\renewcommand{\exp}[1]{{\textup{exp}}\left(#1\right)}

\newcommand{\indep}{\perp\kern-5pt\perp}

\newcommand{\eps}{\varepsilon}

\newcommand{\Ber}[1]{\textmd{\textup{Ber}}\left(#1\right)}

\newcommand{\grid}{G_K}

\newcommand{\thres}{\delta}

\newcommand{\block}{D}

\newcommand{\prof}{\textsc{Profit}}

\newcommand{\fs}{f_s}
\newcommand{\fb}{f_b}
\newcommand{\Fs}{F_s}
\newcommand{\Fb}{F_b}

\newif\ifSparseParameters
\SparseParametersfalse

\makeatletter
\@ifundefined{newcounteralias}{}{%
  \renewcommand\thmt@autorefsetup{%
    \@xa\def\csname\thmt@envname autorefname\@xa\endcsname
    \@xa{\thmt@thmname}}%
}
\makeatother
\theoremstyle{plain}

\newtheorem{theorem}{Theorem}[section]

\newtheorem*{theorem*}{Theorem}
\newtheorem{lemma}[theorem]{Lemma}
\newtheorem{mainresult}{Main Result}
\newtheorem*{warmupresult}{Warm-up Result}

\newtheorem{proposition}[theorem]{Proposition}

\newtheorem{claim}[theorem]{Claim}
\newtheorem{remark}[theorem]{Remark}

\newtheorem{definition}[theorem]{Definition}

\crefname{claim}{claim}{claims}
\Crefname{claim}{Claim}{Claims}
\Crefname{lemma}{Lemma}{Lemmata}

\definecolor{regretOrange}{HTML}{F3A08B}
\definecolor{regretYellow}{HTML}{FFF2B3}
\definecolor{regretGreen}{HTML}{D6ECCD}

\crefname{mainresult}{main result}{main results}
\Crefname{mainresult}{Main Result}{Main Results}
    
\crefname{appendix}{appendix}{appendices}
\Crefname{appendix}{Appendix}{Appendices}

\title{Regret Bounds for Profit Maximization in Bilateral Trade: Learning the Best Posted-Price Mechanism}
\author{
Johannes Br\"ustle\thanks{Department of Computer, Control, \& Management Engineering, Sapienza University of Rome, Rome, Italy. Email: \texttt{\{brustle,digregorio,fuscof,leonardi\}@diag.uniroma1.it}}
\and
Simone Di Gregorio$^*$
\and
Federico Fusco$^*$
\and
Stefano Leonardi$^*$
}
\date{}

\begin{document}
\pagenumbering{roman}

\maketitle
\thispagestyle{empty}

\begin{abstract}
    Bilateral trade models the task of intermediating between two strategic agents, a seller and a buyer, willing to trade a good for which they hold private valuations. We study this problem from the perspective of a profit-maximizing broker in a regret minimization framework.
    At each time step, a new seller and buyer arrive and the broker proposes a trading price to each agent. The trade happens if both agents accept, and the profit of the broker is the difference between the price received from the buyer and the one paid to the seller. At the end of each interaction, the learner only observes whether the trade occurred, and its goal is to achieve sublinear regret relative to the best fixed pair of prices in hindsight.

    We start by investigating the adversarial version of the problem, where buyer and seller valuations are generated by an adversary. In this setting, we construct a polynomial-time learning algorithm with a $\tilde O(T^{\nicefrac 23})$ regret rate, which is nearly tight even if the valuations are drawn i.i.d. from independent distributions, thus matching the rate for one-sided pricing. Our adversarial algorithm combines a localized estimation procedure with an information-based minimax construction.

    We then consider valuations drawn i.i.d. from independent distributions satisfying the monotone-hazard-rate (MHR) property and provide a learning algorithm with nearly tight $\tilde{O}(\sqrt{T})$ regret. The algorithm combines localized estimation with a stochastic convex bandit routine applied to the negative logarithm of expected profit.
    Finally, we complement this result by showing a \emph{separation} between MHR and regular distributions, proving that the minimax regret rate in the latter setting is $\tilde \Theta(T^{\nicefrac 35})$. This last result contrasts with the $\tilde{O}(\sqrt{T})$ regret bound that can be obtained for one-sided pricing with regular distributions.
 
\end{abstract}

\newpage
\tableofcontents
\newpage
\pagenumbering{arabic} 
\section{Introduction}
\label{sec:intro}

    Bilateral trade studies the interaction of two agents---a seller and a buyer---interested in trading a good \citep[e.g.,][]{MyersonS83}. Each agent holds a private valuation for the good, and they strive to maximize their own utility. Designing ``good'' mechanisms for bilateral trade is a fundamental problem in economics and has been studied extensively in the computer science community \citep[e.g.,][]{ColiniBaldeschiKLT16,BrustleCWZ17,BabaioffGG20,BlumrosenD21,DuttingFLLR26,DengMSW21,kang22fixed,LiuR023,CaiW23,BernasconiCCF24,HajiaghayiHPS25}. There are two natural objectives that a broker tasked with designing a mechanism for bilateral trade may be interested in maximizing: social welfare, often measured in terms of its increase, the gains from trade, and profit. The predominant approach considered in prior work is \emph{Bayesian}, as it assumes that both agents' valuations are drawn from known and typically independent distributions. In many practical applications, however, knowledge of these distributions is a non-trivial assumption, and valuations may not be independent.

    In this work, we study profit maximization in repeated bilateral trade through the lens of online learning \citep[e.g.,][]{BlumKRW03,KleinbergL03,Cesa-BianchiGM15}. While a substantial literature studies social welfare maximization \citep[e.g.,][]{Cesa-BianchiCCF24,Cesa-BianchiCCF24jmlr,AzarFF24,BernasconiCCF24,LunghiC026}, only recently has the literature started investigating profit from a regret-minimizing perspective, with \citet{GregorioDFS25,GregorioDFS26} providing a complete characterization of the minimax regret regimes for profit maximization when the learner is allowed to post any mechanism which ensures \emph{dominant-strategy incentive compatibility} and \emph{individual rationality}. While these general mechanisms for bilateral trade are very expressive, they may be fairly complex and, crucially, require agents to submit their actual valuations to the learner.
    
    In this work, we focus on the important and natural family of posted-price mechanisms. In a posted-price mechanism, the broker just posts two prices, one for the seller and one for the buyer, and the trade happens if and only if both agents agree to trade for the proposed price. These \emph{simple} mechanisms are specified by two numbers and require each agent only to accept or reject an offer.

    We consider two families of data-generation models for the online valuations: the adversarial model, where an oblivious adversary fixes the agents' valuations before play, and the stochastic model, where valuation pairs are drawn i.i.d. from a fixed unknown distribution. Our benchmark is the best fixed pair of prices, evaluated on the input sequence in the adversarial setting and in expectation in the stochastic setting. We seek algorithms whose expected cumulative profit falls short of this benchmark by a quantity sublinear in the time horizon $T$.

\subsection{Our Results}
\label{sec:results}

    We characterize the minimax regret rates for the data-generation models below, up to polylogarithmic factors. See Table~\ref{tab:results} for a summary. All upper bounds use one-bit feedback (whether both offers were accepted), whereas the lower bounds hold even with two-bit feedback (the seller's and buyer's acceptance decisions observed separately).
    
\begin{itemize}
    \item For an oblivious adversary, our algorithm guarantees $O(T^{\nicefrac 23}\log^{\nicefrac 73}T)$ regret (Thm.~\ref{thm:adv-regret}). Given the lower bound of $\Omega(T^{\nicefrac 23})$ for one-sided pricing \citep{KleinbergL03}, our algorithm is nearly tight even when the agents' valuations are drawn i.i.d. from generic independent distributions. Indeed, setting the seller's value identically to zero embeds one-sided pricing: the seller always accepts, and a positive seller price only reduces profit without providing additional information.
    \item For valuations drawn i.i.d. from independent regular distributions, we obtain $O(T^{\nicefrac 35}\log^5T)$ regret (Thm.~\ref{thm:regular}), complemented by a matching $\Omega(T^{\nicefrac 35})$ lower bound, up to logarithmic factors (Thm.~\ref{thm:regular_lower}).
    \item For independent MHR valuations we get $O(\sqrt{T}\log^{\nicefrac 52}\!T)$ regret (Thm.~\ref{thm:mhr-regret}), which is nearly tight \citep{LemeSTW23}.\footnote{To reduce from the pricing lower bound from Theorem 3.4 in \citet{LemeSTW23}, we need to reduce bilateral trade to pricing, but here we cannot do it immediately, since assuming densities for MHR prevents atoms: we cannot just consider the instance where the seller's value is $0$ deterministically. This can be solved by considering a relatively straightforward limiting argument, with an appropriate seller distribution weakly converging to an atom at $0$.}
\end{itemize}

    First, it is interesting to compare the bilateral trade regret rates with those for the one-dimensional problem of (one-sided) pricing \citep{KleinbergL03}, where the learner only interacts with the buyer. There, the adversarial rate is the same, but when it comes to regularity there is a clear differentiation: regularity and MHR yield the same $\sqrt{T}$ rate in one-sided pricing \citep{SinglaW24,LemeSTW23}, while bilateral trade exhibits two intriguing rates, i.e., $\sqrt T$ for MHR and the novel (and surprising) rate of $T^{\nicefrac 35}$ for regular distributions.

    Second, a comparison with previous work on regret minimization in bilateral trade is also instructive. While for gain-from-trade maximization (and profit maximization with generic mechanisms) there is typically a separation between the adversarial and the stochastic settings, where the former allows for no-regret only under additional assumptions (such as $\sigma$-smoothness) \citep{BernasconiF26,GregorioDFS25,GregorioDFS26}, this does not happen for our problem. In particular, faster rates are here allowed by classical mechanism design concepts, namely regularity and monotone hazard rate. This is partly due to an exciting and novel connection between online bilateral trade and online convex optimization: MHR makes the negative logarithm of expected profit convex, while regularity yields a family of auxiliary profit functions with convex reciprocals.

\begin{table}[t]
\centering
\renewcommand{\arraystretch}{1.3}
\setlength{\arrayrulewidth}{0.5pt}
\arrayrulecolor{black}
\begin{tabular}{|l|l|c|}
\hhline{|-|-|-|}
Setting & Distribution assumptions & Minimax regret \\
\hhline{|-|-|-|}
Adversarial & ---
    & \cellcolor{regretOrange}$\widetilde{O}(T^{\nicefrac 23})$ -- Thm.~\ref{thm:adv-regret}
    \\
\hhline{|-|-|-|}
\multirow{4}{*}{Stochastic Setting} & Generic
    & \cellcolor{regretOrange}\\
\hhline{|~|-|-|}
& Independent
    & \cellcolor{regretOrange}$\Omega(T^{\nicefrac 23})$ -- \citep{KleinbergL03} \\
\hhline{|~|-|-|}
& Independent \& Regular
    & \cellcolor{regretYellow}$\widetilde{\Theta}(T^{\nicefrac 35})$ -- Thms. \ref{thm:regular} \& \ref{thm:regular_lower} \\
\hhline{|~|-|-|}
& Independent \& MHR
    & \cellcolor{regretGreen}$\widetilde{\Theta}(\sqrt T)$ -- Thms.~\ref{thm:mhr-regret} \& \citep{LemeSTW23} \\
\hhline{|-|-|-|}
\end{tabular}
\caption{Minimax regret rates for profit maximization with posted prices in bilateral trade. Entries with the same color share the same minimax rate. The adversarial positive result applies to all other settings, while the lower bound for i.i.d. distributions with independent marginals also applies to the generic stochastic setting and the adversarial one. }
\label{tab:results}
\end{table}

    \subsection{Challenges}
    \label{sec:challenges}

    Finding profit-maximizing posted prices poses two challenges: exploiting the structure of the profit function and learning from one-bit feedback. Two generic approaches illustrate why additional ideas are needed.
    
    \paragraph{Structure of Profit.}
    
    The profit function exhibits a convenient one-sided Lipschitzness property: raising the seller's price and lowering the buyer's price preserves every previously accepted trade while lowering the profit proportionally to the displacement. Therefore, we can effectively discretize the action space $[0,1]^2$: restricting the algorithm's attention to  the uniform grid $G_K$ of mesh $\nicefrac 1K$, with $\Theta(K^2)$ pairs of prices, induces a discretization error of at most $\approx \nicefrac TK$ (more on this in \Cref{lem:discretization_err}). Treating these pairs as unrelated arms and running a finite-armed bandit algorithm yields
    \[
        R_T\lesssim \sqrt{K^2T}+\frac{2T}{K}
        \approx T^{\nicefrac 34},
    \]
    for the right tuning of $K\approx T^{\nicefrac 14}$. This is the generic argument for Lipschitz bandits in $\R^2$ \citep{KleinbergSU19}. This ``naive'' approach of using a uniform grid and then treating each point independently is tight for (adversarial) one-sided pricing but can be improved for certain classes of objectives using more refined notions of \emph{dimension}, both in the stochastic and adversarial worlds \citep{KleinbergSU19,PodimataS21}. Unfortunately, these off-the-shelf approaches do not help for bilateral trade, since the \emph{zooming dimension} of our problem is still $2$ (see Appendix~\ref{sec:dimension-comparison}).

    \paragraph{One-Bit Feedback.} 
    The one-bit feedback is at least as informative as the bandit feedback (if you know the outcome of the trade and the prices posted, you can reconstruct exactly your profit), but is also somehow richer, as the prices are not isolated arms but share an underlying structure (so that agents accepting price $(p,q)$ allow you to reconstruct the profit for any $(p',q)$ with $p' \ge p$). This idea is exploited in a recent work of \citet{CastiglioniLunghiMarchesi2026} for the stochastic version of our problem. There, they show how to uniformly $\eps$-estimate the two-dimensional CDF, and thus the profit, of the underlying distribution on the uniform grid of mesh $\nicefrac 1\eps$, using as few as $\widetilde O(\nicefrac{1}{\eps^{3}})$ pricing queries (which are equivalent to our one-bit feedback). For context, estimating the reward of $\nicefrac{1}{\eps^2}$ independent arms in the stochastic multi-armed bandit problem requires $\Omega(\nicefrac{1}{\eps^4})$ queries. It follows directly from \citet{CastiglioniLunghiMarchesi2026} that an explore-then-commit approach for our problem would satisfy:
    \[
        R_T\lesssim\underbrace{\nicefrac1{\eps^{3}}}_{\text{exploration}}
        +\underbrace{\eps T}_{\text{exploitation}} \approx T^{\nicefrac 34},
    \]
    by fine-tuning the precision $\eps\approx T^{-\nicefrac 14}$. Although the query complexity of \citet{CastiglioniLunghiMarchesi2026} is optimal up to logarithmic factors, this does not imply a $T^{\nicefrac 34}$ lower bound in the online setting. Indeed, it may be possible to quickly rule out prices that are clearly suboptimal and focus the exploration on ``good'' candidate prices. Indeed, we show that such fine-grained estimation is possible: we propose a localized estimation procedure which allows us to acquire useful information on possibly optimal prices by only querying prices which are not ``too'' suboptimal.

\subsection{Related work}
\label{sec:related}

\paragraph{Learning in bilateral trade.}
Regret minimization for bilateral trade has largely focused on gains from trade, with a range of feedback models, stochastic assumptions, and budget constraints~\citep{Cesa-BianchiCCF24,Cesa-BianchiCCF24jmlr,AzarFF24,BernasconiCCF24,LunghiC026}. Profit differs from gains from trade both in its objective and in the useful action space, {since} a broker benefits from a positive spread between the buyer and seller prices. Recent work studies profit maximization over general dominant-strategy incentive-compatible and individually rational mechanisms~\citep{GregorioDFS25,GregorioDFS26}, {with these mechanisms having a much richer allocation rule} and a different benchmark; {moreover, their} learning protocols can use reported valuations. We {instead} focus on fixed pairs of posted prices and on the single bit that reports whether the proposed trade succeeds. Accordingly, the rates are not directly comparable across these mechanism and feedback classes. For more details on online learning and bilateral trade, we refer to the survey by \citet{BernasconiF26}.

\paragraph{Learning Posted Prices in Bilateral Trade.}

\citet{CastiglioniLunghiMarchesi2026} study uniform approximation of multidimensional CDFs on finite uniform grids from one-bit queries, and are the first to explicitly mention online learning of posted prices for bilateral trade as a direct application of their results, although with a non-tight $O(T^{\nicefrac 34})$ rate.  We build on their use of monotonicity and shared probability estimates, but manage to achieve nearly optimal rates by \emph{localizing} the one-bit queries needed to estimate profit. 
In subsequent work, \citet{CastiglioniLunghi2026Breaking} obtain $\widetilde O(T^{\nicefrac{7}{10}})$ regret for a class of stochastic i.i.d.\ CDF-based objectives that includes posted prices in bilateral trade. Their approach is based on an initial coarse pruning phase, complemented by ideas from the feedback graph literature \citep{AlonCDK15}. 
Both \citet{CastiglioniLunghiMarchesi2026} and \citet{CastiglioniLunghi2026Breaking} only study the stochastic i.i.d. setting with possibly correlated seller and buyer valuations, and do not investigate the adversarial setting or MHR and regular distributions.

\paragraph{Pricing and distributional structure.}
One-sided online pricing is a classical connection between online learning and economics~\citep{BlumKRW03,KleinbergL03,Cesa-BianchiGM15}; see also \citet{Slivkins2019} for the bandit perspective. Pricing-query complexity studies how much can be learned when values are accessible only through acceptance decisions~\citep{LemeSTW23}. For regular distributions, \citet{SinglaW24} exploit half-concavity to obtain square-root regret up to logarithmic factors for sequential posted pricing. Our separation shows that applying this intuition independently on the two sides of a market is insufficient. Our MHR algorithm instead uses concavity of the logarithm of expected bilateral profit as a function of both prices.

\section{The Learning Model} \
\label{sec:model}

        In this paper, we study profit maximization in bilateral trade in an online learning setting. In particular, we focus our attention on the family of posted-price mechanisms. At each time step $t$, a new seller and buyer arrive with private valuations $(s_t,b_t)\in[0,1]^2$. The learner posts prices $(p_t,q_t)\in[0,1]^2$, offering $p_t$ to the seller and charging $q_t$ to the buyer. Trade occurs if $s_t\le p_t$ and $q_t\le b_t$, with ties resolved in favor of acceptance. The learner receives profit
        \[
            \prof_t(p_t,q_t) = (q_t - p_t) \cdot \ind{s_t \le p_t} \ind{q_t \le b_t}.
        \]

        We consider two data-generation models: the \emph{adversarial} setting, where an oblivious adversary fixes an arbitrary sequence $(s_1,b_1),\dots,(s_T,b_T)$ before play, and the \emph{stochastic} setting, where the pairs $(s_t,b_t)$ are sampled i.i.d. from a fixed unknown joint distribution $\cD$. Note that, in the latter model, independence across rounds does not require independence between the two agents in a round.
     
        In each round, the learner receives only the \emph{trade bit} $z_t=\ind{s_t\le p_t}\ind{q_t\le b_t}$. Its future prices may depend on its past prices, observed bits, and private randomness. We allow all pairs in $[0,1]^2$, including $p_t>q_t$, which may produce negative profit. Given an algorithm $\cA$ and a valuation-generating environment $\cS$, the corresponding regret is defined as        \[
            R_T(\cA,\cS) = \sup_{(p,q) \in [0,1]^2} \sum_{t=1}^T \E{\prof_t(p,q) - \prof_t(p_t,q_t)}.
        \]
        Note that the expectation is over the learner's randomness and, in the stochastic setting, the sampled valuations.
        An optimal comparator can always be chosen in the compact triangle $\cT=\{(p,q)\in[0,1]^2:p\le q\}$. Outside $\cT$, profit is nonpositive, while diagonal prices yield zero.
        On $\cT$, each realized profit function is bounded, nonnegative, and upper semicontinuous, so the comparator objective attains a maximum $(p^\star,q^\star)$ on $\cT$. For a class $\mathfrak S$ of environments, the worst-case regret is $\sup_{\cS\in\mathfrak S}R_T(\cA,\cS)$, and the minimax regret is its infimum over learning algorithms. 
        The learning goal is \emph{no regret}: a worst-case regret bound $o(T)$, so the expected shortfall from the benchmark, divided by $T$, vanishes.

        \begin{algorithm}[t]
        \caption*{\textbf{Learning Protocol for Repeated Bilateral Trade}}\label{a:learning-model}
        \begin{algorithmic}
            \For{$t=1,2,\ldots$}
                \State The environment supplies the next private valuation pair $(s_t,b_t)\in[0,1]^2$
                \State 
                 The learner posts prices $(p_t,q_t) \in [0,1]^2$ 
                 \State 
                 The learner observes $z_t = \ind{s_t \le p_t}\cdot \ind{q_t \le b_t}$ and is awarded $ \prof_t(p_t, q_t)\in [-1,1]$
            \EndFor
        \end{algorithmic}
            
        \end{algorithm} 

    \paragraph{
    Expected profit and distributional assumptions.} In the stochastic setting, we denote by $\mu$ the expected profit function, so that
        \(
            \mu(p,q) = \E{\prof(p,q)},
        \)
        where $\prof$ is the profit function when the underlying valuations are drawn according to $\cD$. Let $P(p,q)=\Prb_{(s,b)\sim\cD}(s\le p,b\ge q)$ be the trade probability and $\mu^\star=\mu(p^\star,q^\star)$ the optimal expected profit. Then
        \(
            \mu(p,q) = (q-p) \cdot P(p,q).
        \)
        In particular, the regret in the stochastic setting can equivalently be written as
        \[
            R_T(\cA,\cD)=T\mu^\star-\sum_{t=1}^T\E{\mu(p_t,q_t)}.
        \]
        
        Our improved stochastic guarantees concern independent seller and buyer distributions, $\cD=\cD_s\times\cD_b$, satisfying regularity or the monotone-hazard-rate property \citep{Myerson81,MyersonS83,HartlineR09}. For these two classes, assume strictly positive probability densities $\fs$ and $\fb$ on $[0,1]$, and let $\Fs$ and $\Fb$ denote their cumulative distribution functions. 
        \begin{definition}[Virtual Valuations]The seller's virtual cost $\psi$ and the buyer's virtual value $\phi$ are 
        \[
            \phi(b) = b - \frac{1-\Fb(b)}{\fb(b)}, \forall b \in [0,1], \quad \psi(s) = s + \frac{\Fs(s)}{\fs(s)}, \forall s \in [0,1].
        \]
        \end{definition}

        \begin{definition}[Regular Distribution]
            A seller, respectively buyer, distribution is \emph{regular} if $\psi$, respectively $\phi$, is nondecreasing on $[0,1]$. 
        \end{definition}

        \begin{definition}[MHR]
            A seller distribution satisfies the monotone hazard rate (MHR) property if its hazard rate  $\fs/\Fs$ is nonincreasing on $(0,1]$. A buyer distribution satisfies MHR if its hazard rate $\fb/(1-\Fb)$ is nondecreasing on $[0,1)$.
        \end{definition}
        These monotonicity conditions imply that $\psi$ and $\phi$ are nondecreasing, so MHR implies regularity.

\section{Technical Overview}

In this section, we provide a technical overview of our results, whose detailed proofs are deferred to the technical sections. We start by presenting in \Cref{subsec:stochastic_sketch} an algorithm which enjoys a $T^{\nicefrac 23}$ regret bound in the generic stochastic setting. While this result is subsumed by our adversarial algorithm with the same rate, its clean structure and analysis provide a first application of our ``localized sampling lemma'', which is then used as a building block for our adversarial result (see \Cref{subsec:adversarial_sketch}). We then {highlight} the $\sqrt{T}$ result for monotone-hazard-rate distributions (\Cref{subsec:MHR_sketch}) and finally the matching upper (\Cref{subsec:regular_upper_sketch}) and lower bounds (\Cref{subsec:regular_lower_sketch}) for regular distributions.
    
    \paragraph{A Simple Discretization Lemma.} Before moving to these results, we make a simple observation which follows directly from the one-sided Lipschitzness of the profit function, and is formally stated and proved as \Cref{lem:discretization_err} in \Cref{app:discretization}: uniform grids are a good discretization of the action space $[0,1]^2$. Indeed, for any integer $K$, consider the uniform grid $G_K= \{(\nicefrac{i}K,\nicefrac{j}{K}), \text{ for }i,j = 0,\dots, K\}$ of step-size $\nicefrac 1K$. For any valuation pair and prices $(p,q)\in\cT$, rounding to $(\lceil pK\rceil/K,\lfloor qK\rfloor/K)$ loses at most $2/K$ in profit; if the rounded seller price exceeds the rounded buyer price, use $(0,0)$ instead. This implies that all our algorithms may just play in the discretized action space $\AA = \cT \cap G_K$ (for suitable values of $K$), as long as they account for an overall discretization error of $\nicefrac{2T}K$ in the regret. Another useful consequence of this discretization result is that, if we fix $K$ upfront, we can imagine all agents' valuations to belong to $G_K$, at the cost of the same additive $\nicefrac{2T}K$ term in the regret.

\subsection{Warm-up: \texorpdfstring{$\widetilde O(T^{\nicefrac 23})$}{O-tilde(T\textasciicircum(2/3))} Regret for the Stochastic Setting}
\label{subsec:stochastic_sketch}
    We first sketch a $T^{\nicefrac 23}$ algorithm for the stochastic i.i.d. setting, in which the valuation pairs $(s_t,b_t)$ are drawn independently across rounds from a fixed, unknown distribution $\cD$. This result is less general than the one we prove in the adversarial setting in \Cref{sec:adversarial}, but already contains the crucial Localized Sampling Lemma (\Cref{lem:local_intro}) as an ingredient. In \Cref{subsec:adversarial_sketch} we sketch how to use the Localized Sampling Lemma in the adversarial setting.
    As discussed in Section~\ref{sec:challenges}, treating grid prices as unrelated arms or estimating all trade probabilities before committing gives only a $\widetilde O(T^{\nicefrac 34})$ guarantee. To improve it, exploration must become cheaper as estimates improve. In particular, we would like to estimate profits at promising price pairs using queries whose profit gaps are of the same order as the estimation accuracy. To formalize this intuition, let's introduce some notation: for a generic pair of prices $a \in \AA=G_K \cap \cT$ ($a$ stands for action) and a generic valuation distribution $\cQ$ over $[0,1]^2$, we denote by $\mu_{\cQ}(a)$ the expected profit obtained by posting $a$ when the underlying valuations are drawn according to $\cQ$. Similarly, we can define the suboptimality gap $\Delta_{\cQ}(a)$ of $a$ under $\cQ$, which, for any precision $\eta$, identifies the family of $\eta$-suboptimal actions $A_\eta(\cQ)$:
\[
    \Delta_{\cQ}(a)=\max_{a'\in\AA}\mu_{\cQ}(a')-\mu_{\cQ}(a),
    \qquad A_\eta(\cQ)=\{a\in\AA:\Delta_{\cQ}(a)\le\eta\}.
\]

To formally introduce our ``localized sampling lemma'', we need two valuation distributions $\cD$ and $\cD_0$. The former is the \emph{unknown} distribution that we need to learn from one-bit pricing queries, while the latter is a \emph{known} reference distribution. For now, we do not make any further assumption on these two distributions. 

We have all the ingredients to sketch our sampling routine and its properties. We refer to \Cref{lem:efficient_estimation} in \Cref{sec:adversarial} for the precise statement and proof.

\begin{mainresult}[Localized Sampling Lemma]
\label{lem:local_intro}
Fix $\AA=G_K \cap \cT$, a known reference distribution $\cD_0$, a precision $\eps\in(0,1]$, and a failure probability $\delta\in(0,1)$. There is a routine, depending only on these inputs, that for every unknown distribution $\cD$ uses at most
\[
    N(\eps,\delta)\in\widetilde O\!\left(\frac K{\eps^2}+\frac1{\eps^3}\right)
\]
one-bit pricing queries, all in $A_{2\eps}(\cD_0)$, and returns estimates $\widehat\mu(\cdot)$ satisfying, with probability at least $1-\delta$,
\[
    |\widehat\mu(a)-\mu_{\cD}(a)|\le\nicefrac{\eps}{64}
    \qquad\text{for every }a\in A_\eps(\cD_0).
\]
\end{mainresult}
For any precision $\eps$, the reference distribution $\cD_0$ then individuates two sets $A_\eps(\cD_0) \subseteq A_{2\eps}(\cD_0) \subseteq G_K$, while we would like to estimate up to precision $\eps$ all prices in $A_\eps(\cD_0)$ with respect to the unknown distribution $\cD$. This result tells us that we can do it using only $\tilde O(\nicefrac{1}{\eps^3}+\nicefrac{K}{\eps^2})$ one-bit pricing queries \emph{by only issuing these queries in $A_{2\eps}(\cD_0)$.} Stated differently, our \emph{target prices} are in $A_\eps(\cD_0)$, while our \emph{allowed queries} are in $A_{2\eps}(\cD_0)$.
Before providing a proof sketch of the sampling result, we showcase its applicability by arguing that it immediately induces an optimal algorithm for the stochastic setting.

\begin{warmupresult}[Stochastic Setting]
Consider the problem of learning the profit-maximizing posted prices in bilateral trade in the stochastic setting. There exists a learning algorithm which enjoys $\widetilde O(T^{\nicefrac 23})$ regret.
\end{warmupresult}

\begin{proof}[Proof sketch]
Set $K\approx T^{\nicefrac 13}$, $H\approx \log K$, and initialize precision scales $\eps_h=4^{-h}$ and failure probability $\delta_h=\nicefrac1{T^{2}H}$ for $h=0,\ldots,H$. The algorithm follows a successive-elimination approach and proceeds in phases. In the initial phase, it issues $\tilde \Theta(1)$ queries for each price in $G_K$, to obtain a uniform $\nicefrac 1{256}$-approximation of all profits. At the generic phase $h$, it chooses any valuation distribution $\cQ_h$ consistent with \emph{all} previous estimates and uses it as \emph{a reference} for the Localized Sampling Lemma (\Cref{lem:local_intro}) to estimate profits on $A_{\eps_h}(\cQ_h)$ with precision $\nicefrac{\eps_h}{64}$, \emph{only posting prices in} $A_{2\eps_h}(\cQ_h)$.

The crucial observation now is that, conditioning on the good event that all calls to the Localized Sampling Lemma and the initial concentrations yield correct estimates, in each phase $h$ the algorithm only plays prices in $A_{2\eps_h}(\cQ_h) \subseteq A_{C \cdot \eps_h}(\cD)$, for some constant $C$, thus suffering instantaneous regret which is $\tilde O(\eps_h)$ at each time step. Moreover, the clean event also tells us that ``good prices'' do not get discarded along the way, so that $A_{\eps_{h+1}}(\cQ_{h+1}) \supseteq  A_{c \cdot \eps_h}(\cD)$ for some constant $c \le C$.

Conditioning on the clean event (which, for our choice of the $\delta_h$, has probability at least $1-1/T$), we have the following bound on the regret (omitting poly-logarithmic terms):
\[
    \sum_{h=0}^{H}\underbrace{\eps_h \cdot N(\eps_h,\delta_h)}_{\text{Regret in Phase $h$}} \!\!\!+ \!\!\underbrace{\frac{T}{K}}_{\substack{\text{Discretization}\\ \text{Error}}}
   \!\!\!\!\!\! \approx \sum_{h=0}^{H}
        \left(\frac K{\eps_h}+\frac1{\eps_h^2}\right) + \frac{T}{K}\approx K^2 +\frac{T}{K} \approx T^{\nicefrac 23},
\]
where in the last step we set $K\approx T^{\nicefrac 13}$. The complement of the clean event gives constant expected regret, thus concluding the proof.
For simplicity, the full confidence-set construction and analysis appear in Appendix~\ref{sec:stochastic-warmup-proof}.
\end{proof}
\begin{proof}[Proof sketch of \Cref{lem:local_intro}]
Write $\mu_0=\mu_{\cD_0}$ for the expected profit under the known
reference distribution. Let $\cD_{0,s}$ and $\cD_{0,b}$ denote the marginal distributions of $\cD_0$ for the seller and buyer, respectively. With $\mathcal L$ denoting Lebesgue measure, define the auxiliary measure $\nu$ over $[0,1]^2$ as
\(
    \nu=\mathcal L\otimes\cD_{0,b}+\cD_{0,s}\otimes\mathcal L+\cD_0\), so that $\nu([0,1]^2)=3$.
This measure controls how the relative profit of two seller prices changes as the buyer price varies: for any $\ell\le r$ and $q_0\le q_1$, it indeed holds that
\begin{equation}
    \left|[\mu_0(r,q_1)-\mu_0(\ell,q_1)]
           -[\mu_0(r,q_0)-\mu_0(\ell,q_0)]\right|
    \le\nu((\ell,r]\times[q_0,q_1)).  \label{eq:reference_decomp}
\end{equation}
    \paragraph{A seller-price tree.} First, we can organize the seller grid $\{0,\nicefrac 1K, \nicefrac 2K, \dots, 1\}$ into a balanced binary tree, where each node corresponds to an interval: the root corresponds to $\{0,\dots,1\}$, its two children to $\{0,\dots,\nicefrac{1}{2}\}$, $\{\nicefrac{1}{2}+\nicefrac 1K,\dots,1\}$, and so on, up to the $K+1$ leaves which correspond to singletons of the form $\{\nicefrac{i}K\}$. For each node $I$ of the tree, we can use the auxiliary measure $\nu$ to partition the buyer grid into maximal consecutive blocks $D^I_1, D^I_2, \dots,$ satisfying $\nu((\min I,\max I]\times[\min D^I_j,\max D^I_j))\le\eps$. With this construction, each pair of the form $(I,D^I_j)$ corresponds to a rectangle $I\times D^I_j \subseteq [0,1]^2$, and can be associated with a representative pair of prices as follows: choose the smallest seller price $p_{(I,D^I_j)}\in I$ maximizing reference profit at $q_{(I,D^I_j)}=\min D^I_j$. Inequality~\eqref{eq:reference_decomp} ensures that this seller stays within $\eps$ of the best seller in $I$ throughout $D^I_j$. Thus every $\eps$-optimal target price pair $(p,q)\in I\times D^I_j$ has a $2\eps$-optimal queried pair $(p_{(I,D^I_j)},q)$. Figure~\ref{fig:localized-overview} shows a target's seller-tree path, from $I_0 = \{0,\nicefrac 1K, \nicefrac 2K, \dots, 1\}$ to $\{p\}$, along with the associated buyer blocks.

\begin{figure}[t]
        \centering
        \begingroup
        \definecolor{SellerBlue}{RGB}{57,91,124}
        \definecolor{BuyerAmber}{RGB}{160,111,40}
        \definecolor{PartitionGrey}{RGB}{167,170,173}
        
        \begin{tikzpicture}[
          x=1cm,y=1cm,
          line cap=round,line join=round,
          every node/.style={inner sep=1pt,outer sep=0pt},
          font=\fontsize{10}{12}\selectfont,
          partition/.style={draw=PartitionGrey,line width=.45pt},
          buyer partition/.style={
            draw=PartitionGrey,line width=.5pt,
            dash pattern=on 2pt off 1.7pt
          },
          projection/.style={
            draw=black!38,line width=.45pt,
            dash pattern=on .8pt off 1.8pt
          },
          path arrow/.style={draw=black!48,line width=.65pt,->,>=stealth}
        ]

        \newcommand{\PricePanel}[9]{%
          \begin{scope}[shift={(#1,#2)},x=2.4cm,y=2.4cm]
            \fill[white] (0,0) rectangle (1,1);
        
            \fill[SellerBlue!10] (#4,0) rectangle (#5,1);
            \ifnum#3>0
              \draw[partition] (.5,0) -- (.5,1);
            \fi
            \ifnum#3>1
              \draw[partition] (.25,0) -- (.25,1);
              \draw[partition] (.75,0) -- (.75,1);
            \fi
        
            \fill[BuyerAmber!25] (#4,#6) rectangle (#5,#7);
            \ifnum#3=0
              \foreach \cut in {.15625,.3125,.46875,.625,.78125,.9375}
                \draw[buyer partition] (#4,\cut) -- (#5,\cut);
            \fi
            \ifnum#3=1
              \foreach \cut in {.28125,.5625,.84375}
                \draw[buyer partition] (#4,\cut) -- (#5,\cut);
            \fi
            \ifnum#3=2
              \draw[buyer partition] (#4,.59375) -- (#5,.59375);
            \fi
            \draw[SellerBlue!80,line width=.65pt]
              (#4,0) rectangle (#5,1);
            \draw[BuyerAmber,line width=.8pt]
              (#4,#6) rectangle (#5,#7);
        
            \draw[projection] (0,\TargetQ) -- (\TargetP,\TargetQ);
            \draw[projection] (\TargetP,0) -- (\TargetP,\TargetQ);
            \fill[white] (\TargetP,\TargetQ) circle[radius=3.25pt];
            \fill[black] (\TargetP,\TargetQ) circle[radius=1.9pt];
        
            \draw[black!68,line width=.65pt] (0,0) rectangle (1,1);
            \draw[black!68,line width=.55pt]
              (\TargetP,0) -- (\TargetP,-.025);
            \draw[black!68,line width=.55pt]
              (0,\TargetQ) -- (-.025,\TargetQ);
            \node[anchor=north] at (\TargetP,-.036) {$\TargetPLabel$};
            \node[anchor=east] at (-.036,\TargetQ) {$\TargetQLabel$};
        
            \draw[SellerBlue,line width=.85pt]
              (#4,1.035) -- (#4,1.06) -- (#5,1.06) -- (#5,1.035);
            \node[text=SellerBlue,anchor=south]
              at ({(#4+#5)/2},1.075) {$#8$};
            \draw[BuyerAmber,line width=.85pt]
              (1.025,#6) -- (1.05,#6)
              -- (1.05,#7) -- (1.025,#7);
            \node[text=BuyerAmber,anchor=west,fill=white,inner sep=1.4pt]
              at (1.065,{(#6+#7)/2}) {$#9$};
          \end{scope}%
        }
        
        \node[text=black!68] at (1.2,6.83) {depth $0$ (root)};
        \node[text=black!68] at (5.4,6.83) {depth $1$};
        \node[text=black!68] at (9.6,6.83) {depth $2$};
        
        \node[anchor=east,align=center] at (-.65,4.85) {Target\\[2pt]$(p,q)$};
        \node[anchor=east,align=center] at (-.65,1.2) {Target\\[2pt]$(p',q')$};
        
        \def\TargetP{.328125}
        \def\TargetQ{.703125}
        \def\TargetPLabel{p}
        \def\TargetQLabel{q}
        \PricePanel{0}{3.65}{0}{0}{1}{.625}{.78125}{I_0}{D_0}
        \PricePanel{4.2}{3.65}{1}{0}{.5}{.5625}{.84375}{I_1}{D_1}
        \PricePanel{8.4}{3.65}{2}{.25}{.5}{.59375}{1}{I_2}{D_2}
        
        \def\TargetP{.671875}
        \def\TargetQ{.890625}
        \def\TargetPLabel{p'}
        \def\TargetQLabel{q'}
        \PricePanel{0}{0}{0}{0}{1}{.78125}{.9375}{I_0}{D'_0}
        \PricePanel{4.2}{0}{1}{.5}{1}{.84375}{1}{I'_1}{D'_1}
        \PricePanel{8.4}{0}{2}{.5}{.75}{.59375}{1}{I'_2}{D'_2}
        
        \foreach \cy in {4.85,1.2} {
          \draw[path arrow] (2.85,\cy) -- (3.55,\cy);
          \draw[path arrow] (7.05,\cy) -- (7.75,\cy);
          \draw[black!48,line width=.65pt] (11.15,\cy) -- (11.43,\cy);
          \node[text=black!48] at (11.75,\cy) {$\cdots$};
          \draw[path arrow] (12.07,\cy) -- (12.40,\cy);
        }
        \node[anchor=west] at (12.55,4.85) {$I_h=\{p\}$};
        \node[anchor=west] at (12.55,1.2) {$I'_h=\{p'\}$};
        
        \begin{scope}[shift={(1.9,-.80)}]
          \filldraw[fill=SellerBlue!10,draw=SellerBlue,line width=.65pt]
            (0,-.105) rectangle (.13,.105);
          \node[anchor=west] at (.25,0) {seller interval};
          \filldraw[fill=BuyerAmber!25,draw=BuyerAmber,line width=.65pt]
            (3.3,-.065) rectangle (3.64,.065);
          \node[anchor=west] at (3.76,0) {buyer block};
          \fill[black] (6.9,0) circle[radius=1.9pt];
          \node[anchor=west] at (7.08,0) {target};
        \end{scope}
        
        \end{tikzpicture}
        \endgroup
        \caption{A possible sequence of seller and buyer blocks involved in the estimation of the probability function of trade for two specific targets in $\AA$, according to \Cref{eq:telescoping_overview}.}
        \label{fig:localized-overview}
    \end{figure}
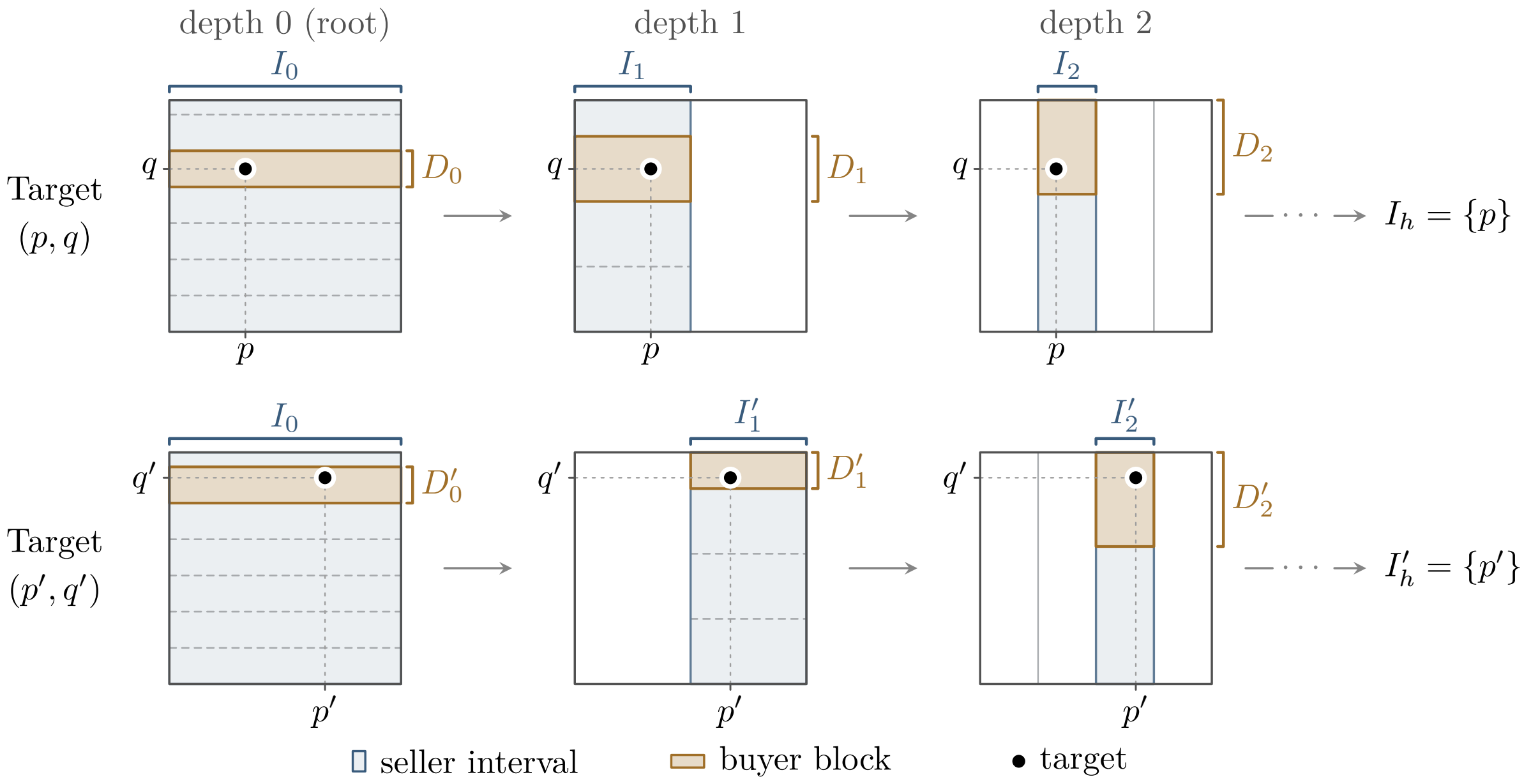

\paragraph{Sharing estimates.}
Let $P$ be the unknown probability function of both seller and buyer accepting a trade. For a target $(p,q)$ in the action space $\AA=G_K \cap \cT$, follow its seller-tree path to the leaf $\{p\}$, and let $p_j$ be the representative seller in the block containing $q$ at depth $j$. We have a convenient telescoping decomposition
\begin{equation}\label{eq:telescoping_overview}
 P(p,q)=\underbrace{P(p_0,q)}_{\text{root term}}
 +\sum_{j=1}^{h}
   \underbrace{\bigl[P(p_j,q)-P(p_{j-1},q)\bigr]}_
              {\text{parent--child function}}.
\end{equation}

Each root buyer block defines one function $q\mapsto P(p_0,q)$. For each parent--child edge, consider the overlap $D_{j-1}\cap D_j$ of the buyer blocks illustrated in Figure~\ref{fig:localized-overview}: both representatives are fixed on this overlap, so it defines one function $q\mapsto P(p_j,q)-P(p_{j-1},q)$.
We call these \emph{shared functions}: targets using the same root block or the same parent--child block overlap reuse the same estimates. Each difference is a signed seller-strip probability, monotone in $q$, and can be observed without bias using two fresh trade bits. 
When estimating a shared function, retain only buyer coordinates $q$ for which some seller price $p$ in the child interval satisfies $(p,q)\in A_\eps(\cD_0)$.
This target also belongs to the parent interval, so the representative guarantee makes both queried pairs $2\eps$-optimal under $\cD_0$.
Apply the analogous restriction at the root. Thus all queries lie in the permitted set, and every desired target
can still be reconstructed using the telescoping decomposition in Equation~\eqref{eq:telescoping_overview}.

\paragraph{Controlling the sample count.}
The constant mass of $\nu$ bounds the total number of blocks, and hence the number of shared functions, by $M=O(K+\eps^{-1}\log K)$. So far, we have only used explicitly the reference distribution $\cD_0$, while $\cD$ has only been used in the telescopic decomposition. Let $W$ denote the sum of the variations of \emph{all} shared and root functions induced by the decomposition, where the variation of a monotone function is just the absolute difference between its first and
last values.   Under $\cD$, the sum of variations $W$ belongs to $O(\log K)$: at each tree depth, each unit of probability mass contributes to at most two child corrections. Thus the reference law controls query profitability and the number of functions, while the unknown law controls their total variation.
Monotonicity allows us to estimate a given shared function by sampling endpoints and refining only intervals with appreciable variation. For $M$ functions of total variation $W$, this costs $\widetilde O(\nicefrac{M}{\alpha^2}+\nicefrac{W}{\alpha^3})$ observations at accuracy $\alpha$ (\Cref{cl:monotone_estimation}). Taking $\alpha\in\Theta(\nicefrac{\eps}{\log K})$ small enough makes the errors along every path sum to at most $\nicefrac{\eps}{64}$, at the claimed cost. Multiplication by the known spread $q-p\in[0,1]$ gives the profit estimates. The full construction is in \Cref{lem:efficient_estimation}; the smaller error constant and confidence $1-\delta$ are obtained by increasing sampling precision, without changing query locality.
\end{proof}

    \subsection{\texorpdfstring{$\widetilde O(T^{\nicefrac 23})$}{O-tilde(T\textasciicircum(2/3))} Regret for the Adversarial Setting}
    \label{subsec:adversarial_sketch}
    In this section, we sketch the adversarial $\widetilde O(T^{\nicefrac 23})$ algorithm, whose full description and analysis are deferred to \Cref{sec:adversarial}. Note that here the sequence of valuations $(s_1,b_1),\ldots,(s_T,b_T)$ is fixed in advance, so that there is no stationary distribution to learn. Instead, our algorithm maintains weights on possible prices to compare against, uses them to choose the price to post, and updates them using each trade bit. We describe how these weights are computed. To this end, we first introduce an auxiliary  one-round game, between the learner and an adversary. Then, we analyze this game using the Localized Sampling Lemma, and finally we show how the resulting insight can be used to update the actual adversarial algorithm.

\paragraph{An Auxiliary One-Round Game.} We present here a one-round game between the learner and an adversary. This adversary chooses a valuation pair in $G_K$ and a comparator $a' \in \AA$, while the learner receives in input probability weights $w$ on $\AA$ and makes a \emph{proposal}: it chooses a posted price $a \in \AA$ and two possible update vectors $Q_0,Q_1$, one for each possible feedback it may receive. Observing the trade bit $z$ then sets the next weights to $Q_z$. We denote with $z_a(v)$ the trade bit for valuation $v$ and action $a$. The goal of the learner is to make a randomized proposal which respects the following inequality, uniformly over all possible choices of the adversary: 
\begin{equation}
\label{eq:one_round_guarantee_intro}
\max_{v \in G_K}\max_{a' \in \AA}\E{\prof(v,a')-\prof(v,a)
       -\beta\log\frac{Q_{z_a(v)}(a')}{w(a')}}
 \le \frac{c}{K},
\end{equation}
with $c$ universal constant and $\beta \in \tilde{\Theta}(K^2)$.  Note, the left-hand-side of the inequality is a variant of the Algorithmic Information Ratio from \citet{XuZ25}, while the expectation is only over the learner's randomized proposal. 

\paragraph{The Information Gain bounds the Logarithmic Term.} To show that there always exists a learner proposal respecting \Cref{eq:one_round_guarantee_intro}, we adopt a dual perspective on the game. There, the adversary first chooses a distribution $\rho$ over valuation--comparator pairs in $G_K\times\AA$, and the learner responds with a deterministic proposal $(a,Q_0,Q_1)$ that minimizes the expected profit gap minus the weighted logarithmic update in \eqref{eq:one_round_guarantee_intro}, evaluated under $\rho$. 

Let's focus on the logarithmic term in the objective. For a fixed action $a$, consider the ideal posterior updates $Q_z(a')=\Prb_\rho(A=a'\mid z_a(V)=z)$. Thus $Q_z$ is the conditional distribution of the comparator after observing feedback $z$. For this choice,
\begin{equation}
\label{eq:one_round_information_intro}
 \Esub{(V,A)\sim\rho}{\log\frac{Q_{z_a(V)}(A)}{w(A)}}
 =\underbrace{I_\rho(A;z_a(V))}_{\cI_\rho(a)}
   +\KL{\rho_A}{w}
 \ge\cI_\rho(a),
\end{equation}
where $\rho_A$ is the marginal distribution of $A$, and $\cI_\rho(a)$ is the mutual information between $A$ and the feedback bit $z_a(V)$ under $\rho$.
Crucially, the inequality in \eqref{eq:one_round_information_intro} shows that mutual information measures a part of the expected logarithmic weight increase. It quantifies how much the trade bit distinguishes the possible comparison prices under this auxiliary law. All in all, to argue that \Cref{eq:one_round_guarantee_intro} holds, it remains to show that for any distribution $\rho$, there exists $a \in \AA$ such that
\(
    \Delta_\rho(a)-\beta\cI_\rho(a) \le \nicefrac{c}{K},
\)
where the quantity being bounded is the parameterized information ratio from \citet{FosterRSS22} and $\Delta_\rho(a)$ is a shorthand for $\Esub{\rho}{\prof(V,A)-\prof(V,a)}$. 
Indeed, here we are essentially using a known relationship between the parameterized and the algorithmic information ratios (see Lemma 2.7 in \citet{XuZ25}).

\paragraph{Inexpensive Detection by Local Sampling.}
We first show how to detect an increase in optimal profit using queries that are inexpensive under a reference distribution. Fix $\rho$ and let $\cD_0$ be its marginal valuation distribution. We supply $\cD_0$ as the known reference to the Localized Sampling Lemma. For an improvement threshold $x\in(0,1]$, our goal is to distinguish $\cD_0$ from every distribution $\cD$ satisfying
\[
 \max_{a\in\AA}\mu_{\cD}(a)
 \ge \max_{a\in\AA}\mu_{\cD_0}(a)+x.
\]

Apply \Cref{lem:local_intro} at precisions $x,2x,4x,\ldots$, capped at $1$, with failure probability $1/10$ per call. The threshold $x$ remains fixed throughout; only the sampling precision increases. The first call estimates profits on $A_x(\cD_0)$ using queries in $A_{2x}(\cD_0)$, and subsequent calls use the corresponding larger sets. In every call, the reference is $\cD_0$. When feedback comes from $\cD_0$, the estimates approximate $\mu_{\cD_0}$; when feedback comes from $\cD$, they approximate $\mu_{\cD}$. We compare these two executions with the same reference, target sets, and querying rules.
Consider all sampling experiments with feedback generated from $\cD_0$ and randomized as in \Cref{rem:range_extension}, and define
\[
 n_x(a)=2\Esub{\cD_0}{\text{total queries at }a\text{ across all precision scales}}.
\]
We want to bound $\sum_a n_x(a)(\Delta_{\cD_0}(a)+x)$. By linearity of expectation, this is twice the expected sum of $\Delta_{\cD_0}(a)+x$ over all queries. We bound these contributions one call at a time. In the first call, \Cref{lem:local_intro} guarantees that every queried price has reference gap at most $2x$, so each contributes at most $3x$. More generally, the additional term remains $x$, and each query contributes at most three times its call's precision. Multiplying the first call's sample bound from \Cref{lem:local_intro} by $3x$ gives, with logarithmic factors as in its full proof,
\[
 3x\cdot O\!\left(\log^3K
       \left(\frac K{x^2}+\frac{\log^3K}{x^3}\right)\right)
 =
 O\!\left(\log^3K
       \left(\frac Kx+\frac{\log^3K}{x^2}\right)\right).
\]
The corresponding bounds at the larger precisions decrease geometrically. We obtain
\[
 \sum_{a\in\AA}n_x(a)\bigl(\Delta_{\cD_0}(a)+x\bigr)
 \le C\log^3K\left(\frac Kx+\frac{\log^3K}{x^2}\right).
\]
Because the expected query counts are taken under $\cD_0$, these weights depend only on $\cD_0$ and $x$, and are the same for every alternative distribution.

Next, fix an improving distribution $\cD$ and let $a'$ maximize its grid profit. Although $a'$ is optimal under $\cD$, it may lie outside $A_x(\cD_0)$, where the first call guarantees accurate estimates. We therefore choose the first call whose precision is at least $\Delta_{\cD_0}(a')$, so that its target set includes $a'$. The improvement assumption gives
\[
 \mu_{\cD}(a')-\mu_{\cD_0}(a')
 \ge x+\Delta_{\cD_0}(a').
\]
Because we choose the first such call, this increase exceeds twice its estimation error. By \Cref{lem:local_intro}, the returned estimate is close to $\mu_{\cD_0}(a')$ when feedback comes from $\cD_0$, and close to $\mu_{\cD}(a')$ when feedback comes from $\cD$, each with probability at least $\nicefrac9{10}$. These two accuracy intervals are disjoint. Hence the event that the estimate lies within the prescribed error of $\mu_{\cD_0}(a')$ has probability at least $\nicefrac9{10}$ under $\cD_0$, but at most $\nicefrac1{10}$ under $\cD$.

Write $P_0$ and $P$ for the trade probabilities under $\cD_0$ and $\cD$. Since the returned estimates are built from trade observations, their different behavior must come from differences in trade probabilities at the sampled prices. The proof quantifies these differences, giving greater weight to prices queried more often under $\cD_0$, and obtains
\[
 \sum_{a\in\AA}n_x(a)\bigl(P(a)-P_0(a)\bigr)^2\ge1.
\]
Together with the sampling-cost bound, this proves \Cref{lem:inexpensive_detection} in \Cref{sec:adversarial}: the same inexpensive weights detect every distribution whose optimal profit exceeds the reference optimum by at least $x$.

\paragraph{From Detection to the Information Inequality.}
We seek a price whose information gain compensates for its regret. Let $a^\star$ maximize profit under $\cD_0$, and let $X_A$ be the positive increase in optimal grid profit after conditioning on $A$: $X_A=\max_a(\Esub{\rho}{\prof(
         V,a)|A} - \Esub{\rho}{\prof(V,a^\star)})_+$. Since the conditional optimum is at least the profit of playing $A$ itself,
\[
 \Delta_\rho(a^\star)\le\Esub{A}{X_A}.
\]
Thus a large comparator advantage requires substantial conditional profit improvements. Detection forces these improvements to produce differences in trade probabilities. Averaging over $A$ with the same sampling weights, and applying Pinsker's inequality, gives
\[
 \Prb(X_A\ge x)\le\frac12\sum_{a\in\AA}n_x(a)\cI_\rho(a).
\]

Suppose every action violates the desired bound, so $\Delta_\rho(a)-\beta\cI_\rho(a)>c/K$. Then $\cI_\rho(a)<\Delta_\rho(a)/\beta$: information is small relative to regret at every price. Since $\Delta_\rho(a)=\Delta_{\cD_0}(a)+\Delta_\rho(a^\star)$, for $x\ge\Delta_\rho(a^\star)/2$ the sampling charge $\Delta_{\cD_0}(a)+x$ controls $\Delta_\rho(a)$ within a factor of two. Integrating the resulting tail bound above this cutoff, and using probability at most one below it, gives $\Esub{A}{X_A}<\Delta_\rho(a^\star)$ for $\beta=K^2\log^6K$ and sufficiently large $c$, a contradiction. Hence
\[
 \min_{a\in\AA}\bigl\{\Delta_\rho(a)-\beta\cI_\rho(a)\bigr\}\le\frac cK.
\]

\paragraph{Final Regret Bound.}
At this point, we have all the ingredients to present a sketch of the result for the adversarial setting. Note, the one-round game we have analyzed so far is just an analytic tool used in guiding the design and analysis of our algorithm, but it is not (directly) how the algorithm itself works.

\begin{mainresult}[Adversarial Setting]
\label{thm:adversarial-overview}
Consider the problem of learning the profit-maximizing posted prices in bilateral trade in the adversarial setting. There exists a learning algorithm which enjoys $\widetilde O(T^{\nicefrac 23})$ regret.
\end{mainresult}
\begin{proof}[Proof Sketch]
    In the adversarial setting there is no notion of distribution $\rho$ over valuation-comparator pairs which we can optimize against. However, we can maintain the bound in \Cref{eq:one_round_guarantee_intro} by using multiplicative weight updates~\citep{AroraHazanKale2012} to find a good (randomized) action/price against any possible adversarial response (see \Cref{lem:adv-oneround} for the formal statement). The adversarial algorithm maintains weights $w_t$ over the prices in the action space $\AA$, starting from uniform at time $1$. At each time step, it calls the multiplicative weight updates routine underlying the auxiliary one-round game with weights $w_t$, in order to obtain a randomized proposal $(a_t,Q_0,Q_1)$. It then posts $a_t$, observes feedback $z_t$, and sets $w_{t+1}=Q_{z_t}$. Fix now any comparator action $a' \in \AA$. For any time step $t$, and any fixed history up to that point, the bound in \Cref{eq:one_round_guarantee_intro} thus can be rewritten as: %
    \[
        \E{\prof_t(a')-\prof_t(a_t)} \le \beta\,\E{\log\frac{w_{t+1}(a')}{w_t(a')}}+\frac cK.
    \]
    Both expectations are over the proposal's randomness, conditional on the current history. By summing up the contributions of each time step, and by the law of total probability, we then get that:
    \begin{align*}
        \sum_{t=1}^T\E{\prof_t(a')-\prof_t(a_t)} &\lesssim \beta \E{\sum_{t=1}^T\log\frac{w_{t+1}(a')}{w_t(a')}} + \frac{T}{K}\\
        &= \beta \E{\log\frac{w_{T+1}(a')}{w_1(a')}}+ \frac{T}{K} \tag{Telescopic sum}\\
        &\le \beta\log |\AA| + \frac{T}{K},
    \end{align*}
    where the last inequality follows from the uniform initialization of $w_1$, and $w_{T+1}(a') \le 1$ for all $a'$.
Consequently, even the best price in hindsight has expected cumulative advantage at most $O(\nicefrac{T}K)+\beta\log|\AA|$. Adding discretization and taking $K\approx T^{1/3}$ gives $R_T=\widetilde O(T^{\nicefrac 23})$
\end{proof}

\subsection{\texorpdfstring{$\widetilde O(\sqrt{T})$}{O-tilde(sqrt(T))} Regret for Monotone-Hazard-Rate Distributions}
    \label{subsec:MHR_sketch}

        We present and sketch the analysis of our learning algorithm for the special case of independent and MHR distributions. A formal presentation and regret analysis can be found in the main body, in \Cref{sec:MHR}. Crucially, our approach is based on a stochastic convex bandit routine \citep{Lattimore021}, applied to a suitable proxy for the profit objective, and employs an initial filtering procedure to ``well condition'' the convex problem. We start by introducing the stochastic convex bandit problem.

        \paragraph{Stochastic Convex Bandits.} In stochastic convex bandits, a learner is presented with a convex set $\cK \subseteq [0,1]^2$ and a number $n$ of interactions. For each one of these interactions, the learner selects a point $x_\tau \in \cK$ and suffers a loss $\cL(x_\tau)$, while only observing noisy feedback $\cL(x_\tau) + \xi_\tau$. The (unknown) function $\cL$ is continuous and convex, while the noise variables $\xi_1, \xi_2,\dots$ form a sequence of conditionally centered random variables (i.e., $\E{\xi_\tau| \cF_\tau} = 0$) with {bounded $\psi_2$-norm $\lVert\xi_\tau\rVert_{\psi_2} = \inf\{t>0:{\E{\exp{\nicefrac{\xi_\tau^2}{t^2}}|\cF_\tau}<2}
\}$}\footnote{The generic $\cF_{\tau}$ belongs to the natural filtration of the process and represents the $\sigma$-algebra generated by the sequence of points $x_1,\dots, x_\tau$ and the previous noises $\xi_1, \dots, \xi_{\tau-1}$.}. For this problem, we know that there exists a learning algorithm which enjoys the following properties, as an immediate corollary of Theorem 1 of \citet{Lattimore021} (for a proof, we refer to Appendix \ref{app:mhr}).
        
        \begin{restatable}[Stochastic convex bandit]{corollary}{convexmhrroutine}
        \label{prop:mhr-convex-bandit}
        Let $\cC$ be a known convex body of constant dimension, and let $L:\cC\to\mathbb R$ be continuous and convex, with values in a known interval of constant length. Suppose that querying $x_\tau\in\cC$ returns $L(x_\tau)+\xi_\tau$. Conditional on previous queries and observations, and on $x_\tau$, assume that $\xi_\tau$ has mean zero and $\psi_2$ norm at most a known $\gamma\ge0$. Then, for every $n\ge1$, there is an $n$-query algorithm such that
        \[
            \mathbb E\left[\sum_{\tau=1}^n
                \left(L(x_\tau)-\min_{x\in\cC}L(x)\right)\right]
            \in O\!\left((1+\gamma)\sqrt n\,\log^2(n)\right),
        \]
        where the expectation is taken over noise and randomness of the algorithm's actions.
        The hidden constant depends only on the affine dimension and the length of the specified interval.
        \end{restatable}
       
        To translate bilateral trade into the language of stochastic convex optimization, we need to find a convex loss which relates to the expected profit and a suitable $\gamma$!

        \paragraph{A Convex Objective.} While the expected profit function $\mu(p,q)$ is critically not concave or convex, the MHR condition implies that the $\log$ of the expected profit exhibits concavity. Indeed, we have the following claim, whose proof is deferred to \Cref{sec:MHR}:
        \begin{restatable}{proposition}{convex}
        \label{prop:conv_mhr}
            Under the MHR assumption, the loss function $L(p,q) = -\log \mu(p,q)$ is continuous and convex in the interior of $\cT$.
        \end{restatable}        
        The loss $L(p,q)$ is convex, so we can flip the problem from maximizing profit to minimizing this notion of loss. If we look at the bandit convex optimization setup, we still have two critical gaps to fill. First, this $L$ has the disadvantage of being unbounded on its natural domain, as it diverges close to the boundary of $\cT$. Moreover, it is not clear how to use the actual profit obtained by posting prices $(p,q)$ to estimate the corresponding loss: the trivial estimator $-\log [\ind{\text{trade}}(q-p)]$ is clearly out of the question.

        \paragraph{A filtering procedure.} To control the range of the loss function $L$, we start by pruning the action space $\cT$ with a first coarse uniform estimate of the profit function $\mu$. We can safely restrict our attention to the prices in the uniform grid $G_K$ of stepsize $2^{-H} = \nicefrac 1K$ that belong to the triangle $\cT$ (we know that this only costs us an extra $\nicefrac TK$ term in the regret by one-sided Lipschitzianity of the profit function, as in \Cref{lem:discretization_err}), and then proceed through decreasing guesses $1, \nicefrac 12, \nicefrac 14, \dots, 2^{-H}$ of the actual range of the optimal profit $\mu^\star$. For the generic guess $\eps_h=2^{-h}$, the filtering procedure probes $\tilde O(K)$ strategic prices, each one approximately $\nicefrac{1}{\eps_h}$ times, which are then used to provide estimates for \emph{all the prices} in $G_K \cap \cT$. Note that $\nicefrac{1}{\eps_h}$ samples are not enough to guarantee an \emph{additive} precision of $\eps_h$, but suffice for a \emph{constant multiplicative} approximation as soon as the expected profit is approximately $\eps_h$. If there exists at least one price on the grid whose current estimate $\hat \mu(p,q)$ is safely comparable to the current guess $\eps_h$, then the routine terminates and returns the convex hull $\cK$ of all the prices on the grid which are ``close'' to the guess. Note that, by convexity of $L$, we are sure that all prices in the convex hull are comparable to $\eps_h$.
        
        If all the guesses fail up to level $H$, corresponding to precision $\nicefrac1K$, then on the concentration event we have $\mu^\star=O(1/K)$. By contrast, if the filtering procedure terminates at the $h$-th precision, {i.e., it triggers}, then the regret incurred in this first phase is at most $O(\eps_{h})$ times the number of probes exploited so far. Denoting by $h^\star$ this last phase, we can then upper bound the regret suffered in this filtering phase as follows:
        \[
            \sum_{h=1}^{h^\star} \eps_{h^\star} \cdot \frac{K}{\eps_h} = K \cdot 2^{-h^\star} \cdot \sum_{h=1}^{h^\star} 2^{h} \approx K. 
        \]
        The properties of the filtering procedure are summarized in the following qualitative lemma. For the exact statement and proof, we refer to \Cref{lem:mhr-budget}.

        \begin{lemma}
        \label{lem:qual_MHR}
        {There exist constants $0<c<C$ such that, with high probability,
        the filtering procedure either does not trigger and
        $\mu^\star\leq \nicefrac CK$, or returns $\eps_{h^\star}=2^{-h^\star}\geq \nicefrac 1K$ and a convex body
        $\cK\subseteq\operatorname{int}(\cT)$ satisfying}:
        \begin{itemize}
            \item[(i)] Every price $(p,q)\in\cK$ has expected profit $\mu(p,q)\geq c\cdot\eps_{h^\star}$.
        
            \item[(ii)] The returned scale approximates the optimal profit: 
                $\eps_{h^\star}\leq\mu^\star\leq C\cdot\eps_{h^\star}$.
        
            \item[(iii)] The set $\cK$ contains every maximizer of $\mu$ over
            $G_K\cap\cT$ and therefore $
                \mu^\star
                \leq \max_{(p,q)\in\cK}\mu(p,q)+\nicefrac{C}{K}$.
        \end{itemize}
        {On the same high-probability event, if the procedure triggers, the regret
        incurred during filtering is $\tilde O(K)$.}
        \end{lemma}

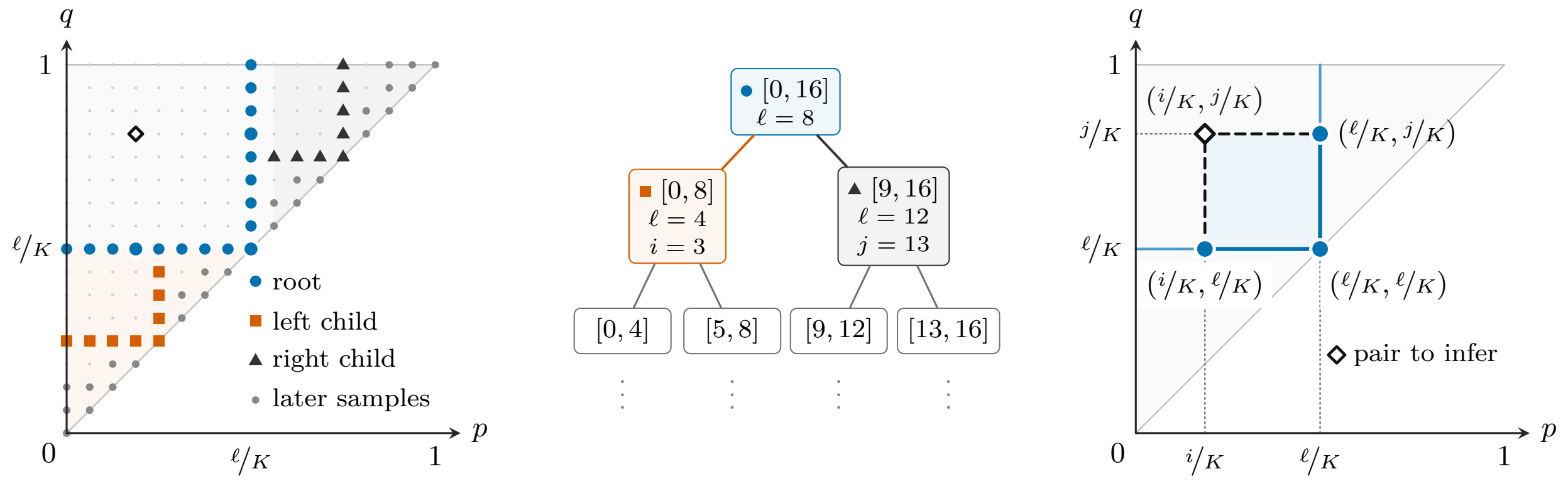
\begin{figure}[t]
\centering
\begingroup
\definecolor{mhrsample}{HTML}{0072B2}
\definecolor{mhrleft}{HTML}{D55E00}
\definecolor{mhrright}{HTML}{333333}
\definecolor{mhrtarget}{HTML}{111111}
\definecolor{mhrmuted}{HTML}{888888}
\resizebox{\linewidth}{!}{%
\begin{tikzpicture}[font=\small, line cap=round, line join=round, >=stealth,
  mhr axis/.style={->, draw=black!85, line width=.6pt},
  mhr guide/.style={draw=black!50, densely dotted, line width=.5pt},
  mhr branch/.style={draw=black!55, line width=.6pt},
  mhr tree node/.style={draw=black!55, fill=white, rounded corners=2pt, minimum width=1.05cm, inner xsep=3pt, inner ysep=3pt, font=\footnotesize, align=center},
  mhr active node/.style={mhr tree node, draw=mhrsample, fill=mhrsample!6}]

\begin{scope}[x=.25cm,y=.25cm]
  \fill[black!2] (0,0) -- (0,16) -- (16,16) -- cycle;
  \fill[mhrleft!6] (0,0) -- (0,8) -- (8,8) -- cycle;
  \fill[mhrright!6] (9,9) -- (9,16) -- (16,16) -- cycle;
  \draw[black!25, line width=.45pt] (0,16) -- (16,16) -- (0,0);

  \foreach \i in {0,...,16} {
    \foreach \j in {\i,...,16} {
      \fill[black!18] (\i,\j) circle[radius=.48pt];
    }
  }

  \foreach \u/\v/\ell in {0/16/8,0/8/4,9/16/12,0/4/2,5/8/6,9/12/10,13/16/14,0/2/1,3/4/3,5/6/5,7/8/7,9/10/9,11/12/11,13/14/13,15/16/15,0/1/0} {
    \foreach \i in {\u,...,\ell} {
      \fill[mhrmuted] (\i,\ell) circle[radius=1.15pt];
    }
    \pgfmathtruncatemacro{\nextindex}{\ell+1}
    \foreach \j in {\nextindex,...,\v} {
      \fill[mhrmuted] (\ell,\j) circle[radius=1.15pt];
    }
  }
  \foreach \i in {0,...,16} {
    \fill[mhrmuted] (\i,\i) circle[radius=1.15pt];
  }

  \foreach \i in {0,...,4} {
    \path[fill=mhrleft] (\i,4) ++(-1.7pt,-1.7pt) rectangle ++(3.4pt,3.4pt);
  }
  \foreach \j in {5,...,7} {
    \path[fill=mhrleft] (4,\j) ++(-1.7pt,-1.7pt) rectangle ++(3.4pt,3.4pt);
  }
  \foreach \i in {9,...,12} {
    \path[fill=mhrright] (\i,12) ++(0,2.3pt)
      -- ++(-2.1pt,-3.7pt) -- ++(4.2pt,0) -- cycle;
  }
  \foreach \j in {13,...,16} {
    \path[fill=mhrright] (12,\j) ++(0,2.3pt)
      -- ++(-2.1pt,-3.7pt) -- ++(4.2pt,0) -- cycle;
  }
  \foreach \i in {0,...,8} {
    \fill[mhrsample] (\i,8) circle[radius=1.75pt];
  }
  \foreach \j in {9,...,16} {
    \fill[mhrsample] (8,\j) circle[radius=1.75pt];
  }
  \foreach \i/\j in {3/8,8/8,8/13} {
    \fill[white] (\i,\j) circle[radius=2.8pt];
    \fill[mhrsample] (\i,\j) circle[radius=2pt];
  }
  \path[draw=mhrtarget, fill=white, line width=.95pt] (3,13.30) -- (3.30,13) -- (3,12.70) -- (2.70,13) -- cycle;

  \draw[mhr axis] (0,0) -- (17.1,0) node[right] {$p$};
  \draw[mhr axis] (0,0) -- (0,17.1) node[above] {$q$};
  \node[below left, inner sep=3pt] at (0,0) {$0$};
  \node[below, inner sep=4pt] at (16,0) {$1$};
  \node[left, inner sep=4pt] at (0,16) {$1$};
  \node[below, inner sep=4pt, text=black] at (8,0) {$\nicefrac{\ell}{K}$};
  \node[left, inner sep=4pt, text=black] at (0,8) {$\nicefrac{\ell}{K}$};
\end{scope}

\begin{scope}[shift={(5.692,5.58)},xscale=.85,yscale=1.5]
\node[mhr active node] (mhr-root) at (2.48,-1.32) {\tikz[baseline=-.6ex,rounded corners=0pt]{\fill[mhrsample] (0,0) circle[radius=1.75pt];}\ $[0,16]$\\[-1pt]{\scriptsize $\ell=8$}};
\node[mhr tree node, draw=mhrleft, fill=mhrleft!6] (mhr-left) at (1.10,-2.15) {\tikz[baseline=-.6ex,rounded corners=0pt]{\fill[mhrleft] (-1.7pt,-1.7pt) rectangle (1.7pt,1.7pt);}\ $[0,8]$\\[-1pt]{\scriptsize $\ell=4$}\\[-1pt]{\scriptsize $i=3$}};
\node[mhr tree node, draw=mhrright, fill=mhrright!6] (mhr-right) at (3.86,-2.15) {\tikz[baseline=-.6ex,rounded corners=0pt]{\fill[mhrright] (0,2.3pt) -- (-2.1pt,-1.4pt) -- (2.1pt,-1.4pt) -- cycle;}\ $[9,16]$\\[-1pt]{\scriptsize $\ell=12$}\\[-1pt]{\scriptsize $j=13$}};
\node[mhr tree node] (mhr-ll) at (.40,-2.98) {$[0,4]$};
\node[mhr tree node] (mhr-lr) at (1.80,-2.98) {$[5,8]$};
\node[mhr tree node] (mhr-rl) at (3.16,-2.98) {$[9,12]$};
\node[mhr tree node] (mhr-rr) at (4.56,-2.98) {$[13,16]$};
\draw[mhrleft, line width=.75pt] (mhr-root) -- (mhr-left);
\draw[mhrright, line width=.75pt] (mhr-root) -- (mhr-right);
\draw[mhr branch] (mhr-left) -- (mhr-ll);
\draw[mhr branch] (mhr-left) -- (mhr-lr);
\draw[mhr branch] (mhr-right) -- (mhr-rl);
\draw[mhr branch] (mhr-right) -- (mhr-rr);
\foreach \x in {.40,1.80,3.16,4.56} {
  \node[text=black!50, font=\footnotesize] at (\x,-3.37) {$\vdots$};
}

\end{scope}

\begin{scope}[xshift=11.6cm,x=.25cm,y=.25cm]
  \fill[black!2] (0,0) -- (0,16) -- (16,16) -- cycle;
  \fill[mhrsample!7] (3,8) rectangle (8,13);
  \draw[black!25, line width=.45pt] (0,16) -- (16,16) -- (0,0);
  \draw[mhr guide] (3,0) -- (3,8);
  \draw[mhr guide] (8,0) -- (8,8);
  \draw[mhr guide] (0,13) -- (3,13);
  \draw[mhrsample!70, line width=.8pt] (0,8) -- (8,8) -- (8,16);
  \draw[mhrsample, line width=1.25pt] (3,8) -- (8,8) -- (8,13);
  \draw[mhrtarget, dash pattern=on 3pt off 2pt, line width=.9pt] (3,8) -- (3,13) -- (8,13);
  \foreach \i/\j in {3/8,8/8,8/13} {
    \fill[white] (\i,\j) circle[radius=3.4pt];
    \fill[mhrsample] (\i,\j) circle[radius=2.4pt];
  }
  \path[draw=mhrtarget, fill=white, line width=1.1pt] (3,13.40) -- (3.40,13) -- (3,12.60) -- (2.60,13) -- cycle;

  \node[anchor=south, text=mhrtarget, inner sep=3pt] at (3,13.35) {$(\nicefrac{i}{K},\nicefrac{j}{K})$};
  \node[anchor=north, text=black, fill=white, inner sep=2.5pt, yshift=-4pt] at (3,8) {$(\nicefrac{i}{K},\nicefrac{\ell}{K})$};
  \node[anchor=north west, text=black, fill=white, inner sep=2.5pt, yshift=-4pt] at (8,8) {$(\nicefrac{\ell}{K},\nicefrac{\ell}{K})$};
  \node[anchor=west, text=black, inner sep=5pt] at (8,13) {$(\nicefrac{\ell}{K},\nicefrac{j}{K})$};

  \draw[mhr axis] (0,0) -- (17.1,0) node[right] {$p$};
  \draw[mhr axis] (0,0) -- (0,17.1) node[above] {$q$};
  \node[below left, inner sep=3pt] at (0,0) {$0$};
  \node[below, inner sep=4pt] at (16,0) {$1$};
  \node[left, inner sep=4pt] at (0,16) {$1$};
  \node[below, inner sep=4pt] at (3,0) {$\nicefrac{i}{K}$};
  \node[below, inner sep=4pt] at (8,0) {$\nicefrac{\ell}{K}$};
  \node[left, inner sep=4pt] at (0,8) {$\nicefrac{\ell}{K}$};
  \node[left, inner sep=4pt] at (0,13) {$\nicefrac{j}{K}$};
\end{scope}

\begin{scope}[shift={(2.05,1.65)},font=\footnotesize]
  \fill[mhrsample] (0,0) circle[radius=1.75pt];
  \node[anchor=west,inner sep=5pt] at (0,0) {root};
  \fill[mhrleft] (0,-.43) ++(-1.7pt,-1.7pt) rectangle ++(3.4pt,3.4pt);
  \node[anchor=west,inner sep=5pt] at (0,-.43) {left child};
  \fill[mhrright] (0,-.86) ++(0,2.3pt)
    -- ++(-2.1pt,-3.7pt) -- ++(4.2pt,0) -- cycle;
  \node[anchor=west,inner sep=5pt] at (0,-.86) {right child};
  \fill[mhrmuted] (0,-1.29) circle[radius=1.15pt];
  \node[anchor=west,inner sep=5pt] at (0,-1.29) {later samples};
\end{scope}
\begin{scope}[shift={(13.78,.85)},font=\footnotesize]
  \path[draw=mhrtarget,fill=white,line width=.9pt]
    (0,.09) -- (.09,0) -- (0,-.09) -- (-.09,0) -- cycle;
  \node[anchor=west,inner sep=5pt] at (0,0) {pair to infer};
\end{scope}
\end{tikzpicture}%
}
\caption{Visualization of Sierpinski sampling.
Left: sampled pairs for $K=16$; color and shape identify the first sampling node in the interval tree (middle), and gray pairs are sampled later.
Right: the first split separating $i$ and $j$ supplies three sampled pairs that determine $P(\nicefrac{i}{K},\nicefrac{j}{K})$ by independence.
Here $(i,j)=(3,13)$ and the splitting midpoint is $\ell=8$.}
\label{fig:mhr-skeleton}
\endgroup
\end{figure}

        \paragraph{Sierpinski Sampling.} Before moving on to the next task, we highlight a cute property of our filtering procedure. Indeed, trivially estimating the expected profit of \emph{all} points on the grid for a certain multiplicative precision would entail probing separately \emph{each one} of those $O(K^2)$  prices, while we are able to perform this task by probing only $ O(K \log K)$ prices. To decrease the number of prices to probe, we perform a tree decomposition of $G_K \cap \cT$ such that for all prices $(\nicefrac iK,\nicefrac jK)$ there exists an $\ell$ such that the three points $(\nicefrac iK,\nicefrac{\ell}{K})$, $(\nicefrac{\ell}K,\nicefrac{\ell}K)$, and $(\nicefrac{\ell}K,\nicefrac{j}K)$ have been sampled, so that estimates $\hat P$ are available for their corresponding probabilities of trade (see the two left-most plots in \Cref{fig:mhr-skeleton}). By independence, whenever $P(\ell/K,\ell/K)>0$, we have the following decomposition of the probability $P(\nicefrac iK,\nicefrac jK)$ of trading (see the right-most plot of \Cref{fig:mhr-skeleton}):
        \[
            P(\nicefrac iK,\nicefrac jK)= \frac{P(\nicefrac iK,\nicefrac \ell K) \cdot P(\nicefrac \ell K,\nicefrac jK)}{P(\nicefrac{\ell}K,\nicefrac{\ell}K)} \approx \frac{\hat P(\nicefrac iK,\nicefrac \ell K) \cdot \hat P(\nicefrac \ell K,\nicefrac jK)}{\hat P(\nicefrac{\ell}K,\nicefrac{\ell}K)} = \hat P(\nicefrac iK,\nicefrac jK).
        \]
        Note that overflow due to a small denominator may be an issue. The analysis in \Cref{sec:MHR} handles this by combining the concentration guarantees with a conveniently tweaked denominator.

        \paragraph{A Suitable Estimator.} At the end of the filtering procedure, {assuming it did trigger}, we have located a convex set of good prices whose expected profit is at least $c\cdot \eps_{h^\star}$ (point (i) of \Cref{lem:qual_MHR}) and at most $C \cdot \eps_{h^\star}$ (point (ii) of \Cref{lem:qual_MHR}), for suitable constants (or a certificate that $\mu^\star=O(1/K)$). The logarithmic loss $L(p,q)$ thus maps to an interval of \emph{constant} length. To apply the bandit convex theorem above, we just need to find a feasible estimator for $L$ and show that the relative ``noise'' $\xi_{\tau}$ has small $\psi_2$-norm.

        Here is the trick: we do not use \emph{one single} round of bilateral trade to simulate one round of the convex problem, but wait for the first success! To estimate $L(p,q)$, we repeatedly post prices $(p,q)$ until we observe the first trade. Let $N_{p,q}$ be the geometric random variable with parameter $P(p,q)$ which counts the number of attempts needed to observe a trade. We then have the following unbiased estimator~\citep{DeGroot1959}:
        \[
            \hat L(p,q) = H_{N_{p,q}-1} - \log (q-p),
        \]
        where $H_i$ is the $i$th harmonic number (with $H_0=0$). At this point, we are just left with bounding the $\psi_2$-norm of the ``noise'' random variable $L(p,q) - \hat L(p,q)$, which can be done by explicitly using the geometric nature of $N_{p,q}$. In particular, it is possible to show that $||L(p,q) - \hat L(p,q)||_{\psi_2} \in O(\sqrt{\log K})$, using the fact that the parameter of the geometric distribution (i.e., the probability of observing a trade) is uniformly at least $\eps_{h^\star}$ due to the guarantees of the filtering round. All in all, we can use the remaining rounds of bilateral trade (i.e., the ones left after the filtering procedure) to simulate this bandit convex problem, obtaining a bound on the (convex-problem) regret of the  following type:
        \[
                \E{\sum_{\tau=1}^n\left(L(p_{\tau},q_{\tau})-\min_{(p,q) \in \cK} L(p,q)\right)} \in O(\sqrt{n \log K} \log^{2}n).
        \]
        We have all the ingredients for the main result of the section, which appears as \Cref{thm:mhr-regret} in \Cref{sec:MHR}. 
         \begin{mainresult}[MHR distributions]
        Consider the problem of learning the best prices in bilateral trade when the agents' valuations are drawn from independent and MHR distributions. There exists a learning algorithm which enjoys $\tilde O(\sqrt{T})$ regret.
    \end{mainresult}
        \begin{proof}[Proof Sketch]
        We are then left with relating the notion of regret for the convex problem above to the \emph{actual} regret for the original bilateral trade problem we have been playing all along. In particular, we show how to relate the instantaneous regret {$L(p_{\tau},q_{\tau}) - \min_{(p, q)\in \cK}L(p, q)$} for the generic time step $\tau$ of the convex problem to the regret suffered during the corresponding $N_{p_{\tau},q_{\tau}}$ rounds in the bilateral trade world, {with respect to the optimal profit $\mu^\star$}. Consider the generic interaction $\tau$ in the bandit convex optimization world, which corresponds to some random $t_{\tau}$ in the bilateral trade world, and in which the bandit convex routine chooses to play prices $(p_{\tau},q_{\tau})$. This means that the bilateral trade algorithm plays $(p_{\tau},q_{\tau})$ from time $t_{\tau}$ to the stopping time $t_{\tau} + N_{p_\tau,q_\tau}-1$. Conditioning on the past, we have the following, {for $(p, q)\in \cK$ and letting $\mu^\star_\cK = \max_{(p,q)\in\cK}\mu(p,q)$}:
        \begin{align*}
            \E{\sum_{t=t_{\tau}}^{t_{\tau}+N_{p_\tau,q_\tau}-1} \left(\prof(p,q)-\prof(p_t,q_t)\right)\Bigg|\cF_{\tau}} &\leq
            \E{N_{p_{\tau},q_{\tau}}|\cF_{\tau}}(\mu^\star_\cK-\mu(p_{\tau},q_{\tau})) \tag{By Wald's equality}\\
            &= {\frac{\mu^\star_\cK-\mu(p_{\tau},q_{\tau})}{P(p_{\tau},q_{\tau})}} \tag{$N_{p_{\tau},q_{\tau}}\!\sim \,$Geom$(P(p_{\tau},q_{\tau}))$}\\
            &\le {\frac{\mu^\star_\cK-\mu(p_{\tau},q_{\tau})}{\mu(p_{\tau},q_{\tau})}}\tag{As $\mu(p,q)\le P(p,q)$}\\
            &\lesssim \log\left(\frac{\mu_\cK^\star}{\mu(p_{\tau},q_{\tau})}\right) \tag{As $z-1\lesssim\log z$ for $z$ bounded}\\
            &= L(p_{\tau},q_{\tau})-\min_{(p,q) \in \cK} L(p,q).
        \end{align*}

        Fix a filtering transcript satisfying the stated guarantees, and let $T_0$ be the remaining horizon. For analysis, continue the same $\COPT$ execution until all $n=T_0$ loss observations are complete. Each observation uses at least one trading round, so the actual execution is a prefix of this run. The regional mean-profit gaps $\mu^\star_\cK-\mu(p_t,q_t)$ are nonnegative, so their sum on this prefix is at most their sum over the full run. Applying the preceding calculation and the fixed-$n$ guarantee, then adding filtering regret and the approximation error in \Cref{lem:qual_MHR}(iii), gives
        \[
            R_T \lesssim  K + {\frac{T}{K}} + \sqrt{T \log K} \log^{2}T \approx \sqrt{T} \log^{\nicefrac 52}T,
        \]
        which follows by choosing $K\approx \sqrt{T}$.
    \end{proof}

     \subsection{\texorpdfstring{$\widetilde O(T^{\nicefrac 35})$}{O-tilde(T\textasciicircum(3/5))} Regret for Regular Distributions}
     \label{subsec:regular_upper_sketch}

    We now extend the approach of \Cref{subsec:MHR_sketch} to independent regular distributions. The same stochastic convex bandit routine remains useful, and we still require an initial filtering procedure to learn the approximate scale of the optimal expected profit, as well as a safe region for the convex bandit routine to query. The main difference is that regularity no longer guarantees convexity of the negative log-profit. We instead construct a small family of convex problems, one of which approximates the original optimum, and progressively discard the less profitable candidates. The formal construction is given in \Cref{sec:Regular}.

\paragraph{A family of convex objectives.}
Under regularity, the reciprocals of the seller's and buyer's acceptance probabilities are convex. 
For each $\lambda>0$ and $(p,q)\in\operatorname{int}(\cT)$, define the auxiliary profit
\[
 r_\lambda(p,q)=
 \frac{4(q-p)}{\bigl(\lambda/\Fs(p)+\lambda^{-1}/(1-\Fb(q))\bigr)^2}.
\]
{First, it is straightforward to show that} $1/r_\lambda$ is convex, and $r_\lambda\le\mu$ everywhere. Moreover, at any interior price pair $x=(p,q)$, choosing the \emph{balancing parameter} $\lambda_x=\sqrt{\Fs(p)/(1-\Fb(q))}$ gives $r_{\lambda_x}(x)=\mu(x)$. Thus optimizing the whole family recovers the optimal profit. Crucially, a small error in the logarithm of this parameter causes only a quadratic profit loss (see \Cref{lem:regular-auxiliary}):
\[
 0\le\mu(x)-r_\lambda(x)
 \le\mu(x)\log^2\nicefrac{\lambda}{\lambda_x}.
\]

\paragraph{Filtering and choosing candidates.}
As in the MHR case, we first estimate the optimal profit scale, denoted by $\thres$, using the sparse grid reconstruction 
{based on Sierpinski sampling}. For a target accuracy $\eta$, take $K\approx \nicefrac 1\eta$. If the filtering procedure certifies that $\mu^\star=O(\eta)$, posting $(0,0)$ incurs only $O(T\eta)$ regret. Otherwise, we group promising grid prices with similar balancing parameters and take each group's convex hull. Convexity of the reciprocal ensures that each hull remains profitable for its associated objectives. Discretizing logarithmic balancing parameters at a spacing of order $\sqrt{\nicefrac{\eta}\thres}$ produces only $N=\tilde O(\sqrt{\nicefrac{\thres}\eta})$ candidates, including one whose auxiliary optimum is within $O(\eta)$ of $\mu^\star$ (\Cref{lem:regular-init}).

\paragraph{Observing the convex losses.}
The MHR algorithm estimates its loss by repeatedly posting a single price pair until trade occurs. Here, the auxiliary loss depends on both marginal acceptance probabilities, while one-bit feedback reveals only their product. We overcome this by combining trade bits at a carefully chosen reference pair and its two crosses with the queried pair: for query $(p,q)$ and reference $(u,v)$, the crossed prices are $(p,v)$ and $(u,q)$. Independence supplies the required identities, and randomized sampling routines {based on Bernoulli factories} turn these bits into bounded, unbiased observations of the reciprocal auxiliary profit (\Cref{lem:regular-observation}). A key geometric property is that, when the reference and query have high auxiliary profit for the same parameter, their crossed pairs also have high actual profit (\Cref{lem:regular-cross}).

\paragraph{Refinement and elimination.}
{We cannot, however, simply apply the stochastic convex bandits bound in \Cref{prop:mhr-convex-bandit} independently to every candidate. First, some candidates may have auxiliary optima too far below $\mu^\star$, so we must discard them before incurring too much regret. Second, producing loss observations also requires posting the reference pair and crossed prices. The reference must therefore improve as refinement proceeds, so that obtaining more accurate observations does not itself incur excessive regret. Third, the candidate parameter $\lambda$ depends on unknown marginal probabilities: using a new reference price $z$ requires estimating the multiplier $\rho$ relating its balancing parameter $\lambda_z$ to the candidate's parameter, by $\lambda = \rho\lambda_z$. This calibration also costs trading rounds and cannot be repeated at every query.}
{Hence, we run successive elimination over the candidate convex problems. In phase $j$, at accuracy $\eps_j=\thres2^{-j}$, we use \Cref{prop:mhr-convex-bandit} to find a nearly optimal price pair within each surviving candidate's region and estimate its best auxiliary profit.} We eliminate candidates whose estimates are too small {compared to the empirical maximum} and update each survivor's reference to the newly found price pair. 
{At the start of each phase, calibration adapts the loss-observation procedure to each candidate's current reference while keeping the auxiliary parameter $\lambda$ close to its fixed intended value.} 
The elimination tolerance is of order $\eta+\eps$, ensuring that a nearly optimal candidate survives (\Cref{lem:regular-refine}). Calibration costs $\widetilde O(\nicefrac{1}\eta)$ per candidate per phase.

\begin{mainresult}[Regular distributions]
        Consider the problem of learning the best prices in bilateral trade when the agents' valuations are drawn from independent and regular distributions. There exists a learning algorithm which enjoys $\tilde O(T^{\nicefrac 35})$ regret.
    \end{mainresult}
    \begin{proof}[Proof Sketch]
    Initialization costs $\widetilde O(K)$ regret. Since only $O(\log T)$ phases can start, calibration costs $\widetilde O(N/\eta)$ in total. Summing the optimizer bounds over the runs started before $T$ gives $\widetilde O(\sqrt{NT})$. The additional $O(\eta)$ per-round term during refinement is charged only over actual trading rounds, contributing $O(T\eta)$.
Combining these bounds gives
\[
 R_T\lesssim
 K+\frac{N}{\eta}+\sqrt{NT}+T\eta
 \lesssim
 \eta^{-3/2}+\sqrt T\,\eta^{-1/4}+T\eta,
\]
using $K\approx \nicefrac{1}\eta$ and $N=\widetilde O(\sqrt{\nicefrac{\thres}\eta})$.
Choosing $\eta\approx T^{-\nicefrac{2}5}$ yields the $\widetilde O(T^{\nicefrac{3}5})$ guarantee. In conclusion, we see that relative to MHR, the additional cost comes from identifying a suitable member of the family of convex objectives.
\end{proof}

    \subsection{Separation between Regular and MHR Distributions}
    \label{subsec:regular_lower_sketch}

        Our last result concerns a separation between MHR and regular distributions. In particular, we prove that no learning algorithm can hope to achieve a regret smaller than $T^{\nicefrac 35}$, even when the valuations are drawn from independent and regular distributions. We provide here a sketch of the lower bound construction, while all the details can be found in \Cref{sec:separation}; in particular, the following result is proved as \Cref{thm:regular_lower}.

        \begin{mainresult}[Regular distributions] Consider the problem of learning the best prices in bilateral trade when the agents' valuations are drawn from independent and regular distributions. For any learning algorithm $\cA$, there exists an independent and regular distribution $\cD$ such that
        $R_T(\cA,\cD) \in \Omega(T^{\nicefrac 35})$.
        \end{mainresult}

        To construct a stochastic lower bound, it is enough to construct a randomized instance (i.e., a mixture of independent and regular distributions) which forces any deterministic algorithm to suffer the desired regret, in expectation with respect to the randomization of the instance. We prove the lower bound even when the learner observes the seller's and buyer's acceptance decisions separately.

        For an integer parameter $K$, we construct a base distribution $\cD_0$ and $K$ auxiliary distributions $\cD_1, \dots, \cD_K$ over $[0,1]^2$. Crucially, these distributions are all independent and regular, and for each buyer price $q$ the optimal seller price is always $\nicefrac q2$. Moreover, under all these distributions the seller valuation is drawn independently from some fixed and known distribution, so that there is no point in deviating from posting $\nicefrac q2$ to the seller to ``gain information''. This property simplifies the argument, so that we can focus our attention --- without loss of generality --- on algorithms that will always post prices of that form. We introduce the functions $h^j(q)$, which specify the expected profit of posting prices $(\nicefrac q2,q)$ when the valuations $(S,B)$ are drawn according to $\cD_j$:
        \[
            h^j(q) = \Eup{j}{\prof(\nicefrac q2,q)}\footnote{\text{We use the superscript $j$ to highlight that the underlying valuations are drawn according to the underlying distribution $\cD_j$.}} \quad j \in \{0,1,\dots, K\}.
        \]

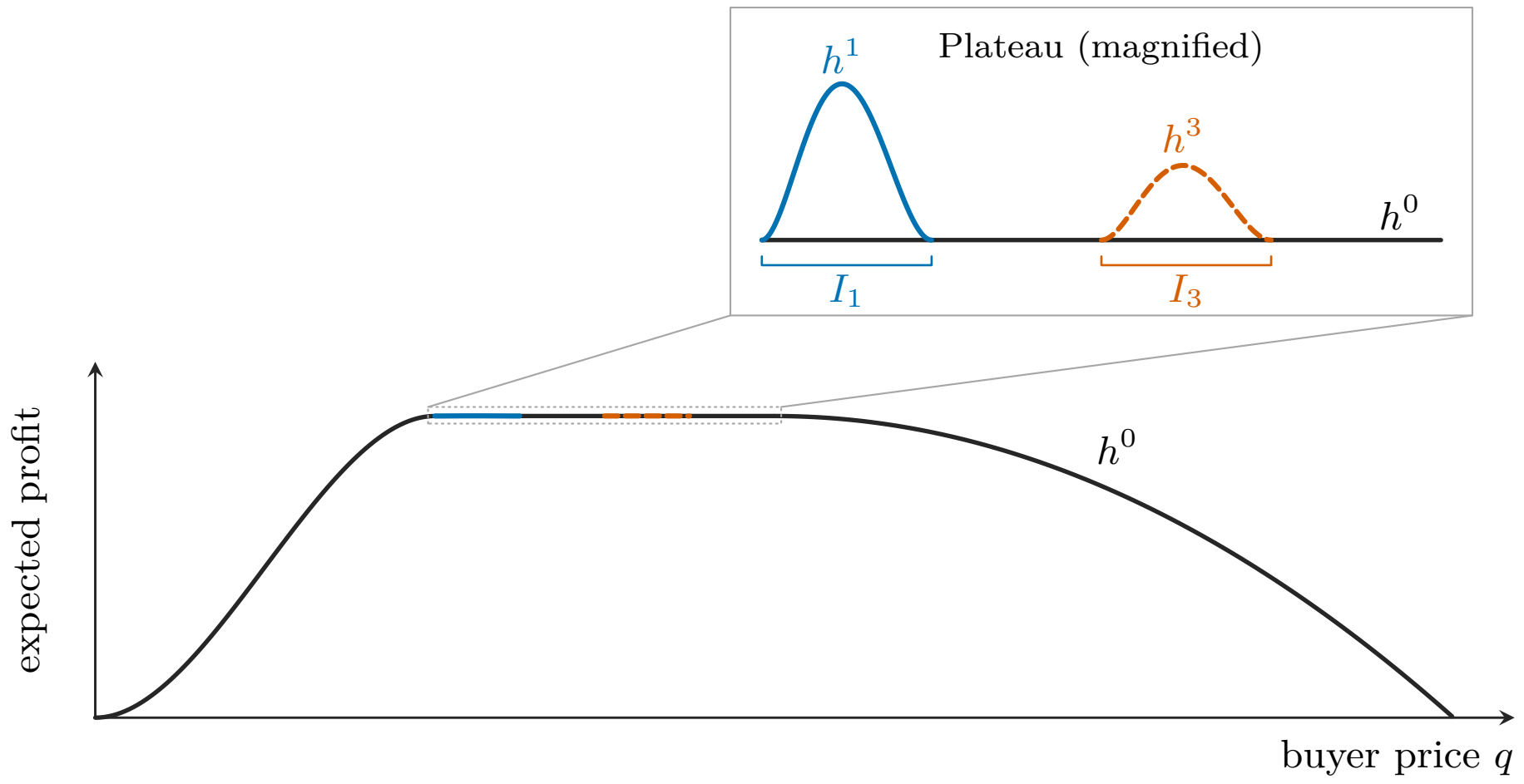
\begin{figure}[t]
        \centering
        \begingroup
\definecolor{lbblue}{HTML}{0072B2}
\definecolor{lborange}{HTML}{D55E00}
\begin{tikzpicture}[
  font=\small, line cap=round, line join=round, >=stealth,
  lb axis/.style={->, line width=.55pt, black!85},
  lb baseline/.style={black!85, line width=1pt},
  lb first/.style={lbblue, line width=1.15pt},
  lb second/.style={lborange, line width=1.15pt,
                    dash pattern=on 3pt off 1.7pt},
  lb guide/.style={black!35, line width=.4pt},
  declare function={
    lbrho(\u)=16*\u*\u*(1-\u)*(1-\u);
    lbexcess(\u,\A)=lbrho(\u)/(96*(\A+\u)^2-lbrho(\u));
    lbzoom(\u,\A)=2400*lbrho(\u)/(96*(\A+\u)^2-lbrho(\u));
  }]

\begin{scope}[x=10.8cm,y=2.4cm]
  \draw[lb axis] (0,0) -- (1.045,0);
  \draw[lb axis] (0,0) -- (0,1.18);
  \node[anchor=north east, inner sep=0pt] at (1.045,-.07)
    {buyer price $q$};
  \node[rotate=90, anchor=south, inner sep=0pt] at (-.035,.59)
    {expected profit};

  \draw[lb baseline, samples=81, domain=0:.25, variable=\q]
    plot (\q,{48*\q*\q*(1-8*\q/3)});
  \draw[lb baseline] (.25,1) -- (.5,1);
  \draw[lb baseline, samples=101, domain=.5:1, variable=\q]
    plot (\q,{4*\q*(1-\q)});

  \draw[lb first, samples=81, domain=0:1, variable=\u]
    plot ({.25+\u/16},{1+lbexcess(\u,4)});
  \draw[lb second, samples=81, domain=0:1, variable=\u]
    plot ({.375+\u/16},{1+lbexcess(\u,6)});
  \node[anchor=south west, inner sep=2pt] at (.73,.81) {$h^0$};

  \draw[lb guide, densely dotted] (.245,.975) rectangle (.505,1.03);
  \coordinate (lb-zoom-left) at (.245,1.03);
  \coordinate (lb-zoom-right) at (.505,1.03);
\end{scope}

\draw[lb guide] (lb-zoom-left) -- (5.05,3.20);
\draw[lb guide] (lb-zoom-right) -- (10.95,3.20);
\filldraw[fill=white, draw=black!35, line width=.4pt]
  (5.05,3.20) rectangle (10.95,5.65);
\node[anchor=north, font=\footnotesize, inner sep=0pt]
  at (8,5.47) {Plateau (magnified)};

\begin{scope}[shift={(5.3cm,3.80cm)},x=5.4cm,y=1cm]
  \draw[lb baseline] (0,0) -- (1,0);
  \draw[lb first, samples=121, domain=0:1, variable=\u]
    plot ({\u/4},{lbzoom(\u,4)});
  \draw[lb second, samples=121, domain=0:1, variable=\u]
    plot ({.5+\u/4},{lbzoom(\u,6)});

  \node[lbblue, anchor=south, inner sep=1.5pt] at (.118,1.27) {$h^1$};
  \node[lborange, anchor=south, inner sep=1.5pt] at (.620,.64) {$h^3$};
  \node[anchor=south, inner sep=2pt] at (.94,.01) {$h^0$};

  \draw[lbblue, line width=.5pt]
    (0,-.13) -- (0,-.20) -- (.25,-.20) -- (.25,-.13);
  \node[lbblue, anchor=north, inner sep=2pt] at (.125,-.20) {$I_1$};
  \draw[lborange, line width=.5pt]
    (.5,-.13) -- (.5,-.20) -- (.75,-.20) -- (.75,-.13);
  \node[lborange, anchor=north, inner sep=2pt] at (.625,-.20) {$I_3$};
\end{scope}
\end{tikzpicture}
\endgroup
        \caption{Expected profit $h^j(q)$ at prices $(q/2,q)$ for the
        lower-bound construction. The baseline $h^0$ has a plateau of maximizers.
        Two alternatives, $h^1$ and $h^3$ (here $K=4$), coincide with $h^0$
        outside $I_1$ and $I_3$, respectively.}
        \label{fig:regular-lower-profit}
        \end{figure}

        The ``baseline'' profit is maximized on a plateau for $q 
        \in [\nicefrac 14, \nicefrac 12]$. This interval is partitioned uniformly into non-overlapping intervals $I_j$ such that each $h^{j}$ exhibits a little ``bump'' on the corresponding $I_j$. Each one of these bumps has magnitude $\Delta \approx \nicefrac{1}{K^2}$. These properties are formalized in the following Claim (whose proof is deferred to \Cref{sec:separation}), and are visualized in \Cref{fig:regular-lower-profit}.

        \begin{restatable}{claim}{plateaux}
        \label{cl:prop_exp_prof}
            For any $j = 1, \dots, K$, it holds that:
        \begin{itemize}
                \item[\textnormal{(i)}] The function $h^0$ has a plateau of maximizers in  $[\nicefrac 14,\nicefrac 12]$, with maximum value $\nicefrac{1}{192}$. 
                \item[\textnormal{(ii)}] Letting $q^\star_j$ be the midpoint of $I_j$, we have $h^{j}(q^\star_{j}) - \max_{q\notin I_j} h^j(q) \ge \Delta$.
        \end{itemize}
        \end{restatable}

        \begin{remark}
            The crux of the argument resides in finding independent and regular distributions which exhibit a structure like the one described in the above claim. In particular, regularity impacts on the magnitude of the ``bumps'' that we can force on the expected profit function.
        \end{remark}

        For any deterministic algorithm $\cA$, we prove that it fails against the uniform mixture of $\{\cD_j\}_{j=1}^K$. To this end, denote by $N_j$ the random variable counting the number of times that $\cA$ posts a buyer price in $I_j$. The crucial ingredient to analyze the regret of $\cA$ under the mixture resides in arguing that the algorithm manages to differentiate between the generic $\cD_j$ and the baseline $\cD_0$ only by posting buyer prices in the ``right'' interval $I_j$. In particular, to discern whether there exists a bump in a given $I_j$ (i.e., if the underlying distribution is actually $\cD_j$ or not, see \Cref{cl:prop_exp_prof}), any learner has to post at least $\nicefrac{1}{\Delta^2} \approx K^4$ prices in $I_j$. This intuition is formalized in \Cref{sec:separation}, where we prove the following claim.

        \begin{restatable}{claim}{pinskerkl}
        \label{cl:KL}
            We have the following inequality:
            \(
                \Ej{N_j} - \Eo{N_j}  \le \frac{T}{22K^2}\sqrt{\Eo{N_j}}  
            \).
        \end{restatable}
   
        By Claim \ref{cl:prop_exp_prof}, the regret of the deterministic algorithm $\cA$ against the generic distribution $\cD_j$ satisfies
        \[
            R_T(\cA,\cD_j) \ge (T-\Ej{N_j})\Delta,
        \]
        so that if we take the expectation with respect to the uniform choice of $j\in [K]$, we get:
        \begin{align*}
            \frac{1}{K} \sum_{j=1}^K R_T(\cA,\cD_j)&\geq \frac{1}{K}\sum_{j=1}^K(T-\Ej{N_j})\Delta\\
            &\ge \frac{1}{K}\sum_{j=1}^K\left(T-\Eo{N_j}- \frac{T}{22K^2}\sqrt{\Eo{N_j}}\right)\Delta \tag{By Claim~\ref{cl:KL}}\\
            &\ge \frac{K-1}{K} T \Delta - \frac{1}{K}\sum_{j=1}^K \frac{T\Delta}{22K^2} \sqrt{\Eo{N_j}} \tag{As $\sum_j \Eo{N_j}\le T$}\\
            &\ge \frac{K-1}{K} T \Delta -  \frac{T\Delta}{22K^2}\sqrt{\frac{1}{K}\sum_{j=1}^K\Eo{N_j}} \tag{By Jensen's Inequality}\\
            &\ge \frac{K-1}{K} T \Delta -  \frac{T\Delta}{22K^2}\sqrt{\frac TK} \tag{As $\sum_j \Eo{N_j}\le T$}\\
            &= T\Delta\left(1 - \frac{1}{K} - \frac{1}{22}\sqrt{\frac {T}{K^5}}\right)\ge \frac{T\Delta}{2} \tag{By choosing $K = \left\lceil T^{\nicefrac{1}{5}}\right\rceil$}, 
        \end{align*}
        where the last inequality follows for all $T$ such that $\left\lceil T^{\nicefrac{1}{5}}\right\rceil \geq 3$. To get the desired $\Omega(T^{\nicefrac 35})$ regret lower bound, we just need to tune the parameter $K$, setting $K \approx T^{\nicefrac 15}$ (recall that $\Delta \approx \tfrac{1}{K^2}$, so that it is proportional to $ {T^{-\nicefrac{2}{5}}}$).

\section{\texorpdfstring{$\tilde O(T^{\nicefrac 23})$}{O-tilde(T\textasciicircum(2/3))} Regret for the Adversarial Setting}
\label{sec:adversarial}
In this section, we provide the complete proof of the regret bound in the adversarial setting. Recall that here we consider an oblivious adversary which fixes a sequence of valuations $v_1,\ldots,v_T\in[0,1]^2$ before play.  We want to prove the following theorem.

\begin{theorem}[Regret bound]\label{thm:adv-regret}
There exists a polynomial time algorithm $\cA$ such that, for every valuation sequence $\cS$ fixed in advance, the sequence of prices posted by the algorithm satisfies the following regret bound:
\begin{equation*}\label{eq:adv-regret}
 R_T(\cA, \cS)\leq O(T^{2/3}\log^{7/3}(T))
\end{equation*}
\end{theorem}
We show the above by first defining and analyzing the algorithm in \Cref{sec:adv-algorithm,subsec:analysisoneround}, proving the regret bound conditional on an estimation lemma which relates to the general i.i.d. setting, see \Cref{lem:inexpensive_detection}. We then dedicate \Cref{subsec:cdf_detection} to prove the lemma via a delicate estimation procedure, which directly implies a $\tilde O(T^{\nicefrac{2}{3}})$ regret rate in the stochastic i.i.d. via a successive elimination routine, as explained in \Cref{subsec:stochastic_sketch}.

\subsection{The adversarial algorithm}\label{sec:adv-algorithm}

The learner maintains positive probability weights $w_t$ on some prices we compare profit against, initially uniform over the action set $\AA$. In each round, the update of the weights as well as the price to play by the algorithm are given by subroutine \oneround, which returns a price $a_t$ and two probability vectors $Q_0,Q_1$ on $\AA$. $Q_z(\cdot)$ is the updated comparison weight vector that the learner will assign upon trade outcome $z = z_{a_t}(v_t) = \ind{s_t\leq p_t}\cdot\ind{b_t\geq q_t}$, recalling that $v_t = (s_t, b_t) $ and $a_t = (p_t, q_t)$. We call the price and its two potential weight updates a \emph{proposal}. 

\begin{algorithm}[H]
\caption{Adversarial bilateral trade}\label{alg:adv-outer}
\begin{algorithmic}[1]
\Require Horizon $T$, resolution $K$
\State $w_1(a')\gets1/|\AA|$ for every $a'\in\AA$
\For{$t=1,\ldots,T$}
  \State $(a_t,Q_0,Q_1)\gets\operatorname{\textsc{OneRound}}(w_t)$, using fresh randomness
  \State Post $a_t$ and observe $Z_t=z_{a_t}(v_t)$
  \State $w_{t+1}(a')\gets Q_{Z_t}(a')$ for every $a'\in\AA$
\EndFor
\end{algorithmic}
\end{algorithm}

\paragraph{Constructing candidate posted prices and updates.}
Each call to \oneround holds $w_t$ fixed and iterates over a generic probability distribution $\rho$ on $G_K \times \AA$, a local variable to this subroutine that is updated at every iteration. Let $(V, A)\sim\rho$ be a valuation/action random vector independently sampled according to this table. At an iteration inside \oneround, for each candidate price $a$ and trade/no trade bit $z$, let $\beta = K^2\log^6K$ and define
\begin{equation}\label{eq:adv-update}
 Q^\rho_z(a')=
 \frac{\beta K\,\Psub{\rho}{A=a'\mid z_a(V)=z}+1/|\AA|}
 {1+\beta K}
\end{equation}
where we leave dependence on $a$ implicit.
Moreover, for a proposal $(a,Q_0,Q_1)$ and weights $w$, define the score of the proposal against comparison price $a'$ and valuation $v$ by
\begin{equation}\label{eq:adv-score}
 h_a(v, a')=\prof(v,a')-\prof(v,a)
          -\beta\log\frac{Q_{z_a(v)}(a')}{w(a')}.
\end{equation}
In words, the score balances exploration with exploitation, with the first term bounding regret with respect to $a'$ and thus representing exploitation, and the second representing posterior information gain and thus exploration, if $Q_z$ is taken to be $Q_z^\rho$.

In the above, $\beta$ is a fixed parameter that controls how much an increase in a comparison price's weight can compensate for the profit gap of the posted price to the comparison price. In particular, the first two terms of $h_a(v, a')$ are the profit gap of candidate price $a$ against $a'$, and if the update $Q_{z_a(v)}(a')$ increases the weight of $a'$, the logarithm is positive and offsets this loss. This score and idea is exactly the concept of Algorithmic Information Ratio from \citet{XuZ25}, when taking expectations over $\rho$ with $Q_z = Q^\rho_z$, and after appropriately translating notation.

Inside one step $j$ of \oneround, we then choose a price $\argmin_{a\in \AA}\mathbb{E}_{\rho}[h_a(V, A)]$ minimizing the expected score with respect to $\rho$, and store its associated updates $Q^\rho_0, Q^\rho_1$. \oneround only uses this data to update the table $\rho$. Only after the final randomly chosen step $j^\star$, is the current proposal, i.e. ($\argmin_{a\in \AA}\mathbb{E}_{\rho}[h_a(V, A)]$, $Q^\rho_0$, $Q^\rho_1$) returned to Algorithm~$\ref{alg:adv-outer}$.

\begin{algorithm}[H]
\caption{$\operatorname{\textsc{OneRound}}(w)$}\label{alg:adv-inner}
\begin{algorithmic}[1]
\Require Probability vector $w$ with $w(a')\ge1/(|\AA|(1+\beta K))$;
         parameters from \Cref{eq:adv-parameters}
\State Draw $j^\star$ uniformly from $\{1,\ldots,J\}$
\State $\rho(v, a')\gets1/(|\AA||G_K|)$ for every $(v, a')\in G_K\times \AA$
\For{$j=1,\ldots,j^\star$}
  \State For each $a\in\AA$, form $Q^\rho_0,Q^\rho_1$ by \Cref{eq:adv-update}
         and compute $\Esub{\rho}{h_a(V, A)}$ using \Cref{eq:adv-score}
  \State Choose a price $a$ minimizing $\Esub{\rho}{h_a(V, A)}$,
         breaking ties by a fixed order, and keep its associated updates $Q_0,Q_1$
  \State Simultaneously update all entries by
    $\displaystyle
      \rho(v, a')\gets
      \frac{\rho(v, a')(1+\eta h_a(v, a'))}
           {1+\eta\Esub{\rho}{h_a(V, A)}}$
\EndFor
\State \Return The last selected price $a$ and its associated updates $Q_0,Q_1$
\end{algorithmic}
\end{algorithm}

\paragraph{Interpretation as MWU.}\label{par:MWU}
\Cref{alg:adv-inner} is a standard application --- see e.g. Theorem 3.1 of \citet{AroraHazanKale2012} --- of the Multiplicative Weight Update algorithm for finding an upper bound of a minimax problem, namely
\begin{equation}\label{eq:adv-minimax}
 \inf_{\text{rand. proposals}}
 \max_{\substack{a'\in\AA\\v\in G_K}}
 \E{h_a(v, a')}.
\end{equation}

The above minimax game can be shown to be equivalent --- in terms of value --- to the inner game from Algorithm 3 in \citet{XuZ25}, so that the upper bound we give on \Cref{eq:adv-minimax} immediately implies an analogous adversarial regret bound via their Theorem 3.1 or 3.3, assuming the possibility of exactly solving their game. The choice of a slightly different game together with a MWU inner routine is to avoid issues when specifying a constructive algorithm: numerically solving their inner game up to some precision does not yield the same precision in regret terms, due to the gradient term appearing in their Theorem 3.4.

To see how MWU provides this direct control, we describe its application to the game from \Cref{eq:adv-minimax}, proving guarantees formally in \Cref{lem:adv-oneround}. The MWU decisions are the pairs $(v, a')\in G_K\times \AA$, and $\rho$ is a distribution over these decisions. At each iteration, the routine constructs a proposal for every candidate
price and selects one minimizing $\Esub{\rho}{h_a(V, A)}$.
The selected proposal then assigns a score to every decision, via \Cref{eq:adv-score}.
Viewing the negative scores as costs, MWU, which plays the role of the adversary in the minimax game, updates
\[
 \rho(v, a')\propto
 \rho(v, a')\bigl(1+\eta h_a(v, a')\bigr),
\]
followed by normalization. Intuitively, this interaction mimics a maximin game over successive inner iterations: the player responds to the current
distribution $\rho$ by selecting a proposal, while MWU adjusts
$\rho$ to place more weight on pairs $(v,a')$ that have received
large scores under past proposals. Outputting a proposal by uniformly mixing over the rounds in $\{1, \ldots, J\}$ then provides a bound on \Cref{eq:adv-minimax}.

In \Cref{alg:adv-inner} and for the following proofs, we set the following hyperparameters: 
\begin{equation}\label{eq:adv-parameters}
 \begin{aligned}
 H&=1+\beta\log\bigl(|\AA|(1+\beta K)\bigr),\qquad
 \eta=\frac1{2KH^2},\\
 J&=\left\lceil4K^2H^2\log(|\AA||G_K|)\right\rceil.
 \end{aligned}
\end{equation}
Here $H$ bounds the absolute value of the score, $\eta$ is the MWU learning rate, and $J$ is a sufficient number of iterations to have a bound of at most $O(1/K)$ \sedit{on \Cref{eq:adv-minimax}}. 

\subsection{Analysis of OneRound}
\label{subsec:analysisoneround}
\subsubsection{A tradeoff between profit and information}
\label{subsubsec:tradeoff}
To analyze \oneround, we first consider the dual formulation of the learning game \Cref{eq:adv-minimax}: a deterministic strategy against a fixed randomized distribution $\rho$ over $G_K\times \AA$. The analysis of \oneround then allows relating the maximin game to the minimax one, yielding a bound on \Cref{eq:adv-minimax}, for any considered probability vector $w$, and thus score $h$ definition from \Cref{eq:adv-score}. 

Before proceeding, we introduce a few relevant concepts. First, given a reference action $a \in \AA$, we can define the expected gap $\Delta_\rho(a)$ when the valuation $V$ and the action $A$ are drawn jointly from $\rho$:
\begin{equation}
\label{eq:delta_def}
    \Delta_{\rho}(a) = \Esub{\rho}{\prof(V, A) - \prof(V,a)}. 
\end{equation}
Similarly, let $\Delta_\rho(a', a) = \Esub{\rho}{\prof(V, a') - \prof(V,a)}$, the expected difference between fixed action $a', a\in \AA$ in terms of profit when the valuations are drawn according to the marginal probability of $\rho$ on the valuation space.

The quantity in \Cref{eq:delta_def} may well be positive if the dependence between valuation $V$ and action $A$ under $\rho$ can be exploited to gain more profit. In the following, we will argue that there must exist an action $a$ which has either small gap $\Delta_{\rho}(a)$ or which reveals a lot of dependency between $z_a(V)$ and $A$, or both. In other words, not all actions in $\AA$ can show substantial sub-optimality with respect to $\rho$ while not revealing much about the ``strength'' of the coupling between $V$ and $A$.

To quantify this dependency, we consider the notion of mutual information, i.e. the KL divergence between the joint distribution of $\{z_a(V), A\}$ under $\rho$ and their joint distribution if we assumed it was the product of the corresponding marginals. More formally: 
\[
    \cI_\rho(a) = \sum_{a' \in \AA}\sum_{z \in \{0,1\}} \Psub{\rho}{A=a',z_a(V)=z} \log \frac{\Psub{\rho}{A=a',z_a(V)=z}}{\Psub{\rho}{A=a'} \cdot \Psub{\rho}{z_a(V)=z}}
\]
In the rest of this subsection, we prove the following statement, which bounds the parameterized information ratio~\citep{FosterRSS22} for our problem.
\begin{theorem}[Information Inequality]\label{thm:adv-information}
        There exists a universal constant $C_0$ such that, for any distribution $\rho$ over $G_K\times \AA$, the following inequality holds for $\beta = K^2 \log^6K$:
        \[
            \min_{a \in \AA} \left\{\Delta_{\rho}(a) - \beta \cdot \cI_\rho(a)\right\} \le \frac{C_0}{K}.
        \]
\end{theorem}
\begin{proof}
    Assume towards contradiction that there exists a distribution $\rho$ which violates the statement, i.e., such that $\Delta_{\rho}(a) - \beta \cdot \cI_\rho(a) > \nicefrac{C_0}{K}$ for all actions $a \in \AA$, and for some constant $C_0\in \mathbb R^+$. We focus on such $\rho.$
    We denote with $a^\star$ the optimal fixed action under the marginal valuation distribution induced by $\rho$. 
    
    We are under the assumption that $\Delta_{\rho}(a) - \beta \cdot \cI_\rho(a) > \nicefrac{C_0}{K}$ for all actions $a \in \AA$, in particular for $a=a^\star$, this implies that $\Delta_\rho(a^\star)$ is at least $\nicefrac {C_0}K$ (as the information term is non-negative). Note, this does not violate the optimality of $a^\star$ under $\rho$, as the random $A$ may adapt to the conditional behavior of $V$, as explained at the beginning of the subsection. We have then proved the following claim. 

    \begin{claim}
    \label{cl:Delta_star}
        Under the assumption towards contradiction, the following inequality holds:
        \(
            \Delta_{\rho}(a^\star)\in (\nicefrac{C_0}{K},1].
        \)
    \end{claim}

    For each action $a$, we can introduce an auxiliary distribution over $G_K$ with the following associated probability trade function, recalling that $v=(s, b)$: 
    \begin{equation}
    \label{eq:trade_func_prob_def}
        P_a(v) = \Psub{\rho}{V \in [0,s] \times [b,1]|A=a}.
    \end{equation}
    The marginal distribution is then the corresponding mixture of all these auxiliary distributions:
    \begin{equation}
    \label{eq:p_0_def}
        P_0(v) = \sum_{a \in \AA} \Psub{\rho} {A=a}P_a(v) = \Psub{\rho}{V \in [0,s] \times [b,1]}.
    \end{equation}
    
    For any action $a$, we can conveniently decompose the information $\cI_{\rho}(a)$ accordingly. In particular, the random feedback $z_{a}(V)$ under $\rho$ is a Bernoulli random variable whose probability of success is $P_0(a)$ under $\rho$, or $P_{a'}(a)$ conditioning on $A=a'$.
    \begin{align}
        \cI_\rho(a) &= \sum_{a' \in \AA}\sum_{z \in \{0,1\}} \Psub{\rho}{A=a',z_a(V)=z} \log \frac{\Psub{\rho}{A=a',z_a(V)=z}}{\Psub{\rho}{A=a'} \cdot \Psub{\rho}{z_a(V)=z}} \notag\\
        &=\sum_{a'\in \AA} \Psub{\rho}{A=a'} \left[\sum_{z \in \{0,1\}} \Psub{\rho}{z_a(V)=z|A=a'} \log \frac{\Psub{\rho}{z_a(V)=z|A=a'} }{\Psub{\rho}{z_a(V)=z}}\right]\label{eq:alternative_information}\\
        &= \sum_{a'\in \AA} \Psub{\rho}{A=a'} \KL{\operatorname{Bern}(P_{a'}(a))}{\operatorname{Bern}(P_{0}(a))} \tag{Definition of KL divergence}\\
        &\ge 2\sum_{a'\in \AA} \Psub{\rho}{A=a'} (P_{a'}(a) - P_0(a))^2 \tag{Pinsker's Inequality}
    \end{align}
    
    We can then combine the above inequality with the fact that we are under the assumption that $\Delta_{\rho}(a) - \beta \cI_{\rho}(a)>0$ to derive the following claim. 
    \begin{claim}
        \label{cl:information}
        Under the assumption towards contradiction, the following inequalities hold:
        \[
            \Delta_{\rho}(a) \ge 2\beta \sum_{a'\in \AA} \Psub{\rho}{A=a'} (P_{a'}(a) - P_0(a))^2, \quad \forall a \in \AA.
        \]
    \end{claim}
    To conclude the argument, we need the following delicate lemma, which is the actual enabler of our claimed regret bound. We defer its proof to a dedicated part of the paper, \Cref{subsec:cdf_detection}, where we explain the estimation routine behind the statement.
    \begin{restatable}[Inexpensive Detection]{lemma}{inexpdetect}
    \label{lem:inexpensive_detection}
            For every distribution $\rho$ on $G_K \times \AA$, and $x \in (0,1]$, there are non-negative weights $n_x(a)$ (which depend only on $\rho$ and $x$) such that the following inequality is verified, for some constant $C$:
            \[
                \sum_{a \in \AA} n_x(a) (\Delta_\rho(a^\star,a) + x) \le C \log^3 K\left(\frac{K}{x} + \frac{\log^3K}{x^2}\right),
            \] 
            where $a^\star$ is optimal according to $\rho$. 
            Moreover, for any random valuation $\tilde V$ with probability function of trade $P$ such that $\max_{a \in \AA}\E{\prof(\tilde V,a)} \ge \Esub{\rho}{\prof(V,a^\star)}+x$, then the same weights respect, with $P_0$ induced by $\rho$:
            \[
                \sum_{a \in \AA} n_x(a)(P(a) - P_0(a))^2 \ge 1.    
            \]
    \end{restatable}
    Crucially, the choice of weights $\{n_x(a)\}_{a\in \AA}$ works for any other considered probability function of trade $P$, once $\rho$ and $x$ are fixed.

    Now fix any value $x \in (\nicefrac{\Delta_{\rho}(a^\star)}2,1]$, and let $n_x(a)$ be the weights as in the lemma above. For each action $a$, denote with $X_a$ the indicator variable on $\AA$ of whether it holds that $\max_{a \in \AA}\E{\prof(\tilde V,a)} \ge \Esub{\rho}{\prof(V,a^\star)}+x$, under $\tilde V$ following the (conditional) probability trade function $P_{a}$. Note, we can apply the second part of the lemma if and only if $X_a$ is equal to $1$. We then have:
     \begin{align*}
         &\Psub{A}{\max_{a \in \AA}\Esub{P_A}{\prof(
         \tilde V,a)} \ge \Esub{\rho}{\prof(V,a^\star)}+x}\\
         \quad &= \sum_{a \in \AA} \Psub{\rho}{A=a} X_a \\
        \quad &\le \sum_{a \in \AA} \Psub{\rho}{A=a} X_a\sum_{a' \in \AA} n_x(a')(P_a(a') - P_0(a'))^2 \tag{By \Cref{lem:inexpensive_detection}}\\
         &\le \frac{1}{2\beta}\sum_{a' \in \AA}n_x(a') \Delta_{\rho}(a')  \tag{By \Cref{cl:information}}\\
         &\le  \frac{1}{2\beta}\sum_{a' \in \AA}n_x(a') (\Delta_\rho(a^\star,a') + 2x)  \tag{As $\Delta_{\rho}(a)= \Delta_{\rho}(a^\star)+\Delta_\rho(a^\star,a)$ and $\nicefrac{\Delta_{\rho}(a^\star)}2 < x$}\\
         &\le \frac{C}{\beta} \log^3 K\left(\frac{K}{x} + \frac{\log^3K}{x^2}\right) \tag{By \Cref{lem:inexpensive_detection}}, 
     \end{align*}
     
    Finally, we can use the fact that the expectation of a positive-valued random variable is the tail integral of its survival function, to obtain: 
     \begin{align*}
         \Delta_{\rho}(a^\star) &=\Esub{A}{\Esub{\rho}{\prof(
         V,A)|A} - \Esub{\rho}{\prof(V,a^\star)}}\\
         &\le \Esub{A}{\left(\max_{a \in \AA}\Esub{P_A}{\prof(
         V,a)} - \Esub{\rho}{\prof(V,a^\star)}\right)_+}\\
         &=\int_0^1 \Psub{A}{\max_{a \in \AA}\Esub{V}{\prof(
         V,a)|A} \ge \Esub{\rho}{\prof(V,a^\star)}+x} dx\\
         &\le \frac{\Delta_{\rho}(a^\star)}2 + \int_{\nicefrac{\Delta_{\rho}(a^\star)}2}^1\frac{C}{\beta} \log^3 K\left(\frac{K}{x} + \frac{\log^3K}{x^2}\right) dx\\
         &=\frac{\Delta_{\rho}(a^\star)}2 + \frac{C}{\beta} \log^3 K\left[{K} \log(x) - \frac{\log^3K}{x}\right]_{\nicefrac{\Delta_{\rho}(a^\star)}2}^1
     \end{align*}
    Rearranging the above, we get, assuming $C_0\geq 1$: 
    \begin{equation*}
        \frac{1}{2}\leq \frac{C}{\beta}\left(\frac{K\log^3K\log(\nicefrac{2}{\Delta_\rho(a^\star)})}{\Delta_\rho(a^\star)}+\frac{2\log^6 K}{\Delta^2_\rho(a^\star)}\right)\leq \frac{C}{\beta}\left(\frac{K^2\log^3K\log(\nicefrac{2K}{C_0})+2K^2\log^6 K}{C_0}\right), 
    \end{equation*}
    The second inequality follows from \Cref{cl:Delta_star}.
    With $\beta=K^2\log^6K$, its right-hand side is at most \[ \frac{C}{C_0} \left(\frac{\log(2K/C_0)}{\log^3K}+2\right) \le\frac{7C}{C_0}<\frac14 \] when $C_0\ge32(C+1)$, since $K\ge2$, $C_0\ge1$, and $\log(2K/C_0)/\log^3K\le2/(\log2)^2<5$. This contradicts the lower bound $1/2$.
\end{proof}

\subsubsection{Proving the regret bound}
With \Cref{thm:adv-information} proven, we can shift our focus to bounding the value of the minimax game from \Cref{eq:adv-minimax}. The Multiplicative Weight Update strategy behind \oneround designs a randomized strategy for proposals which works well against the worst-case choice of a pair $(v, a')\in G_K\times \AA$, in terms of the score definition from \Cref{eq:adv-score}. Specifically, we show the following, recalling \Cref{alg:adv-inner,eq:adv-update}.

\begin{lemma}[\oneround guarantee]\label{lem:adv-oneround}
Let $w$ be a probability vector with
$w(a')\ge1/(|\AA|(1+\beta K))$ for all $a'\in\AA$ and let $(a,Q_0,Q_1)$ be the proposal returned by $\operatorname{OneRound}(w)$. Its score satisfies
\begin{equation}\label{eq:adv-oneround}
 \max_{\substack{a'\in\AA\\v\in G_K}}\E{h_a(v, a')}
 \le\frac{C_0+2}{K}.
\end{equation}
\end{lemma}

\begin{proof}[Proof of Lemma~\ref{lem:adv-oneround}]

\medskip

Consider the sequence Algorithm~\ref{alg:adv-inner} would
generate over all $J$ iterations. At $j\in [J]$, write $\rho_j$ for its table and $a_j$ for its selected price, with $h_{a_j}$ denoting the
score using that iteration's updates, i.e. $Q^{\rho_j}_0, Q_1^{\rho_j}$.

We apply the MWU bound from e.g. \citet{AroraHazanKale2012}, which we report below, with decisions $(v, a')\in G_K\times \AA$, costs $-h_{a_j}(v, a')/H$, and learning rate $\eta H$.
\begin{proposition}[Theorem 2.1 from \citet{AroraHazanKale2012}]
\label{prop:MWURegretBound}
    Let $\{\ell_{i, t}\}_{i\in \cX, t\in [T]}$ be a set of losses over an action space $\cX$ and time horizon $T$, bounded in $[-1, 1]$. Then the set of probability distributions $\{\mathbf{p}_t\}_{t\in [T]}$ generated by the Multiplicative Weights Algorithm, with learning rate $\gamma\leq \nicefrac{1}{2}$ and against the sequence $\{\ell_{i, t}\}_{i, t}$, satisfies, for any fixed $j\in \cX$:
    $$
    \sum_{t=1}^T \langle \mathbf{p}_t, \{\ell_{i, t}\}_i\rangle \leq \sum_{t=1}^T \ell_{j, t} + \gamma \sum_{t=1}^T\lvert\ell_{j, t} \rvert + \frac{\log \lvert \cX \rvert}{\gamma}
    $$
\end{proposition}
\noindent Considering \Cref{eq:adv-parameters} and the bounds on $w$ and $Q_z$, together with $\prof\in[0,1]$, imply $|h_{a_j}(v, a')|\le H$. Thus the costs lie in $[-1,1]$, and our choice of $\eta$ ensures $\eta H\le1/2$, as required by the theorem.

With these choices, the standard MWU rule multiplies each weight by
\[
 1-(\eta H)\left(-\frac{h_{a_j}(v, a')}{H}\right)
 =1+\eta h_{a_j}(v, a').
\]
After normalization, this is precisely our update of $\rho_j$, starting from the uniform distribution as required. Moreover, each factor is at least $1-\eta H\ge1/2$, so the tables remain positive and all conditional updates are well defined.

Substituting these costs in \Cref{prop:MWURegretBound} and rescaling gives,
for every fixed $(v, a')$:
\begin{align}
 \frac1J\sum_{j=1}^J h_{a_j}(v, a')
 &\le\frac1J\sum_{j=1}^J\Esub{\rho_j}{h_{a_j}(V, A)}
       +\frac{\log(|\AA||G_K|)}{\eta J}+\eta H^2\notag\\
 &\le\frac1J\sum_{j=1}^J\Esub{\rho_j}{h_{a_j}(V, A)}+\frac1K,
 \label{eq:adv-mwu}
\end{align}
where the last inequality only uses our choices in \Cref{eq:adv-parameters}.
Therefore, in order to get the claimed bound, it suffices to bound each weighted average by $(C_0+1)/K$.

By \Cref{eq:adv-update}, for every $a'\in\AA$ and $z\in\{0,1\}$, $ Q^{\rho_j}_z(a')\ge\tfrac{\beta K}{1+\beta K}\, \Psub{\rho_j}{A=a'\mid z_{a_j}(V)=z}$. To identify the mutual-information term, insert the marginal
probability of $A$ into the conditional log ratio:
\[
 \log\frac{\Psub{\rho_j}{A=a'\mid z_{a_j}(V)=z}}{w(a')}
 =
 \log\frac{\Psub{\rho_j}{A=a'\mid z_{a_j}(V)=z}}
          {\Psub{\rho_j}{A=a'}}
 +
 \log\frac{\Psub{\rho_j}{A=a'}}{w(a')}.
\]
Averaging the first term over $A$ and its feedback gives
$\cI_{\rho_j}(a_j)$, using the alternative definition for mutual information considered also in \Cref{eq:alternative_information}. Combining this identity with the preceding
lower bound on $Q_z$ thus yields:
\begin{align*}
 \Esub{\rho_j}{\log\frac{Q^{\rho_j}_{z_{a_j}(V)}(A)}{w(A)}}
 &\ge \cI_{\rho_j}(a_j)
   +\sum_{a'\in\AA}\Psub{\rho_j}{A=a'}
      \log\frac{\Psub{\rho_j}{A=a'}}{w(a')}
   -\log\left(1+\frac1{\beta K}\right)\\
 &\ge \cI_{\rho_j}(a_j)-\frac1{\beta K},
\end{align*}
where the second inequality follows because the sum
is nonnegative and
$\log(1+1/(\beta K))\le1/(\beta K)$.
Consequently,
\[
 \Esub{\rho_j}{h_{a_j}(V, A)}
 \le\Delta_{\rho_j}(a_j)-\beta\cI_{\rho_j}(a_j)+\frac1K.
\]

Therefore, \Cref{thm:adv-information} and the algorithm's
minimization over candidate prices imply
$\Esub{\rho_j}{h_{a_j}(V, A)}\le(C_0+1)/K$ for all $j \in [J]$, for some constant $C_0$. The actual algorithm stops at a uniformly at random chosen index $j^\star \in [J]$. Therefore, for every fixed $(v, a')$, \Cref{eq:adv-mwu} yields, taking the expectation over the random sampling of an index in $\{1, \ldots, J\}$:
\[
 \Esub{a, Q_0, Q_1}{h_a(v, a')}=\frac1J\sum_{j=1}^J h_{a_j}(v, a')
 \le\frac{C_0+1}{K}+\frac1K=\frac{C_0+2}{K}.
\]
\end{proof}

Using the guarantee for a single round, we can complete the proof for the adversarial regret bound, in a way identical to Theorem 3.1 in \citet{XuZ25}.
\begin{proof}[Proof of Theorem~\ref{thm:adv-regret}]

We use Lemma~\ref{lem:adv-oneround} to bound $h_a$ in each round: the expected profit gap plus the change in log weights. The changes in weight then telescope when summed over rounds. The lower-bound condition on $w_t$ of Lemma~\ref{lem:adv-oneround} holds initially and is preserved by each update prescribed as in \Cref{eq:adv-update}. 
Now fix a comparison price $a'\in\AA$.

Condition on all observations and internal random choices from rounds $(1,\ldots,t-1)$. This fixes $w_t$, while $v_t$ is fixed in advance. Applying the lemma to the fresh randomness used by \oneround in round $t$, we obtain:
\[
 \mathbb E\!\left[
 \prof(v_t,a')-\prof(v_t,a_t)
 -\beta\log\frac{w_{t+1}(a')}{w_t(a')}
 \,\middle|\,\text{History from rounds $1,\ldots,t-1$}\right]
 \le\frac{C_0+2}{K}.
\]
Taking expectations and summing over $t$, the logarithms
telescope:
\begin{align*}
 \sum_{t=1}^T\prof(v_t,a')
 -\E{\sum_{t=1}^T\prof(v_t,a_t)}
 &\le \frac{(C_0+2)T}{K}
      +\beta\E{\log\frac{w_{T+1}(a')}{w_1(a')}}\\
 &\le \frac{(C_0+2)T}{K}+\beta\log|\AA|,
\end{align*}
where the last inequality uses $w_1(a')=1/|\AA|$ and
$w_{T+1}(a')\le1$.

Restricting to $v_t\in G_K$ is without loss of generality since rounding seller values up and buyer values down to the grid preserves the profit and feedback for any action in $\AA$. Moreover, comparing against a price $(p,q)\in\cT \setminus G_K$ can add at most $2T/K$ regret, by \Cref{lem:discretization_err}. This proves the regret bound from \Cref{eq:adv-regret}, factoring in the choice of $\beta$, and setting $K = \Theta\!\left((T/\log^7(T))^{1/3}\right)$.

To conclude, we prove that the algorithm runs in polynomial time. Notice that, for each of its iterations, \oneround needs to scan over the whole grid for the candidate posted price to use for the updates in \Cref{eq:adv-update}, and for each candidate price it needs to scan $G_K\times\AA$ again to compute the conditional probabilities in the updates, for a $O(K^6)$ complexity. Factoring in the values of $J$ and $H$, which determine the maximum number of iterations of \oneround, and the fact that \oneround is called once for each time step, we obtain a crude $\tilde O(TK^{12})$ bound.
\end{proof}

\subsection{Low Regret Estimation in the i.i.d. Setting}
\label{subsec:cdf_detection}
This section is dedicated to proving the crucial estimation lemma of the regret bound, namely \Cref{lem:inexpensive_detection}. For this section, we consider the stochastic i.i.d. setting, meaning that the distribution of the valuation pairs satisfies a specific and stationary probability function of trade $P$, on $[0, 1]^2$. Given a choice of distribution $\rho$ on $G_K\times \AA$, we also have a reference marginal probability of trade $P_0$ on $G_K$, according to \Cref{eq:p_0_def}. 

Even though we assign it a specific meaning in \Cref{subsec:analysisoneround}, from now on and for the rest of the section, $\rho$ can be regarded as any valid distribution and $P_0$ as its probability trade function. Indeed, the estimation routine presented here can be immediately used to obtain $\tilde O(T^{\nicefrac{2}{3}})$ regret rate in the stochastic i.i.d. setting via successive elimination, as explained in \Cref{subsec:stochastic_sketch}, setting $P_0$ to some probability function of trade compatible with the confidence intervals considered at a specific stage. 

At a very high level, what we want to prove is that there exists an estimation routine for the stochastic i.i.d. setting satisfying the following two properties:
\begin{enumerate}[label=\roman*)]
    \item The regret of the routine is small with respect to the probability function of trade $P_0$.
    \item For any other probability function of trade $P$ with maximum achievable profit substantially larger than $P_0$, the probability of trade under the two distributions must differ significantly at the prices posted by the routine (when receiving feedback drawn from $P_0$).
\end{enumerate}
        \inexpdetect*
        To claim the above, we need a delicate construction that lets us estimate $P$ while minimizing regret induced under $P_0$.
        We can do this by designing an algorithm which fully exploits the access to $P_0$ itself. This --- together with more standard concentration arguments --- gives the first bound of the claim, while an information theoretic argument gives the second once we have the first. More specifically, the first bound of the claim follows by the fundamental lemma below, which we prove in a sequence of steps. 
        \begin{lemma}\label{lem:efficient_estimation}
        For every precision $\eps\in (0, 1]$, there exists a routine estimating $\Esub{P}{\prof(\tilde V, a)}$ up to precision $\nicefrac{\eps}{4}$, with probability at least $\nicefrac{9}{10}$ and simultaneously for all actions in $\AA$ satisfying $\Delta_\rho(a^\star, a)\leq \eps$. Moreover, for some constant $C$, this routine makes at most $C\log^3(K)\left(\frac{K}{\eps^{2}}+\frac{\log^3 K}{\eps^3}\right)$ pricing queries in total, and each played action $a\in \AA$ satisfies $\Delta_\rho(a^\star, a)\leq 2\eps$.
        \end{lemma}
        \begin{proof}
    \noindent \textbf{Step 1: Build a tree over the seller domain}
    We start by arranging the seller domain $G_{K, s} = \{0, \nicefrac{1}{K}, \ldots, 1\}$ in a binary tree. We do this in the usual recursive fashion: starting from the whole $G_{K, s}$, split a node $I$ if $\lvert I\rvert\geq 2$, and assign to the left child the first $\lfloor\nicefrac{\lvert I\rvert}{2}\rfloor$ elements in increasing order, while leaving the rest to the other child.

    \medskip
    \noindent \textbf{Step 2: Control profit differences}
    We now control the variation of profit when fixing seller coordinates and varying buyer coordinates. Let $l$ and $r$ be two seller prices in a seller node $I$ with $l\leq r$ and let $q_0$ and $q$ be two buyer prices with $q_0\leq q$. We want to control the following difference: 
    \[
    \Delta_{\rho}((r,q),(l,q))-\Delta_\rho((r,q_0),(l,q_0))
    \]
    We analyze the above by decomposing the space where trade is allowed to happen under the considered posted prices, and conditioning correspondingly. First, if the valuation $V= (S, B)$ belongs to $[0, l]\times [q, 1]$, all four considered pairs of prices trade, and the above difference can be shown to be null. If $S\in[0,l]$ and $B\in[q_0,q)$, then posting $q$ prevents trade. Only the second term is nonzero: it equals $l-r$, so subtracting it gives the overall difference $r-l$. If $S\in(l,r]$ and $B\in[q_0,q)$, only $(r,q_0)$ trades, and the overall difference is $-(q_0-r)=r-q_0$. If $S\in (l, r]$ and $B\in [q, 1]$, posting $l$ gives null profit, and the above difference evaluates to $q-q_0$. Averaging, we thus obtain that the absolute value of the difference is bounded by: 
    \begin{align}
    &\lvert r-l\rvert\cdot \Psub{P_0}{[0, l]\times[q_0, q)} + \lvert q-q_0\rvert\cdot \Psub{P_0}{(l, r]\times[q, 1]}+\lvert q_0-r\rvert\cdot \Psub{P_0}{(l, r]\times[q_0, q)}\notag\\
    &\leq \lvert r-l\rvert\cdot \Psub{P_0}{[0, 1]\times[q_0, q)} + \lvert q-q_0\rvert\cdot \Psub{P_0}{(l, r]\times[0, 1]} + \Psub{P_0}{(l, r]\times[q_0, q)} \label{eq:square_decomposition}
    \end{align}
    
    Accordingly, define the positive measure $\nu$ on $[0, 1]^2$ in the following way:
    \begin{equation}
    \label{eq:nu_definition}
    \nu = \mathcal L \otimes \mathbb{P}_{P_{0, b}} + \mathbb{P}_{P_{0, s}}\otimes \mathcal L + \mathbb{P}_{P_0},
    \end{equation}
    where $\mathcal L$  is the Lebesgue measure and $\mathbb{P}_{P_{0, b}}$, $\mathbb{P}_{P_{0, s}}$ are the two marginals stemming from the distribution $\mathbb{P}_{P_0}$ induced by the trade probability function $P_0$.
    Combining \Cref{eq:square_decomposition,eq:nu_definition}, we then get that:
    \begin{equation}
    \label{eq:bound_measure}
    \lvert\Delta_{\rho}((r,q),(l,q))-\Delta_\rho((r,q_0),(l,q_0))\rvert\leq \nu((l, r]\times[q_0, q))
    \end{equation}
    Importantly, $\nu$ is finite, and its total mass is equal to $3$.
        \begin{figure}[t]
        \centering
        \resizebox{0.5\columnwidth}{!}{
            \begin{tikzpicture}[
            x=1bp,y=-1bp,
            line cap=butt,line join=miter,
            every node/.style={inner sep=0pt,outer sep=0pt},
            split/.style={
              draw=black,line width=1.2bp,
              dash pattern=on 6bp off 4.8bp,
              preaction={draw=white,line width=3.3bp}
            }
          ]
          \path[use as bounding box] (0,0) rectangle (439.856,440.856);
          \fill[white] (0,0) rectangle (439.856,440.856);
        
          \def\Left{67.76}
          \def\Right{434.096}
          \def\Top{5.76}
          \def\Bottom{372.096}
          \def\SellerL{159.344}
          \def\SellerR{258.25472}
          \def\BuyerZero{170.6112}
          \def\BuyerQ{88.1856}
        
          \begin{scope}
            \clip (\Left,\Top) rectangle (\SellerL,\BuyerQ);
            \foreach \row in {0,...,15} {
              \pgfmathsetmacro{\offset}{3*mod(\row,2)}
              \foreach \col in {10,...,27}
                \fill[black] ({6*\col+\offset},{6*\row})
                  circle[radius=.875bp];
            }
          \end{scope}
        
          \begin{scope}
            \clip (\SellerL,\Top) rectangle (\SellerR,\BuyerQ);
            \foreach \row in {0,...,15}
              \draw[black,line width=.55bp]
                (\SellerL,{3+6*\row}) -- (\SellerR,{3+6*\row});
          \end{scope}
        
          \begin{scope}
            \clip (\Left,\BuyerQ) rectangle (\SellerL,\BuyerZero);
            \foreach \col in {10,...,27}
              \draw[black,line width=.55bp]
                ({3+6*\col},\BuyerQ) -- ({3+6*\col},\BuyerZero);
          \end{scope}
        
          \begin{scope}
            \clip (\SellerL,\BuyerQ) rectangle (\SellerR,\BuyerZero);
            \foreach \row in {14,...,29} {
              \pgfmathsetmacro{\offset}{3*mod(\row,2)}
              \foreach \col in {25,...,44}
                \draw[black,line width=.67bp]
                  ({6*\col+\offset},{6*\row}) circle[radius=1.14bp];
            }
          \end{scope}
        
          \begin{scope}
            \clip (\Left,\Bottom) -- (\Right,\Bottom) -- (\Right,\Top)
              -- (\SellerR,\Top) -- (\SellerR,\BuyerZero)
              -- (\Left,\BuyerZero) -- cycle;
            \foreach \i in {0,...,25} {
              \pgfmathsetmacro{\intercept}{-1+.12*\i}
              \draw[black,line width=.85bp]
                (\Left,{\Bottom-366.336*\intercept})
                -- (\Right,{\Bottom-366.336*(\intercept+1)});
              \draw[black,line width=.85bp]
                (\Left,{\Bottom-366.336*\intercept})
                -- (\Right,{\Bottom-366.336*(\intercept-1)});
            }
          \end{scope}
        
          \begin{scope}
            \clip (\Left,\Top) rectangle (\Right,\Bottom);
            \foreach \seller in {\SellerL,\SellerR}
              \draw[split] (\seller,\Bottom) -- (\seller,\Top);
            \foreach \buyer in {\BuyerZero,\BuyerQ}
              \draw[split] (\Left,\buyer) -- (\Right,\buyer);
          \end{scope}
        
          \foreach \seller in {\SellerL,\SellerR} {
            \foreach \buyer in {\BuyerZero,\BuyerQ} {
              \fill[white] (\seller,\buyer) circle[radius=4.6bp];
              \fill[black] (\seller,\buyer) circle[radius=2.8bp];
            }
          }
        
          \draw[black,line width=1.2bp]
            (\Left,\Top) rectangle (\Right,\Bottom);
          \foreach \seller in {\SellerL,\SellerR}
            \draw[black,line width=1bp]
              (\seller,\Bottom) -- (\seller,{\Bottom+5});
          \foreach \buyer in {\BuyerZero,\BuyerQ}
            \draw[black,line width=1bp]
              (\Left,\buyer) -- ({\Left-5},\buyer);
        
          \node[anchor=base west,font=\fontsize{20bp}{24bp}\selectfont\itshape]
            at (156.544,398.75225) {l};
          \node[anchor=base west,font=\fontsize{20bp}{24bp}\selectfont\itshape]
            at (254.35472,398.9085) {r};
          \node[anchor=base west,font=\fontsize{20bp}{24bp}\selectfont\itshape]
            at (37.76,177.9237) {q};
          \node[anchor=base west,font=\fontsize{14bp}{17bp}\selectfont]
            at (47.759985,180.9237) {0};
          \node[anchor=base west,font=\fontsize{20bp}{24bp}\selectfont\itshape]
            at (45.76,95.4981) {q};
          \node[anchor=base,font=\fontsize{19bp}{23bp}\selectfont]
            at (250.928,430.346) {seller};
          \node[anchor=base,rotate=90,font=\fontsize{19bp}{23bp}\selectfont]
            at (20.01,188.928) {buyer};
        \end{tikzpicture}
        }
        \caption{The decomposition considered in Step 2 of the proof of \Cref{lem:efficient_estimation}: the area filled with squares is the region of the space when no trade happens under any of the four posted prices; each area where some trade happens and thus considered in the decomposition is filled with a distinct pattern.}
        \label{fig:measure_decomposition}
    \end{figure}
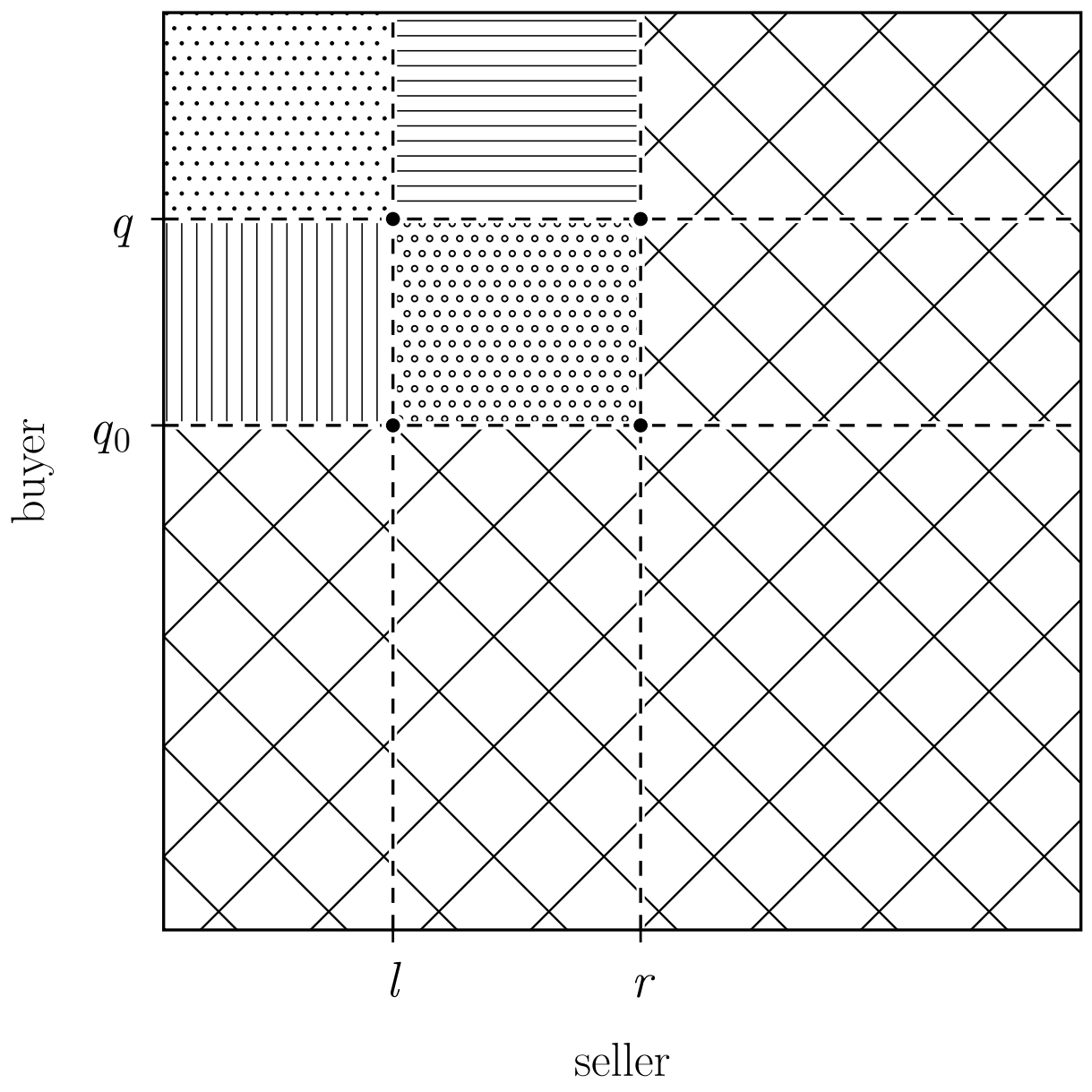

    \medskip
    \noindent \textbf{Step 3: Split the buyer coordinates into blocks} In this step, for each seller node $I$, we partition the buyer grid into a small number of blocks and choose a representative seller price $p_{I,\block}\in I$ for each block $\block$. Our goal is to ensure that
    \[
     \Esub{P_0}{\prof(V,(p_{I,\block},q))}
     \ge \max_{p\in I}\Esub{P_0}{\prof(V,(p,q))}-\eps \quad \forall q\in \block
     \qquad
    \]
    We use the definition of $\nu$ to partition the set of buyer coordinates $G_{K, b}$, for any given node $I\subseteq G_{K,s}$ of seller coordinates. Fixing $I$, the partition $\{\block_{\ell,I}\}_\ell$ is then defined iteratively, according to the following simple algorithm.
    \begin{algorithm}[H]
    \caption{Partition buyer coordinates}\label{alg:buyer-blocks}
    \begin{algorithmic}[1]
    \Require Node $I$, ordered grid $G_{K,b}$, measure $\nu$, threshold $\eps$
    \State $\block_{1,I}\gets\{\min G_{K,b}\}$, $\ell = 1$
    \For{each $q\in G_{K,b}$ in increasing order}
        \If{$\nu((\min I,\max I]\times[\min \block_{\ell,I},q))\leq\eps$}
            \State $\block_{\ell,I}\gets \block_{\ell,I}\cup\{q\}$
        \Else
            \State $\ell\gets \ell+1$, $\block_{\ell,I}\gets\{q\}$
        \EndIf
    \EndFor
    \State \Return $(\block_{1,I},\ldots,\block_{\ell,I})$
    \end{algorithmic}
    \end{algorithm}
    It follows that we are splitting $G_{K, b}$ in contiguous blocks which are maximal according to $\nu$: adding the next coordinate in increasing order to any of the blocks would violate the condition $\nu((\min I,\max I]\times[\min \block_{\ell,I},\max \block_{\ell,I}))\leq \eps$. Notice that we can bound the cardinality of $\lvert \{\block_{\ell, I}\}_\ell\rvert$ according to $\nu$. Specifically, each element of the partition is maximal with respect to $\nu$, i.e. we add a new block only when adding one more coordinate to the current block would make mass greater than $\eps$. Therefore the cardinality of the partition cannot exceed one plus the total mass of $\nu$ on $(\min I,\max I]\times[0,1]$ divided by $\eps$. In other words, we can have at most 
    \begin{equation}
    \label{eq:buyer_coords_blocks_card}
    1+\frac{\nu((\min I, \max I]\times [0, 1])}
    {\eps}
    \end{equation} 
    buyer coordinate blocks for each $I$.

    Now, for each seller node $I$ and buyer block $\block$, choose the smallest maximizer at buyer price $\min \block$:
    \begin{equation}
    \label{eq:representative_def}
     p_{I,\block}
     =\min\argmax_{p\in I}
           \bigl\{(\min \block-p)P_0(p,\min \block)\bigr\}.
    \end{equation}
    The crucial property of this construction is that, taking any $p\in I$ and $q\in \block$ and assuming $p\leq p_{I, \block}$, we have $(p, p_{I, \block}]\times[\min \block, q)\subseteq (\min I,\max I]\times[\min \block,\max \block)$, so that $\nu((p, p_{I, \block}]\times[\min \block, q))\leq \eps$ by definition of the blocks. In turn, using \Cref{eq:bound_measure}, this implies the following: 
    \begin{equation}
    \label{eq:row_stability_guarantee}
    \Delta_{\rho}((p_{I, \block},q),(p,q))
    \geq \Delta_{\rho}((p_{I, \block},\min \block),(p,\min \block))-\eps \\
    \geq -\eps ,
    \end{equation}
    where the last inequality follows by the definition of $p_{I, \block}$ as maximizer seller price in $I$ at buyer price $\min \block$. An identical reasoning goes through if $p > p_{I, \block}$, by reversing the interval extremes accordingly and using that $\Delta_{\rho}((p,q),(p_{I, \block},q)) = -\Delta_{\rho}((p_{I, \block},q),(p,q))$. 
    Therefore, each block $\block$ has a single seller price $p_{I, \block}$ which is approximately optimal --- under $P_0$ and over $I$ --- for any considered buyer price $q\in \block$. 

    \medskip
    \noindent \textbf{Step 4: Reconstruct the probability of trade}
    We want to use the previous constructions to efficiently recover $P(p, q)$ for any $(p, q)$ satisfying $\Delta_\rho(a^\star, (p, q))\leq \eps$, while satisfying the constraint of playing only $2\eps$-optimal actions, i.e. prices in $\{a\in \AA: \Delta_\rho(a^\star, a)\leq 2\eps\}$. 
    
    Due to the partitioning from Step 1, there must exist a leaf in the binary tree with $p$ as its value; correspondingly, we must have a path in the tree $I_0\supseteq I_1\supseteq\ldots \supseteq I_h$ with $I_0 = G_{K, s}$, $I_h = \{p\}$ and $h\leq 1+\log_2(K+1)$. We must also have a sequence of non-disjoint blocks $\block_0, \block_1, \ldots, \block_h$ of buyer coordinates, one for each node in the root-to-leaf path, for which $q\in \block_j$ for any $j\in \{0, 1, \ldots, h\}$. See \Cref{fig:localized-overview} in \Cref{subsec:stochastic_sketch} for an illustration. We can then consider the following telescoping decomposition of $P(p, q)$: 
     \begin{equation}
    \label{eq:tree_telescoping}
    P(p, q) = \underbrace{P(p_{I_0, \block_0}, q)}_{(\#)}+\sum_{j=1}^h \underbrace{P(p_{I_j, \block_j}, q)-P(p_{I_{j-1}, \block_{j-1}}, q)}_{(\#)}
    \end{equation}
    
    \begin{claim}
    \label{cl:telescope_properties}
        Any of the prices considered in the telescopic decomposition are contained in $\AA$ and are $2\eps$-optimal according to $P_0$, if $(p, q)$ is $\eps$-optimal. Moreover, each term highlighted with $(\#)$ is monotone in $q$, for $q\in D_j\cap D_{j-1}$ for the j-th parent to child difference, and for $q\in D_0$ for the root node value.
    \end{claim}
    \begin{proof}
        Notice that $(p, q)$ is assumed to be $\eps$-optimal according to $P_0$, meaning that \Cref{eq:row_stability_guarantee} immediately yields $2\eps$-optimality, since $p\in I_j$ and $q\in \block_j$ by construction. To prove that each belongs to $\AA$, we proceed by contradiction, using that in \Cref{eq:representative_def} ties are broken by taking minima. 
        
        First, if $(p_{I,\block},q)\notin \AA$, then $p_{I,\block}> q$. By assumption then there exists $p\in I$ with $(p,q)\in\AA$ and thus $p\leq q<p_{I, \block}$, with the profit obtained at $(p, \min \block)$ lower bounded by the one obtained at $(p_{I, \block}, \min \block)$: this follows since posting $(p_{I, \block}, \min \block)$ always results in negative profit when trading, due to $\min \block\leq q$ and $(p_{I,\block},q)\notin \AA$. Therefore, $p$ is also a seller price maximizer at buyer price $\min \block$, and since $p<p_{I, \block}$, this contradicts the fact that $p_{I, \block}$ should be the smallest maximizer.
        
        Finally, monotonicity follows by definition of the probability function of trade, see e.g. \Cref{eq:trade_func_prob_def}, and since $p_{I_j, \block_j}$ and $p_{I_{j-1}, \block_{j-1}}$ --- or $p_{I_0, D_0}$, for the root node value --- remain fixed, provided $q$ is only allowed to vary in the ranges mentioned in the claim.
    \end{proof}

    Therefore, in order to use \Cref{eq:tree_telescoping} for the routine guarantee, we need to estimate one monotone function in $q$ for any parent to child edge $\{I_{j-1}, I_{j}\}$ and intersection of the type $\block_{j}\cap \block_{j-1}$, together with one for each buyer block of the root node. To satisfy the constraint of considering only $2\eps$-optimal actions, it then suffices to restrict the domain of each function to $\block_j\cap \block_{j-1}\cap \{q\in G_{K,b}: \exists\,p\in I_j \text{ such that }(p,q)\in\AA,\ \Delta_\rho(a^\star,(p,q))\leq\eps\}$, and analogously for the root functions. This guarantees that for every $q$ in the resulting domain we have the same certificate of optimality granted by \Cref{cl:telescope_properties}. Since we are interested in estimating $P$ only for $\eps$-optimal points of the grid, this restriction does not impact the telescoping decomposition in \Cref{eq:tree_telescoping}.

    We now count the number of monotone functions in the class resulting from this procedure, which we denote as $\cF$. By the bound on the number of buyer coordinates blocks from \Cref{eq:buyer_coords_blocks_card} and considering that at each depth of the seller tree the ranges of the nodes are pairwise disjoint, we get that the overall number of possible pairs $(I, \block)$ of seller node and buyer block is bounded by, for $K\geq 40$:
    \begin{equation}
    \label{eq:count_pairs_node_block}
    2K+1 + \frac{\nu([0, 1]^2)}{\eps}\cdot 2\log K \leq 3K + \frac{3}{\eps}\cdot 2\log K. 
    \end{equation}
    In the above we have used the bound on the total mass of $\nu$ and the fact that the number of nodes in a full binary tree with $K+1$ leaves is $2K+1$, over a number of levels bounded by $2+\log_2(K+1)$, considering the halving rule from Step 1.
    
    It remains to factor in that we have to estimate a different function for every distinct and non-empty buyer block intersection of the type $\block_j\cap \block_{j-1}$, for any edge connecting a parent node $I_{j-1}$ to a child node $I_j$. To do this, notice that the buyer blocks of $I_{j-1}$ and $I_j$ form two different partitions of $G_{K, b}$, with contiguous coordinates in each block, so that the number of nonempty intersections is at most the total number of blocks in the two partitions. Since each node in the tree has at most two children, we can then conclude that its block count must appear once in its own contribution and at most twice in those of its children, when counting all the possible parent-child buyer block intersections and thus when bounding $\lvert \cF\rvert$. Therefore, we need to estimate a number of functions which is at most three times the bound from \Cref{eq:count_pairs_node_block}:
    \begin{equation}
    \label{eq:monotone_functions_count}
    \lvert\cF\rvert\leq \min\left(9K + \frac{18\log K}{\eps}, 3K^2\right),
    \end{equation}
    where $\lvert\cF\rvert\leq (2K+1)(K+1)\leq 3K^2$, for $K\geq 4$, follows trivially by counting the number of edges in the tree and the number of possible buyer coordinates.

    \medskip
    \noindent \textbf{Step 5: Bound total variation and estimate}
    We approach the conclusion of the proof of the lemma by bounding the total variation of each function $f\in\cF$. Intuitively, each function considered is monotone, meaning that we can provide a uniform estimation of the function values by querying and estimating only a number of points which is logarithmic in the size of the domain. We can then use the fact that the function is non-increasing or non-decreasing to ``fill the gaps'' between points. Lower variation then further decreases the number of points in the domain of the functions that need to be queried to obtain uniform control. Formally, we have the following claim, whose proof is deferred to Appendix \ref{app:adversarial}.

    \begin{restatable}{claim}{monotoneestimation}
    \label{cl:monotone_estimation}
        Let $\mathcal G$ be a set of monotone functions mapping from some subset of finite domain $\cY$ to $[-1, 1]$. For every $g\in \cG$ and each point $y\in \cY$ in the domain of $g$, suppose we have a sequence of mutually independent random variables \[ Z_{g,y}^{(1)},Z_{g,y}^{(2)},\ldots,
        \] 
        with expectation $g(y)$ and taking values in a common interval of length at most $4$. Each query at $g(y)$ reveals the next unused variable in its sequence. Suppose that $\sum_{g\in \cG} \text{\normalfont TV}(g)\leq W$, where $W$ is a known constant and $\text{\normalfont TV}(g)$ stands for the total variation of $g$. 
        
        Then, for $\alpha,\delta'\in(0,1)$, there exists a procedure which estimates every function uniformly to error $\alpha$, simultaneously with probability at least $1-\delta'$, using at most
        \[C\frac{\log(2|\cG||\cY|/\delta')}{\alpha^2}
        \left(|\cG|+\frac{W}{\alpha}\log(2|\cY|)\right)
        \]
        queries, where $C$ is an absolute constant.
    \end{restatable}

    In the above claim, we can plug in \Cref{eq:monotone_functions_count} as the cardinality bound, and use $K+1$ as a trivial upper bound on the size of the domain of each function. It remains to bound $\sum_{f\in \cF} \text{TV}(f)$. To do this, importantly, we now use the actual underlying probability function of trade $P$, consistently with \Cref{eq:tree_telescoping}. We use $P_0$ to define each function in $\cF$ and guarantee approximate optimality with respect to $P_0$, but every actual evaluated term refers to $P$, which is our estimation target.

    A function $f\in \cF$ that comes from the combination of a root node $I$ and one of its buyer blocks $\block$ is just $f(q) = P(p_{I, \block}, q)$, for $q$ in some subset of $\block$, with $p_{I, \block}$ the block representative seller price. It follows that: 
    $$
    \text{TV}(f)\leq \Psub{P}{[0, p_{I, D}]\times [\min D, \max D]}\leq \Psub{P}{[0, 1]\times [\min D, \max D]},
    $$
    by definition of the trade probability function, see e.g. \Cref{eq:trade_func_prob_def}. Moreover, the blocks of buyer coordinates are disjoint, so that the sum of the total variations over the functions associated to the root node must be bounded by $1$.

    For a function $f\in \cF$ that resulted from a tree edge between parent node $I_{j-1}$ and child node $I_{j}$, we instead have $f(q) = P(p_{I_j, \block_j}, q)-P(p_{I_{j-1}, \block_{j-1}}, q)$, with $\block_j$ and $\block_{j-1}$ as the corresponding buyer blocks and $q$ in some subset of their intersection. It follows that:
    $$
    \text{TV}(f)\leq \Psub{P}{[\min I_{j-1}, \max I_{j-1}]\times [\min (D_j\cap D_{j-1}), \max (D_j\cap D_{j-1})]}.
    $$
    Moreover, the collection of non-empty intersections $\block_{j}\cap\block_{j-1}$ associated with the edge $\{I_{j-1}, I_j\}$ is pairwise disjoint, and each intersection is a contiguous subset of $G_{K, b}$, since the partitioning of the buyer coordinate always consists of contiguous blocks, by \Cref{alg:buyer-blocks}. Therefore, the sum of the total variations of the functions associated to a specific edge $\{I_{j-1}, I_j\}$ must be bounded by $\Psub{P}{[\min I_{j-1}, \max I_{j-1}]\times [0, 1]}$. To conclude, each parent has two children and the seller nodes are disjoint at each depth, so that each parent covers a different subset of $G_{K, s}$; since the maximum depth is bounded by $1+\log_2(K+1)$, we all in all get, for $K\geq 10$ and factoring in the bound for the root functions:
    \begin{equation}
        \label{eq:tv_bound}
        \sum_{f\in \cF}\text{\normalfont TV}(f)\leq 1+ 4\log K\leq 8\log K
    \end{equation}

    We can then use the above and apply \Cref{cl:monotone_estimation}, with $\alpha = \nicefrac{\eps}{4\cdot 2\log K}$ and $\delta'=1/10$, so as to target error $\nicefrac{\eps}{4}$ and compensate the accumulation of the estimation error over the telescoping decomposition in \Cref{eq:tree_telescoping}, when targeting $P(p, q)$, for some $(p, q)$ satisfying $\Delta_\rho(a^\star, (p, q))\leq \eps$. 

    \medskip
    \noindent \textbf{Step 6: Wrapping up}    
    At this point, with probability at least $\nicefrac{9}{10}$, for each $\eps$-optimal target $(p, q)$, we obtain an estimator $\hat P(p, q)$ satisfying $\lvert \hat P(p, q)-P(p, q)\rvert\leq \nicefrac{\eps}{4}$. Now, for any such target $(p, q)$, let $\mu(p, q) = \Esub{\tilde V\sim P}{\prof(\tilde V, (p, q))}$, with associated estimator $\hat{\mu}(p, q) = (q-p)\hat P(p, q)$. We immediately get: 
    $$\left\lvert \mu(p, q)-\hat\mu(p, q)\right\rvert = \left\lvert (q-p) \left(P(p, q) -\hat P(p, q)\right)\right\rvert \leq \left\lvert \left(P(p, q) -\hat P(p, q)\right)\right\rvert\leq \frac{\eps}{4} $$
    The above finally proves the claimed statement, since by the construction of $\cF$ and of the function domains in Steps 3 and 4, every queried action is \(2\eps\)-optimal under \(P_0\).
\end{proof}
\begin{remark}\label{rem:sampling_lem_better_guarantee}\normalfont For any $\delta'\in(0,1)$, the same construction gives error $\eps/64$ with probability at least $1-\delta'$ without changing the target or query regions. Keep the tree and blocks at precision $\eps$, and apply \Cref{cl:monotone_estimation} with $\alpha=\eps/[64(2+\log_2(K+1))]$. Each telescoping sum has at most $2+\log_2(K+1)$ terms, so its error is at most $\eps/64$. The resulting query bound is
\[ C\log^2(K)\log(K/\delta') \left(\frac K{\eps^2}+\frac{\log^3(K)}{\eps^3}\right). \]
\end{remark}
\begin{remark}
\label{rem:range_extension}
\normalfont
    The routine of \Cref{lem:efficient_estimation} applies invariably, and with the same guarantees, if the feedback received is not the random trade bit, but a generic random variable with the same average and taking values in a fixed range of bounded length. More specifically, for the above estimation guarantee, we only needed \Cref{cl:monotone_estimation} for random variables with range of length $2$, since we were estimating either the average of a random bit, or the average difference of two random bits. 
    
    For the proof of \Cref{lem:inexpensive_detection} which follows, we need that the success probability of the queried bits are bounded away from $0$, so we add some bias to each and rescale the feedback to preserve the original average, resulting in a random variable which takes values in $\{-\nicefrac{1}{2}, \nicefrac{3}{2}\}$. This does not impact in any way the guarantees of \Cref{lem:efficient_estimation} or the procedure of its proof, since it just suffices to apply \Cref{cl:monotone_estimation} with random variables over a common range of length at most $4$.
\end{remark}

\begin{proof}[Proof of \Cref{lem:inexpensive_detection}]
In this proof, we need to consider profit gaps $\Delta_\rho(a^\star, a)$ at different scales, so we run separate instances of \Cref{lem:efficient_estimation}, at precisions
\[
 \eps_j=\min\{2^jx,1\},
 \qquad j=0,\ldots,\lceil\log_2(1/x)\rceil.
\]

We modify the observations so as to prevent arbitrarily small success probabilities, which would be problematic when bounding from above the KL divergences of the associated Bernoulli variables. At each queried action $a$, we add a small bias to the success probability by independently flipping the trade bit with probability $1/4$, obtaining a bit $Z$ with
\[
 \Psub{P}{Z=1}= \frac{1}{4}\left(1-P(a)\right)+\left(1-\frac{1}{4}\right)P(a) = \frac14+\frac12P(a).
\]
Supply the estimation routine with $2Z-\tfrac12$.
This observation has mean $P(a)$ and takes values in
$\{-\tfrac12,\tfrac32\}$: notice that \Cref{lem:efficient_estimation} immediately applies to these tweaked observations, by \Cref{rem:range_extension}.

Now, let $n_j(a)$ be the expected number of queries at $a$ in
experiment $j$ under $P_0$. Then set: 
\[
 n_x(a)=2\sum_j n_j(a), 
\]
which depends only on $\rho$ and $x$. In the following, we bound the cost of each experiment by multiplying its sample bound by the largest cost of one query. Specifically, by the guarantee on actions played of \Cref{lem:efficient_estimation} and the definition of $\eps_j$, every query in experiment $j$ satisfies:
$$
 \Delta_\rho(a^\star,a)+x
 \le2\eps_j+x\le3\eps_j.
$$
Hence 
\begin{align*}
 \sum_{a\in\AA}n_x(a)
       \bigl(\Delta_\rho(a^\star,a)+x\bigr)
 &\le6\sum_j\eps_j\sum_{a\in\AA}n_j(a)\\
 &\le 6C\log^3K\sum_j
       \left(\frac{K}{\eps_j}
             +\frac{\log^3K}{\eps_j^2}\right)\\
 &\le C'\log^3K
       \left(\frac{K}{x}+\frac{\log^3K}{x^2}\right),
\end{align*}
for some other constant $C'$, where the last inequality sums two geometric series.
This proves the cost bound.

\medskip

Next, we show that, for every improving distribution $P$, there is a precision $\eps_j$ at which the estimation routine distinguishes $P$ from $P_0$ through its returned profit estimates. Fix any $P$ satisfying the improvement condition. 

Choose an action $a'\in \AA$ maximizing $\Esub{P}{\prof(\tilde V,a)}$ and choose $j$ as the first scale with
$\eps_j\ge\Delta_\rho(a^\star,a')$. We get
\begin{equation}\label{eq:profit_gap}
 \Esub{P}{\prof(\tilde V,a')}
 -\Esub{\rho}{\prof(V,a')}
 \ge x+\Delta_\rho(a^\star,a')
 >
 \frac{\eps_j}{2}.
\end{equation}
The first inequality follows from the improvement assumption and the second inequality follows by minimality of $j$, when $j\neq 0$. If $j=0$, then $\Delta_{\rho}(a^\star, a')\leq x$ and $\eps_j = x$, and the same inequality holds.
Thus the difference in profit exceeds twice the estimation error $\eps_j/4$. 
Write $\Psub{P_0}{\cdot}$ and $\Psub{P}{\cdot}$ for probabilities
when the routine's trade observations are generated under
$P_0$ and $P$, respectively, including the independent
bit-flipping coins. In both cases, the routine uses the same
reference $P_0$ and precision $\eps_j$. Let $E$ be the event that the returned estimate at $a'$ is within $\eps_j/4$ of $\Esub{\rho}{\prof(V,a')}$.
Then \Cref{eq:profit_gap} and the estimation guarantee of \Cref{lem:efficient_estimation} give
\[
 \Psub{P_0}{E}\ge0.9,
 \qquad
 \Psub{P}{E}\le0.1.
\]

Finally, we analyze the KL divergence between the distributions
of the estimation procedure's transcript under $P_0$ and $P$
to prove the weighted separation bound. Let $H$ be the sequence of processed bits received by the estimation  routine at precision $\eps_j$. Write $\mathcal L_{P_0}(H)$ and $\mathcal L_P(H)$ for its laws under $P_0$ and $P$, respectively, keeping the routine and its reference fixed.
Since the returned estimates are determined by $H$, so is
the event $E$. Pinsker's inequality therefore gives
\begin{equation}\label{eq: transcriptH_lower_bound}
 \operatorname{KL}\!\left(
 \mathcal L_{P_0}(H)\,\middle\|\,\mathcal L_P(H)
 \right)
 \ge2\bigl(\Psub{P_0}{E}-\Psub{P}{E}\bigr)^2
 \ge2(0.8)^2>1.
\end{equation}

For the upper bound, we express this divergence as the sum of the contributions from individual observations.
Write $\operatorname{Bern}(u)$ for the Bernoulli law with mean $u$. Conditional on any processed history, the query and stopping rules are identical under both laws, so only the next observation contributes to KL divergence. The chain rule consequently gives, with $\tau$ being the random number of queries effectively performed by the procedure, $N$ being the deterministic cap on them given by \Cref{lem:efficient_estimation} and $a_t$ the action played at round $t$:
\begin{equation}\label{eq:transcriptH_upper_bound}
\begin{aligned}
 \operatorname{KL}\!\left(
 \mathcal L_{P_0}(H)\,\middle\|\,\mathcal L_P(H)
 \right) &= \sum_{t=1}^N\sum_{a\in \AA}\Psub{P_0}{t\leq \tau, a_t = a}\,
 \operatorname{KL}\!\left(
 \operatorname{Bern}\!\left(\frac14+\frac12P_0(a)\right)
 \,\middle\|\,
 \operatorname{Bern}\!\left(\frac14+\frac12P(a)\right)
 \right)\\
 &=
 \sum_{a\in\AA}n_j(a)\,
 \operatorname{KL}\!\left(
 \operatorname{Bern}\!\left(\frac14+\frac12P_0(a)\right)
 \,\middle\|\,
 \operatorname{Bern}\!\left(\frac14+\frac12P(a)\right)
 \right)\\
 &\le \sum_{a\in \AA}n_j(a)\frac{\nicefrac{1}{4}\cdot(P(a)-P_0(a))^2}{(\nicefrac{1}{4}+\nicefrac{1}{2}\cdot P(a))\cdot (\nicefrac{3}{4}-\nicefrac{1}{2}\cdot P(a))}\\
 &\le
 \frac43\sum_{a\in\AA}n_j(a)
                 \bigl(P(a)-P_0(a)\bigr)^2, 
\end{aligned}
\end{equation}
where the second to last inequality uses the chi-squared bound, see \Cref{prop:chi_squared}.

Finally, first using $n_x(a)\ge2n_j(a)$, and then combining \Cref{eq: transcriptH_lower_bound} and \Cref{eq:transcriptH_upper_bound}, we obtain
\[
\sum_{a\in\AA}n_x(a)\bigl(P(a)-P_0(a)\bigr)^2
\ge2\sum_{a\in\AA}n_j(a)\bigl(P(a)-P_0(a)\bigr)^2
>\frac32\ge1.
\]
\end{proof}

\section{\texorpdfstring{$\widetilde O(\sqrt{T})$}{O-tilde(sqrt(T))} Regret for Monotone-Hazard-Rate Distributions}
\label{sec:MHR}
\begingroup
\let \cC \cK

In this section, we show that the additional structure supplied by the monotone-hazard-rate assumption improves the regret from $\widetilde O(T^{\nicefrac 23})$ to $\widetilde O(\sqrt{T})$.  Notably, the algorithm still requires only one-bit feedback.

\begin{theorem}[Regret for MHR distributions]
\label{thm:mhr-regret}
There exists a learning algorithm $\cA$ with one-bit feedback such that, for every $T$ and every pair $\cD$ of independent MHR buyer and seller distributions on $[0,1]$ with strictly positive densities,
\[ R_T(\cA,\cD) \le O\left(\sqrt{T}\,(\log(T))^{5/2}\right).\]
\end{theorem}

To prove the above, the first fundamental fact is the convexity of the surrogate loss $L(p, q) = -\log\mu(p, q)$, which is established in the following proposition. 
\convex*
\begin{proof}
    Notice that we can write explicitly the function $L(p, q) = - \log (q-p) - \log \Fs(p) - \log (1-\Fb(q))$, which is well defined and continuous in the interior of $\cT$ (indeed, the probability density functions are strictly positive by the MHR assumption).  The first term is jointly convex on $p<q$. The remaining two terms depend on separate coordinates and have nondecreasing derivatives $-\fs(p)/\Fs(p)$ and $\fb(q)/(1-\Fb(q))$, respectively, by MHR. Thus all three terms are convex, and so is their sum. %
\end{proof}
The convexity of $L(p,q)$ suggests an algorithm based on stochastic convex optimization, which we unravel in the next few subsections. 
Specifically, \Cref{subsec:convex_bandits} presents the underlying stochastic convex bandit result, and \Cref{subsec:convex_optimization} describes our approach and algorithm. Next, \Cref{subsec:mhr-safe-region} analyzes the construction of a safe convex region, and \Cref{subsec:mhr-oracle} explains how to estimate $L$ from one-bit feedback. Finally, \Cref{subsec:mhr-analysis} proves the regret bound in \Cref{thm:mhr-regret}.

\subsection{Stochastic Convex Bandits}
\label{subsec:convex_bandits}

We introduce the central statement for the proof of \Cref{thm:mhr-regret}, which is a corollary of a known theorem on stochastic convex bandits by \citet{Lattimore021} and gives us an algorithm we denote as $\COPT$. We defer the proof of the following claim to \Cref{app:mhr}.

\convexmhrroutine*

\subsection{Convex Optimization Algorithm}
\label{subsec:convex_optimization}

Given the result of the previous section, we describe our online algorithm, namely how to use stochastic convex bandit routine $\COPT$ to minimize the convex loss $L(p,q)=-\log\mu(p,q)$ using one-bit feedback. In order to reduce our problem to a setting where the $\COPT$ applies, several conditions need to be met. We require a convex region containing a nearly optimal price pair, on which $-\log \mu$ takes a known range of constant size. We must learn this region with high probability and $\widetilde O(\sqrt{T})$ regret. We also need unbiased loss observations of $-\log \mu$ with bounded noise, obtained from $1$-bit feedback.
In this subsection, we give a detailed description of our algorithm.

\par\addvspace{\medskipamount}

\range takes as input the horizon $T$ and grid resolution $K$.
It builds a sparse set of grid pairs by recursive bisection and
samples them at successively halved scales $\thres\ge1/K$, using
independence to estimate every grid profit from three empirical
trade frequencies, as explained in the proof of \Cref{prop:mhr-concentration}. 
At the first scale with $\max_e\widehat\mu(e)\ge3\thres/2$,
the procedure returns the convex hull of the grid pairs with
$\widehat\mu(e)\ge3\thres/4$, with these thresholds leaving slack for
the estimation error.

\begin{algorithm}[H]
\captionsetup{labelformat=default,font=normalsize}
\caption{\textsc{Initialize mhr}}
\label{alg:mhr-bootstrap}
\small
\begin{algorithmic}
\Statex \textbf{Input:} Horizon $T$ and grid resolution $2\le K\le T$.
\Statex \textbf{Output:} A scale $\thres$ and a convex region $\cC$,
or ``no trigger''.

\State $G_K\gets\{(i/K,j/K):0\le i\le j\le K\}$.
\State Recursively bisect $\{0,\ldots, K\}$ down to singletons and for every resulting discrete interval $I$, let $k_I = \lfloor \tfrac{\max I + \min I}{2}\rfloor $
\State Let $\mathcal S_I = \{(r, k_I):r\in I, k_I\geq r\} \cup \{(k_I, r): r\in I, k_I < r\}$ and $\mathcal S = \cup_I \mathcal S_I $

\For{$\thres=1,1/2,1/4,\ldots$ with $\thres\ge1/K$}
    \State $N\gets\lceil 2^{16}\log(2T)/\thres\rceil$.
    \For{$(i,j)\in\mathcal S$}
        \State Post $(i/K,j/K)$ for $N$ fresh rounds; let
        $\overline P_{ij}$ be the fraction with trade.
    \EndFor

    \State Set $\widehat\mu(e)\gets0$ for every $e\in G_K$.
    \For{$1\le i<j\le K-1$}
        \State Let $k$ be the midpoint of the first recursive split
        separating $i$ and $j$.
        \State $\displaystyle \widehat\mu(i/K,j/K)\gets
            \frac{j-i}{K}\,
            \frac{\overline P_{ik}\overline P_{kj}}
            {\max\{\overline P_{ik},\overline P_{kj},\overline P_{kk}\}}$,
            with $0/0=0$.
    \EndFor
    \If{$\max_{e\in G_K}\widehat\mu(e)\ge3\thres/2$}
        \State $\cC\gets\operatorname{conv}
        \{e\in G_K:\widehat\mu(e)\ge3\thres/4\}$.
        \State \Return $(\thres,\cC)$.
    \EndIf
\EndFor
\State \Return ``no trigger''.
\end{algorithmic}
\end{algorithm}

\par\addvspace{\medskipamount}
\optimize receives the output $(\thres,\cC)$ of \range, grid resolution $K$ and $T_0$, the number of remaining trading rounds after \range. By MHR, we have the additional property that the loss function $L(p,q) = -\log(\mu(p,q))$ is bounded on $\cC$. Moreover, it is possible to obtain an unbiased estimator of $L(p,q)$ by querying $(p,q)$ until the first time trade occurs, and this estimator can be shown to have $\psi_2$ noise at most $4\sqrt{1+\log(2K)}$, where $\psi_2$ is the standard subgaussian norm, defined below for completeness.

\begin{definition}[Subgaussian norm, see e.g. Definition 2.6.4 in \citet{vershynin2026highdimensional}]
\label{def:subgaussian_norm}
A random variable $X$ is called subgaussian if $\exists K > 0$ such that $\E{\exp{\nicefrac{X^2}{K^2}}}\leq 2$. $\lVert X\rVert_{\psi_2}$ is the infimum value of $K$ satisfying the condition, which defines a norm on the space of subgaussian random variables.
\end{definition}

Since the conditions explained above are satisfied, \optimize can run the convex stochastic bandit subroutine \COPT{} with respect to the loss function $L(p,q)$ on $\cC$, noise bound $4\sqrt{1+\log(2K)}$ and loss range $[-\log6\thres,\log \nicefrac{2}{\thres}]$, using our unbiased estimator for each query. 
The mechanism runs \COPT{} for at most $T_0$ query rounds, and the actual query rounds performed correspond to a stochastic number of bilateral trade trading rounds, due to the randomness of the unbiased estimator. The procedure does anyway halt when the number of trading rounds reaches $T_0$, in order to respect the time horizon.

\begin{algorithm}[H]
\captionsetup{labelformat=default,font=normalsize}\caption{\textsc{Optimize}}
\label{alg:mhr-optimization}
\small
\begin{algorithmic}
\Statex \textbf{Input:} \range output $(\thres,\cC)$, grid resolution $K$,
and $T_0$ remaining trading rounds.

\textbf{Subroutine:} \COPT
\Statex \hspace{\algorithmicindent}\textbf{Input:} $\cC$, $n$, loss range, $\psi_2$ noise bound, and past query--observation pairs.
\Statex \hspace{\algorithmicindent}\textbf{Output:} Next query point in $\cC$.
\State $n\gets T_0$; $b\gets T_0$.

\State Initialize \COPT{} on $\cC$ for $n$ queries, with
\Statex \hspace{\algorithmicindent} loss range $[-\log6\thres,\log\nicefrac{2}{\thres}]$ and
$\psi_2$ noise bound $4\sqrt{1+\log(2K)}$.

\For{$\tau=1,\ldots,n$}
    \If{$b=0$} \textbf{stop}
    \EndIf
    \State Receive the next query $x_\tau=(p_\tau,q_\tau)$ from \COPT.
    \State Post $x_\tau$ until the first trade or until $b$ rounds have been used.
    \State Let $N_\tau$ be the number of rounds used; set $b\gets b-N_\tau$.
    \If{no trade occurred}
        \State Post $(0, 0)$ in every remaining trading round; \textbf{stop}.
    \EndIf
    \State Send $\displaystyle
        \sum_{k=1}^{N_\tau-1}\frac1k-\log(q_\tau-p_\tau)$
    to \COPT.
\EndFor
\end{algorithmic}
\end{algorithm}
\par\addvspace{\medskipamount}


\par\addvspace{\medskipamount}
\mhr combines \range and \optimize within a horizon of $T$
trading rounds. It sets
\(
    K=\left\lfloor c_0\frac{\sqrt T}{\log(2T)}\right\rfloor,
\)
where $c_0\in(0,1/2]$ is a sufficiently small universal constant, with this choice balancing the approximation error of the grid against the cost of the initial phase. First, it runs \range with inputs $T$ and $K$. If no scale triggers, it posts $(0,0)$
in every remaining round.
When \range returns $(\thres,\cC)$, the mechanism passes this
output, together with $K$ and the number $T_0$ of remaining
trading rounds, to \optimize. 

\begin{algorithm}[H]
\captionsetup{labelformat=default,font=normalsize}
\caption{\textsc{Bilateral mhr}}
\label{alg:mhr}
\small
\begin{algorithmic}
\Statex \textbf{Input:} Horizon $T$ and access to one-bit feedback.
\Statex \textbf{Output:} Price pairs $(p_t,q_t)$ posted in rounds $t=1,\ldots,T$.
\State Fix a sufficiently small universal constant $c_0\in(0,1/2]$.
\State $K\gets\left\lfloor c_0\sqrt T/\log(2T)\right\rfloor$.
\State Run \Call{\range}{$T,K$}.
\If{the procedure returns ``no trigger''}
    \State Post $(0,0)$ in every remaining round; \textbf{stop}.
\EndIf
\State Let $(\thres,\cC)$ be the returned pair and $T_0$ the number of
remaining rounds.
\State $\Call{Optimize}{\thres,\cC,K,T_0}$.
\end{algorithmic}
\end{algorithm}
\par\addvspace{\medskipamount}

\subsection{Learning a Safe Convex Region}
\label{subsec:mhr-safe-region}

In this subsection, we show that $\cC$ contains a near optimal price, that the expected profit of all prices in $\cC$ is at a constant factor of the current scale $\delta$, and give an upper bound on the regret incurred by \range.

\begin{lemma}[\range guarantees]
\label{lem:mhr-safe-body}
\label{clm:mhr-simultaneous-confidence}
\label{lem:mhr-budget}
Let $2\le K\le T$. Whenever \range triggers, the region $\cC$ returned is nonempty, compact, convex, and contained in the interior of $\cT$.
With probability at least $1-\nicefrac{1}{T}$, the following hold:
\begin{enumerate}[label=(\roman*)]
    \item If \range returns ``no trigger'', then $\mu^\star<6/K$;
    otherwise, its returned scale satisfies $\thres\le \mu^\star<6\thres$.

    \item Every returned region $\cC$ satisfies
    \begin{equation*}\label{eq:mhr-safe-level}
        P(p,q)\ge\mu(p,q)\ge\thres/2
        \qquad\text{for all }(p,q)\in\cC.
    \end{equation*}

    \item Every returned region $\cC$ contains all grid
    maximizers, implying, via \Cref{lem:discretization_err}:
    \begin{equation*}\label{eq:mhr-body-comparator} \mu^\star_{\cC}=\max_{(p,q)\in\cC}\mu(p,q)\ge \mu^\star-2/K.
    \end{equation*}
\end{enumerate}
    Moreover, for a universal constant $c_0$, choosing
    \begin{equation}\label{eq:mhr-grid-size}
        K=\left\lfloor c_0\frac{\sqrt T}{\log(2T)}\right\rfloor
    \end{equation}
    ensures that \range uses at most $T$ rounds on every history and incurs $O(\sqrt T\log(T))$ expected regret. If it returns ``no trigger'', posting $(0,0)$ thereafter gives the same regret bound over the entire horizon.
\end{lemma}

The following proposition deals with the concentration and reconstruction guarantees of estimators in \range. We defer its proof to Appendix \ref{app:mhr}.
\begin{restatable}[Simultaneous concentration]{proposition}{mhrconcentration}
\label{prop:mhr-concentration}
\label{lem:mhr-recovery}
Let $2\le K\le T$. With probability at least $1-\nicefrac{1}{T}$,
simultaneously at every scale $\thres$ reached by \range,
the estimates computed at that scale satisfy
\begin{equation}\label{eq:mhr-lower-estimates}
    \bigl|\widehat\mu(e)-\mu(e)\bigr|
    \le \frac{\mu(e)+\thres}{16}
    \qquad\text{for every }e\in G_K.
\end{equation}
\end{restatable}
\noindent For the remainder of this section, let $\clean$ be the event that
\Cref{eq:mhr-lower-estimates} holds for every $e\in G_K$ at every
scale reached by \range. Using the associated accuracy guarantee, we can now prove the main statement.

\begin{proof}[Proof of \Cref{lem:mhr-safe-body}]
Whenever a scale triggers, at least one grid pair has estimate
at least $\nicefrac{3\thres}{2}$, so the set selected at $\nicefrac{3\thres}{4}$ is nonempty. Boundary and diagonal estimates are zero,
so every selected pair lies in $\{(p,q):0<p<q<1\}$.
The convex hull of this finite set is therefore nonempty, compact,
convex, and contained in the same open triangle, on every history. For the remaining conclusions, assume $\clean$ holds and write
$\mu^\star_K=\max_{e\in G_K}\mu(e)$. 

Suppose \range triggers at scale $\thres$. If $\mu^\star_K<\thres$,
then $\clean$ gives
\[
    \max_{e\in G_K}\widehat\mu(e)
    \le \mu^\star_K+\frac{\mu^\star_K+\thres}{16}
    <\frac{9\thres}{8}<\frac{3\thres}{2},
\]
contradicting the trigger and thus implying $\mu^\star_K\ge\thres$.
If $\thres<1$, the preceding scale $2\thres$ did not trigger.
If $\mu^\star_K\ge4\thres$, the estimate of any grid maximizer at that
preceding scale needs to be at least
\(
    \mu^\star_K-\frac{\mu^\star_K+2\thres}{16}
    \ge\frac{29\thres}{8}>3\thres,
\)
which contradicts the fact that the algorithm did not trigger at scale $2\thres$.
For $\thres=1$, the bound $\mu^\star_K<4\thres$ follows from $\mu^\star_K\le1$.
Hence
\(
    \thres\le \mu^\star_K<4\thres.
\)
By the grid approximation bound \Cref{lem:discretization_err} and $\thres\ge1/K$,
$\thres\le \mu^\star_K\leq  \mu^\star\le \mu^\star_K+2/K<6\thres$, proving (i).

At the triggering scale, $\clean$ also gives
\[
\begin{aligned}
    \mu(e)\ge\thres
    &\quad\Longrightarrow\quad
    \widehat\mu(e)\ge\frac{15\mu(e)-\thres}{16}
    \ge\frac{7\thres}{8}>\frac{3\thres}{4},\\
    \mu(e)<\thres/2
    &\quad\Longrightarrow\quad
    \widehat\mu(e)\le\frac{17\mu(e)+\thres}{16}
    <\frac{19\thres}{32}<\frac{3\thres}{4}.
\end{aligned}
\]
Consequently,
\begin{align}
    \{e\in G_K:\mu(e)\ge\thres\}&\subseteq
    \{e\in G_K:\widehat\mu(e)\ge\nicefrac{3\thres}{4}\}\notag\\
    &\subseteq
    \{e\in G_K:\mu(e)\ge\nicefrac{\thres}{2}\}\notag\\ 
    &\subseteq \{(p,q)\in [0, 1]^2: 0<p<q<1,\, \mu(p,q)\ge\nicefrac{\thres}{2}\} \label{eq:inclusion_mhr_convex}
\end{align}
By log-concavity of $\mu$, the superlevel set
$\{(p,q): 0<p<q<1,\, \mu(p,q)\ge\thres/2\}$ is convex. Taking the convex hull over the above chain of inclusions therefore proves that the convex hull $\cK$ of selected prices is contained in the superlevel set:
together with $P(p,q)\ge\mu(p,q)$, this proves (ii).
Since $\mu^\star_K\ge\thres$, the first inclusion in \Cref{eq:inclusion_mhr_convex}
shows that $\cC$ contains every grid maximizer.
Thus $\mu^\star_{\cC} \ge \mu^\star_K\ge \mu^\star-2/K$, via \Cref{lem:discretization_err}, proving (iii).

Finally, if no scale triggers, let $\thres$ be the last scale tested.
Then
\(1/K\le\thres<2/K.\)
If $\mu^\star_K\ge2\thres$, a grid maximizer would have estimate at least
\[
    \mu^\star_K-\frac{\mu^\star_K+\thres}{16}
    \ge\frac{29\thres}{16}>\frac{3\thres}{2},
\]
forcing a trigger. Therefore, $\mu^\star_K<2\thres<4/K$, and \Cref{lem:discretization_err} gives $\mu^\star<6/K$.

It remains to prove the round and regret bounds for the grid choice
in \Cref{eq:mhr-grid-size}. For the round bound, consider an
arbitrary observation history and restrict $c_0\le1/2$, so that
$K\le T$.
Each interval in the recursive construction of $\mathcal S$
contributes at most as many pairs as it has indices. At any fixed
depth the intervals are disjoint, so they contribute at most $K+1$
pairs in total. There are at most
$\lceil\log_2(K+1)\rceil+1$ depths. Thus
\[
    |\mathcal S|
    \le(K+1)\bigl(\lceil\log_2(K+1)\rceil+1\bigr)
    \le6K\log(2T).
\]

At each tested scale $\thres$, \range posts every pair in
$\mathcal S$ for
\[
    \left\lceil\frac{2^{16}\log(2T)}{\thres}\right\rceil
\]
fresh rounds. Constructing $\mathcal S$, reconstructing the grid
estimates, and checking the trigger require no additional trading
rounds. Letting $\thres$ denote the last tested scale, notice that the reciprocals of the tested scales sum to $1+2+\cdots+\nicefrac{1}{\thres}=\nicefrac{2}{\thres}-1<\nicefrac{2}{\thres}$. 
Therefore, writing $C=12(2^{16}+1)$, we get that the total number of rounds is bounded by at most:
\begin{equation}\label{eq:bound_numberof_bootstrap_rounds} 
    C\frac{K\log^2(2T)}{\thres}
    \le CK^2\log^2(2T)
    \le Cc_0^2T,
\end{equation}
where we used $\thres\ge1/K$ and the choice of $K$ from \Cref{eq:mhr-grid-size}.
Choosing $c_0=\nicefrac{1}{\sqrt C}$ gives a bound of $T$ on every history.

For the regret bound, assume $\clean$ holds again. The bounds already
proved give $\mu^\star<6\thres$ if \range triggers at $\delta$.
\Cref{eq:bound_numberof_bootstrap_rounds} then yields a cumulative regret bound of
\[  
    6C\delta\frac{K\log^2(2T)}{\delta}
    \le6CK\log^2(2T)
    \le6Cc_0\sqrt T\log(2T).
\]
Similarly, if no scale triggers and $(0,0)$ is posted thereafter, the regret
over the entire horizon is at most
\[
    \sum_{t=1}^{T}\bigl(\mu^\star-\mu(p_t,q_t)\bigr)
    \le T\mu^\star<\frac{6T}{K}
    \le\frac{12}{c_0}\sqrt T\log(2T), 
\]
where the second to last inequality follows by $\mu^\star\leq \nicefrac{6}{K}$ if \range does not trigger.

We conclude the proof by noticing that $\P{\clean}\geq 1-\nicefrac{1}{T}$, by \Cref{prop:mhr-concentration}, which also implies that the expected regret associated to $\clean^c$ can be at most constant.
\end{proof}

For the remainder of the section, consistently with \Cref{alg:mhr}, we use the choice of $K$ in
\Cref{eq:mhr-grid-size}.

\subsection{A One-Bit Log-Profit Oracle}
\label{subsec:mhr-oracle}

The previous subsection proved that assuming $\clean$, if \range triggers at some scale $\thres$, it returns a convex set $\cC$ which approximately contains the optimum and on which expected profit is always within a constant factor of the optimum. We now need to show that \optimize efficiently finds $\max_{(p,q)\in\cC}\mu(p,q)$  with one-bit feedback. 

The first ingredient is building an estimator of the convex loss $L(p, q)$ from $1$-bit observations. For $m\ge1$, let $\mathsf H_m=\sum_{k=1}^m1/k$, and set
$\mathsf H_0=0$. Recall that for a query $x=(p,q)\in\cC$, \optimize posts $x$ until the first trade. Let $N$ be the number of rounds used, then
\begin{equation}\label{eq:mhr-oracle}
    Z=\mathsf H_{N-1}-\log(q-p)
\end{equation}
is the input given to $\COPT$ in \optimize. We now show that $Z$ is indeed an unbiased estimator of the loss $L(p,q) = -\log(\mu(p,q))$, with the needed tail control. While the unbiasedness of this estimator is a well-known fact (see Theorem 4.1 in \citet{DeGroot1959} for a more general result), we provide a proof in Appendix \ref{app:mhr} for completeness and to establish the subgaussian norm bound required to apply \Cref{prop:mhr-convex-bandit}.

\begin{restatable}[Estimating a logarithm from a geometric waiting time]{claim}{logestimation}
\label{lem:mhr-harmonic}
Let $Q\in(0,1]$ and let $N\sim \text{\normalfont Geom}(Q)$. Then
\[
    \E{\mathsf H_{N-1}}=-\log Q,
    \qquad
    \|\mathsf H_{N-1}+\log Q\|_{\psi_2}
    \le 4\sqrt{1-\log Q}.
\]
\end{restatable}

We now have all the ingredients to verify  that the region and loss observations used by \optimize satisfy the conditions of \Cref{prop:mhr-convex-bandit}. The next lemma
gives a regret bound for the queries, assuming each query is allowed to run until trade.

\begin{lemma}[Convex optimization from trade bits]
\label{lem:mhr-convex-oracle}
Fix a complete transcript compatible with the clean event $\clean$ on which \range triggers and returns $(\thres,\cC)$.
For any integer $n\ge1$, initialize \COPT{} as in \optimize for $n$ queries and answer each query using the estimator in \Cref{eq:mhr-oracle}, without a trading-round limit.
Conditional on this transcript, the resulting query points $x_1,\ldots,x_n\in\cC$ satisfy

\begin{equation}\label{eq:mhr-virtual-regret}
\begin{aligned}
    \E{\sum_{\tau=1}^n\left(L(x_\tau)-\min_{x\in\cC}L(x)\right)}
    &=\E{\sum_{\tau=1}^n\log\frac{\mu^\star_{\cC}}{\mu(x_\tau)}}
    &=O\bigl(\sqrt{n\log(T)}\,\log^2(2n)\bigr).
\end{aligned}
\end{equation}
\end{lemma}

\begin{proof}
By \Cref{lem:mhr-safe-body}, if \range triggers, the set $\cC$ is nonempty, compact, and convex. By \Cref{prop:conv_mhr}, $L$ is continuous and convex on $\cC$, since $\cC$ is a subset of the interior of $\cT$.
Moreover, \Cref{lem:mhr-safe-body} with $\clean$ gives
\[
    \mu(p, q)\in [\nicefrac{\thres}{2}, 6\thres]
    \qquad((p,q)\in\cC),
\]
which implies that $L$ on $\cK$ takes values in a fixed (given $\thres$) interval of constant length:
\begin{equation}\label{eq:mhr-loss-range}
    L(p,q)\in [-\log 6\thres, \log \nicefrac{2}{\thres}]
    \qquad((p,q)\in\cC).
\end{equation}

We next verify the observation and noise bounds.
Fix a query $(p,q)$. The algorithm holds these prices fixed until a trade occurs.
Since future valuation pairs are i.i.d. given the choice of $(p, q)$, the waiting time $N$ is geometric with success probability which can be lower bounded by, using $\thres\geq \nicefrac{1}{K}$:
\begin{equation}\label{eq:mhr-trade-floor}
    P(p,q)=\frac{\mu(p,q)}{q-p}
    \ge\mu(p,q)\ge\frac \thres 2\ge\frac1{2K}.
\end{equation}

By \Cref{lem:mhr-harmonic}, the estimator $Z$ in \Cref{eq:mhr-oracle} is unbiased for $L(p, q)$, and we have: 
$$
\lVert Z-L(p, q)\rVert_{\psi_2}\leq 4\sqrt{1-\log P(p,q)}\le4\sqrt{1+\log(2K)},
$$
where the inequality follows from \Cref{eq:mhr-trade-floor}.

Finally, $\cC$ has affine dimension at most two.
Applying \Cref{prop:mhr-convex-bandit} with the above guarantees yields:
\[
    \E{\sum_{\tau=1}^n\left(L(x_\tau)-\min_{x\in\cC}L(x)\right)}
    =O\bigl(\sqrt{n(1+\log(2K))}\,\log^2(2n)\bigr).
\]
Since $K\le T$ and
\[
    L(x_\tau)-\min_{x\in\cC}L(x)
    =\log\frac{\mu^\star_{\cC}}{\mu(x_\tau)},
\]
this proves \Cref{eq:mhr-virtual-regret}.
\end{proof}

\subsection{Proving the Regret Bound}
\label{subsec:mhr-analysis}
Using the guarantees of \Cref{lem:mhr-convex-oracle}, we can now proceed to the proof of the main result of this section, the regret bound in \Cref{thm:mhr-regret}. As in \Cref{alg:mhr,alg:mhr-optimization}, let $T_0$ be the number of bilateral trade rounds remaining after the execution of \range. The main idea is that we can pretend that \COPT{} always runs until the maximum number of convex optimization query rounds $T_0$, even though we may have to stop earlier due to the number of bilateral trade pricing rounds having reached $T_0$. Indeed, the regret of the bandit routine restricted to $T_0$ bilateral trade rounds is for sure bounded by the regret of the unrestricted routine which stops after $T_0$ convex bandits queries.

\begin{proof}[Proof of \Cref{thm:mhr-regret}]
Recall that the regret is given by
\begin{equation}\label{eq:mhr-pseudo-regret}
    \E{\sum_{t=1}^T\bigl(\mu^\star-\mu(p_t,q_t)\bigr)}.
\end{equation}
where every summand belongs to $[0,1]$. We bound this sum assuming the clean event $\clean$ from \range. This is possible because the contribution of $\clean^c$ to the regret is at most $T\P{\clean^c}$, and \Cref{lem:mhr-safe-body} gives that $\P{\clean^c}\leq \nicefrac{1}{T}$. Conditionally on $\clean$, if \range did not trigger, \Cref{lem:mhr-safe-body} gives total regret over the horizon of $O(\sqrt{T}\log T)$.

Now, fix a complete \range transcript on which $\clean$ holds and the procedure triggers. This implies the remaining horizon $T_0$, scale $\thres$, and region $\cC$ are fixed. Recall that  \Cref{lem:mhr-safe-body} bounds the regret of \range by $O(\sqrt T\log T)$. It remains to bound the regret of \optimize over $T_0$ rounds.
For the analysis, continue the same $\COPT$ execution until all $n=T_0$ convex optimization queries finish, and let $N_\tau$ be the number of trading rounds used by query $\tau$. Since $N_\tau\ge1$, the actual remaining $T_0$ bilateral trade rounds form a prefix of this execution. Splitting off the approximation error gives
\begin{equation}\label{eq:mhr-prefix-regret}
\begin{aligned}
 \sum_{t=T-T_0+1}^T\bigl(\mu^\star-\mu(p_t,q_t)\bigr)
 &=T_0\bigl(\mu^\star-\mu^\star_{\cC}\bigr)
   +\sum_{t=T-T_0+1}^T
       \bigl(\mu^\star_{\cC}-\mu(p_t,q_t)\bigr)\\
 &\le\frac{2T_0}{K}
   +\sum_{\tau=1}^{T_0}N_\tau
       \bigl(\mu^\star_{\cC}-\mu(x_\tau)\bigr).
\end{aligned}
\end{equation}
The inequality uses \Cref{lem:mhr-safe-body}(iii) for the approximation error.

Now, let $\cF_{\tau-1}$ be the sigma algebra generated by the history before query $\tau$ (waiting times for \Cref{eq:mhr-oracle} and randomization of \COPT, including the choice of $x_\tau$); also let $N_\tau$ be the waiting time for the estimator associated to query $\tau$. Conditional on $\cF_{\tau-1}$, $N_\tau$ is a geometric random variable with success probability $P(x_\tau)$, and therefore, using (ii) from \Cref{lem:mhr-safe-body}: 
$$
\E{N_\tau\mid \cF_{\tau-1}} = \frac{1}{P(x_\tau)}\leq \frac{2}{\thres}
$$
Taking the outer expectation and using that $x_\tau$ is measurable according to $\cF_{\tau-1}$ (see e.g. Chapter 9 in \citet{Williams_1991}), the instantaneous bilateral trade regret suffered for query $\tau$ is thus at most:
\begin{align*}
\E{\E{N_\tau(\mu^\star_\cC-\mu(x_\tau))\mid \cF_{\tau-1}}} &= \E{(\mu^\star_\cC-\mu(x_\tau))\E{N_\tau\mid \cF_{\tau-1}}} \tag{$(\mu^\star_\cC-\mu(x_\tau))$ bounded and $\E{N_\tau}<\infty$}\\
&= \E{\frac{\mu^\star_\cC-\mu(x_\tau)}{P(x_\tau)}}\\
&\leq \E{\frac{\mu^\star_\cC}{\mu(x_\tau)}-1} \tag{$P(x_\tau)\geq \mu(x_\tau)$} \\
&\leq 5\E{\log\frac{\mu^\star_\cC}{\mu(x_\tau)}} \tag{Point (i)-(ii) of \Cref{lem:mhr-safe-body}, $z-1\leq 5\log z$ for $z\in [1, 12]$},
\end{align*}

Finally, we can combine the above with \Cref{eq:mhr-prefix-regret} and \Cref{lem:mhr-convex-oracle} to get a regret bound of:
    $$
    \frac{2T}{K}+\sum_{\tau=1}^{T_0}\E{N_\tau(\mu^\star_\cC-\mu(x_\tau))} = O\left(\sqrt{T}\log T\right) + O\left(\sqrt{T_0\log(2T)}\log^2(2T_0)\right) \leq O\left(\sqrt{T}\log^{\nicefrac{5}{2}}T\right),
    $$
    where we used the value of $K$ from \Cref{eq:mhr-grid-size}.
\end{proof}

\endgroup

\begingroup
\let \cC \cK
\section{\texorpdfstring{$\widetilde O(T^{\nicefrac 35})$}{O-tilde(T\textasciicircum(3/5))} Regret for Regular Distributions}\label{sec:Regular}

\begin{theorem}[Regret for regular distributions]
\label{thm:regular}
There exists a learning algorithm $\cA$ with one-bit feedback such that, for every $T$ and every pair $\cD$ of independent regular buyer and seller distributions on $[0,1]$ with strictly positive densities,
\[
 R_T(\cA,\cD)\leq O\!\left(T^{3/5}\log^5(T)\right).
\]
\end{theorem}

Throughout the proof, let $\eta\in(0,1)$ denote the target additive
profit accuracy we aim for. Set the discretization parameter $K=\lceil1/\eta\rceil$ and assume $K\le T$. We choose $\eta$ as a function of $T$ at the end.

\subsection{A family of convex auxiliary objectives}

As opposed to \Cref{sec:MHR}, regularity need not make $-\log\mu$ convex. We instead define \emph{auxiliary}
profits with convex reciprocals that match profit at suitable parameters.
For $x=(p,q)\in\operatorname{int}(\cT)$ and $\lambda>0$, let

\begin{equation}\label{eq:regular-auxiliary}
 r_\lambda(p,q)=
 \frac{4(q-p)}{\bigl(\lambda/\Fs(p)+\lambda^{-1}/(1-\Fb(q))\bigr)^2},
 \qquad
 \lambda_x=\sqrt{\frac{\Fs(p)}{1-\Fb(q)}}.
\end{equation}
We call $\lambda_x$ the \emph{balancing parameter} of $x$. The following lemma justifies optimizing profit through this family.

\begin{lemma}[Auxiliary profits]\label{lem:regular-auxiliary}
For every $\lambda>0$, the function $x\mapsto1/r_\lambda(x)$ is
convex on $\operatorname{int}(\cT)$. Writing $\operatorname{sech}(t)=2/(e^t+e^{-t})$, we have for every
$x\in\operatorname{int}(\cT)$,
\begin{equation}\label{eq:regular-balance}
 r_\lambda(x)=\mu(x)\operatorname{sech}^2\!\left(\log\frac{\lambda}{\lambda_x}\right),
 \qquad
 0\le\mu(x)-r_\lambda(x)
 \le\mu(x)\left(\log\frac{\lambda}{\lambda_x}\right)^2.
\end{equation}
In particular, $0<r_\lambda(x)\le\mu(x)$ and
$r_{\lambda_x}(x)=\mu(x)$. Moreover, for every $\theta\in\mathbb R$,
\begin{equation}\label{eq:regular-stability}
 \opt -r_{\lambda e^\theta}(x)
 \le2\bigl(\opt -r_\lambda(x)\bigr)+2\opt \theta^2.
\end{equation}
\end{lemma}

\begin{proof}
Regularity implies that $1/(1-\Fb)$ is convex
\cite[Lemma 3.4]{regular-link}; applying the same result to the
reflected seller value $1-s$ then gives convexity of $1/\Fs$. The map $(u,t)\mapsto u^2/t$ is convex for $t>0$ and non-decreasing in $u\ge0$. Composing it with the positive convex sum $\lambda/\Fs(p)+\lambda^{-1}/(1-\Fb(q))$ and the positive affine function $4(q-p)$ proves convexity of $1/r_\lambda$.

Now set $\xi=\log(\lambda/\lambda_x)$. By definition, $\lambda = e^\xi \sqrt{F_s(p)/(1-F_b(q))}$, so that: 
$$
\frac{\lambda}{F_s(p)} = \frac{e^\xi}{\sqrt{F_s(p)(1-F_b(q))}} \qquad \frac{\lambda^{-1}}{1-F_b(q)} = \frac{e^{-\xi}}{\sqrt{F_s(p)(1-F_b(q))}},
$$
which implies $r_\lambda (x) = \mu(x)\sech^2(\xi)$, by definition of $\sech$.

Now let $\tanh(t)=(e^t-e^{-t})/(e^t+e^{-t})$ be the hyperbolic tangent. By the standard identity $\tanh^2(t) = 1-\sech^2(t)$, we immediately also get:
\begin{equation}
\label{eq:tanh_equality}
    \mu(x)-r_{\lambda}(x)=\mu(x)\tanh^2(\xi).
\end{equation}
Finally, since $\tanh(0)=0$ and $0\leq\tanh'\leq 1$ everywhere, the mean value theorem implies that $\lvert\tanh(\xi)\rvert\leq \lvert\xi\rvert$, which in turn yields $\mu(x)-r_{\lambda}(x)\leq \mu(x)\xi^2$ as claimed. 

To conclude, notice that \Cref{eq:tanh_equality} and $\tanh$ being Lipschitz give:
 \begin{align*}
 \opt-r_{\lambda e^\theta}(x)
 &=\opt-\mu(x)+\mu(x)\tanh^2\left(\log \frac{\lambda e^{\theta}}{\lambda_x}\right)\\
 &=\opt-\mu(x)+\mu(x)\tanh^2\left(\xi+\theta\right)\\
 &\le\opt-\mu(x)+2\mu(x)\tanh^2(\xi)+2\mu(x)\theta^2 \\
 &\le2\bigl(\opt-r_\lambda(x)\bigr)+2\opt \theta^2,
 \end{align*}
where the last step uses $\mu(x)\le\opt$.
\end{proof}

What the above lemma tells us is essentially that, if we did know $\lambda_{x^\star}$, where $x^\star\in \arg\max_{(p, q)\in [0, 1]^2}\mu(p, q)$, then the corresponding auxiliary profit $r_{\lambda_{x^\star}}$ would be exactly the right objective to optimize. Moreover, an approximation to $\lambda_{x^\star}$ in log-scale would suffice for the same purpose, due to \Cref{eq:regular-balance}. \Cref{eq:regular-stability} instead certifies that if a price has auxiliary profit close to $\mu^\star$ for a certain $\lambda$, then its auxiliary profit cannot decrease too much if $\lambda$ gets a small additive perturbation in log-scale.

We now need to find a way to estimate $r_\lambda$. Notice that one-bit feedback means that the learner observes only the joint trade bit, whereas $r_\lambda$ depends on the two marginal acceptance probabilities. To address this, let $x=(p, q)$ and $z=(u, v)$ be two prices in $\operatorname{int}(\cT)$, and call $(p,v)$ and $(u,q)$ their \emph{crossed} pairs. We describe how knowing the joint trade probabilities at $z$ and its two crosses lets us compute $r_\lambda$ in $x$. 
Specifically, for a known multiplier $\rho>0$ and $\lambda=\rho\lambda_z$,
we can estimate the auxiliary profit and its convex reciprocal using queries at $z$ and its two crosses, since independence implies
\begin{equation}\label{eq:regular-observable}
 \frac1{r_\lambda(x)}
 =\frac{P(z)}{4(q-p)}
 \left(\frac{\rho}{P(p,v)}+\frac{\rho^{-1}}{P(u,q)}\right)^2.
\end{equation}

Using this identity for estimation requires trading at the crossed prices, and the following lemma ensures that this is compatible with low regret. When $x$ and $z$ have high auxiliary profit for the same $\lambda$, their crosses belong to $\cT$ and also have high profit, with comparable trade probabilities.

\begin{lemma}[Crossed prices]\label{lem:regular-cross}
If $x=(p,q)$ and $z=(u,v)$ are two pairs of prices in $\operatorname{int}(\cT)$ satisfying
$r_\lambda(x),r_\lambda(z)\ge0.99\opt $, then $(p,v)$ and $(u,q)$
belong to $\cT$, and either crossed pair $y$ satisfies
\[
 \opt -\mu(y)\le
 4\bigl(\opt -r_\lambda(x)+\opt -r_\lambda(z)\bigr).
\]
Moreover, the four trade probabilities are within a factor of 3 of each other.
\end{lemma}

\begin{proof}
Let
\begin{equation*}
\label{eq:def_hx_hz}
 h_x=\frac12\left(\frac{\lambda}{\Fs(p)}
       +\frac{\lambda^{-1}}{1-\Fb(q)}\right),\qquad
 h_z=\frac12\left(\frac{\lambda}{\Fs(u)}
       +\frac{\lambda^{-1}}{1-\Fb(v)}\right).
\end{equation*}
First, we obtain a bound between each term in the average defining $h_x$ and $h_x$:
\begingroup
\allowdisplaybreaks
\begin{alignat}{2}
 \left|\frac{\lambda}{\Fs(p)}-h_x\right|
 &=\left|\frac{\lambda^{-1}}{1-\Fb(q)}-h_x\right|
 &&\quad\text{(the two terms sum to }2h_x\text{)}
 \notag\\
 &=\sqrt{h_x^2-\frac1{P(x)}}
 &&\quad\text{($(\lambda/\Fs(p)-h_x)^2=h_x^2-1/P(x)$)}
 \notag\\
 &=\sqrt{h_x^2-\frac{q-p}{\mu(x)}}
 &&\quad\text{(}\mu(x)=(q-p)P(x)\text{)}
 \notag\\
 &=h_x\sqrt{1-\frac{r_\lambda(x)}{\mu(x)}}
 &&\quad\text{(}q-p=r_\lambda(x)h_x^2\text{, by \Cref{eq:regular-balance})}
 \notag\\
 &\le h_x\sqrt{\frac{\opt-r_\lambda(x)}{\opt}}
 &&\quad\text{(}\mu(x)\le\opt\text{)}
 \notag\\
 &\le0.1h_x
 &&\quad\text{(}r_\lambda(x)\ge0.99\opt\text{)}.
 \label{eq:regular-cross-factors}
\end{alignat}
\endgroup
In particular, this implies a bound on the crossed price probabilities
\begin{equation}\label{eq:regular-cross-probabilities}
 P(p,v)=\left(\frac{\lambda}{\Fs(p)}
              \frac{\lambda^{-1}}{1-\Fb(v)}\right)^{-1}
 \le\frac1{(0.9h_x)(0.9h_z)}=\frac1{0.81h_xh_z}.
\end{equation}
The same holds for $P(u,q)$.

Now write $\Delta=(\opt-r_\lambda(x))+(\opt-r_\lambda(z))\le0.02\opt$. Then
\begin{align}
0&\le\frac{\opt-\mu(p,v)}{P(p,v)}
       +\frac{\opt-\mu(u,q)}{P(u,q)}
\notag\\
&= \mu^\star\left(\frac{1}{P(p,v)}+\frac{1}{P(u,q)}\right)
-\left(\frac{\mu(p,v)}{P(p,v)}
+\frac{\mu(u,q)}{P(u,q)}\right)
\notag\\
&= \mu^\star\left(
\frac{\lambda\lambda^{-1}}{F_s(p)(1-F_b(v))}
+\frac{\lambda\lambda^{-1}}{F_s(u)(1-F_b(q))}
\right)-((v-p)+(q-u))
\tag{Independence}
\\
&= \mu^\star\left(
\frac{\lambda}{F_s(p)}
\left(2h_z-\frac{\lambda}{F_s(u)}\right)
+\frac{\lambda}{F_s(u)}
\left(2h_x-\frac{\lambda}{F_s(p)}\right)
\right)-((q-p)+(v-u))
\notag\\
&=2\opt h_xh_z
-2\opt\left(\frac{\lambda}{\Fs(p)}-h_x\right)
       \left(\frac{\lambda}{\Fs(u)}-h_z\right)
-r_\lambda(x)h_x^2-r_\lambda(z)h_z^2
\tag{\(q-p=r_\lambda(x)h_x^2\)}
\\
&\le 2\opt h_xh_z-r_\lambda(x)h_x^2-r_\lambda(z)h_z^2+2h_xh_z\sqrt{\opt-r_\lambda(x)}\sqrt{\opt-r_\lambda(z)}     \tag{\(\ref{eq:regular-cross-factors},\,
\left\lvert\frac{\lambda}{F_s(p)}-h_x\right\rvert
\leq h_x\sqrt{\frac{\mu^\star-r_\lambda(x)}{\mu^\star}}\)}\notag\\
&=\left(
h_x\sqrt{\opt-r_\lambda(x)}
+h_z\sqrt{\opt-r_\lambda(z)}
\right)^2-\opt(h_x-h_z)^2 \tag{Completing the square}
\\
&\le\Delta(h_x^2+h_z^2)-\opt(h_x-h_z)^2
\tag{Cauchy--Schwarz}
\\
&=2\Delta h_xh_z-(\opt-\Delta)(h_x-h_z)^2. \label{eq:regular-cross-weighted-gap}
\end{align}
Since $\opt-\Delta>0$, applying first the bound above and then \Cref{eq:regular-cross-probabilities} verifies our claim, with $y$ as one of the crossed pairs:
\begin{equation}\label{eq:crossed profit bound}
 \opt-\mu(y)\le P(y)\,2\Delta h_xh_z
 \le\frac{2\Delta}{0.81}\le4\Delta.
\end{equation}
Thus both crossed profits are at least $0.92\opt>0$, proving that both crossed prices are in $\cT$.

For the probability comparison, the penultimate line of
\Cref{eq:regular-cross-weighted-gap} also gives
\[
 |h_x-h_z|\le\sqrt{0.02(h_x^2+h_z^2)}
 \le0.2\max\{h_x,h_z\}.
\]
Hence $\min\{h_x,h_z\}\ge0.8\max\{h_x,h_z\}$.
Consider the original pairs $x=(p,q)$, $z=(u,v)$ and their crosses
$(p,v)$, $(u,q)$. For each of these four pairs, the reciprocal of the trade probability is a
product of two factors bounded in \Cref{eq:regular-cross-factors}.
Each factor is thus at least $0.9\min\{h_x,h_z\}$ and at most
$1.1\max\{h_x,h_z\}$, so that the reciprocal of the probability of trade is in $[(0.9\min\{h_x, h_z\})^2, (1.1\max\{h_x, h_z\})^2]$.
Taking maxima and minima over the four pairs therefore gives
\[
 \begin{aligned}
 \frac{\max P}{\min P}=\frac{\max(1/P)}{\min(1/P)}
\le\left(\frac{1.1\max\{h_x,h_z\}}
                 {0.9\min\{h_x,h_z\}}\right)^2
 \le\left(\frac{1.1}{0.9\cdot0.8}\right)^2<3.
 \end{aligned}
\]
\end{proof}

\subsection{Constructing the convex candidate problems}\label{sec:Constructing the convex candidate problems}

In \Cref{sec:MHR}, \range finds a suitable convex region $\cC$ for the \emph{single} loss function $L$. In this section, \init must also select candidate objectives
$1/r_\lambda$, which is nontrivial since the balancing parameters $\lambda_x$ depend on unknown marginal distributions. Recalling \Cref{lem:regular-auxiliary}, we seek a small family containing a candidate $\lambda$ whose auxiliary profit nearly attains $\opt$, together with convex regions on which
crossed queries (see \Cref{lem:regular-cross}) have controlled regret.

Initialization starts from empirical estimates $\widehat P(e)$ of
the trade probabilities $P(e)$ for $e\in\AA = \grid\cap\cT$, constructed
in Step~\ref{step:regular-init-estimates}, and retains pairs with
high estimated profit. It groups these pairs around representative retained pairs,
called \emph{roots}, by comparing their balancing parameters.
Although the parameters are unknown, their ratios can be estimated
from trade bits at crossed pairs: for $x=(p,q)$ and $z=(u,v)$,
independence gives
\begin{equation}\label{eq:regular-balance-ratio}
 \log\frac{\lambda_x}{\lambda_z}
 =\frac12\log\frac{P(p,v)}{P(u,q)}.
\end{equation}

Each group contains pairs with high profit and similar balancing
parameters, so they have a common high auxiliary-profit bound at the
root's parameter. Convexity of the reciprocal in
\Cref{lem:regular-auxiliary} extends this bound to the group's
convex hull. \Cref{lem:regular-cross} then controls the regret of
crossed queries throughout that hull.

For each hull, we discretize a short interval of logarithmic offsets
around the root's balancing parameter. The quadratic approximation
bound in \Cref{lem:regular-auxiliary} ensures that this small mesh suffices to preserve a nearly optimal choice.
Each hull--offset pair defines one candidate: if $\cC_i$ is its hull, $w_i$ its root, and $\theta_i$ its offset, its parameter is
\(
 \lambda_i=e^{\theta_i}\lambda_{w_i}.
\)
Each candidate stores $\cC_i$, $w_i$, $\theta_i$, and the root estimate
$\widehat{P}(w_i)$, all of which remain fixed. In the following, we use again the routine \range from \Cref{sec:MHR}.

The proof of \Cref{lem:regular-init}, which summarizes the properties of \init, is deferred to \Cref{app:missing_regular}. 
\begin{algorithm}[H]
\captionsetup{labelformat=default,font=normalsize}
\caption{\textsc{Initialize regular}}
\label{alg:regular-initialization-sketch}
\small
\begin{algorithmic}[1]
\Statex \textbf{Input:} Horizon $T$, target accuracy $\eta$ with
$K=\lceil1/\eta\rceil\le T$, and access to one-bit feedback.
\Statex \textbf{Output:} A scale $\thres$ and candidates, or ``no trigger''.
\Statex Stop at trading round $T$, even if initialization is unfinished.

\State \label{step:regular-init-scale}\textbf{Find the profit scale.} Run \range unchanged, with
inputs $T,K$, and retain its scale $\thres$ and sampled index set $\cS$.
Return ``no trigger'' if \range does so or if $\thres<2^{16}/K$.
\State \label{step:regular-init-estimates}\textbf{Sample more accurately at the fixed scale.} After \range
returns, post each $(i/K,j/K)$ with $(i,j)\in\cS$ for
$n=\lceil2^{38}\log(2T)/\thres\rceil$ new rounds, and let
$\overline P_{ij}$ be the empirical trade frequency.
Use the same reconstruction as \range: for $1\le i<j\le K-1$, set
\begin{equation}\label{eq:regular-empirical-estimate}
 \widehat P(i/K,j/K)=
 \frac{\overline P_{ik}\overline P_{kj}}
      {\max\{\overline P_{ik},\overline P_{kj},\overline P_{kk}\}},
 \qquad 0/0=0,
\end{equation}
where $k$ is the midpoint of the first recursive split separating $i,j$.
Use $\widehat P(i/K,i/K)=\overline P_{ii}$ for $1\le i\le K-1$,
and set probabilities with $i=0$ or $j=K$ to zero.
\State \label{step:regular-init-retain}\textbf{Retain promising pairs.} Set
$\widehat\mu(p,q)=(q-p)\widehat P(p,q)$ and let
$\widehat\mu_K^\star$ be its maximum over $\AA$.
Keep pairs with $\widehat\mu\ge(1-2^{-13})\widehat\mu_K^\star$.
\State \label{step:regular-init-group}\textbf{Group prices.} Process retained pairs in a fixed order.
Assign $x=(p,q)$ to an existing root $w=(u,v)$ if both crosses
$(p,v)$ and $(u,q)$ lie in $\cT$,
have estimated profit at least $0.99\widehat\mu_K^\star$, and satisfy
\[
 \left|\frac12\log\frac{\widehat P(p,v)}{\widehat P(u,q)}\right|
 \le0.01.
\]
If no root passes all tests, start a new group with $x$ as its root.
\State \label{step:regular-init-hulls}\textbf{Take convex hulls.} Use each group's convex hull as a region.
\State \label{step:regular-init-offsets}\textbf{Discretize the parameter.} For
$D=\operatorname{arctanh}(1/\sqrt{2000})$, divide $[-D,D]$ into the
fewest equal intervals of length at most $\tfrac12\sqrt{\eta/\thres}$.
For each hull, use these endpoints as offsets and create one candidate
per offset, saving its root and the root's probability estimate.
\State \Return $\thres$ and the candidates
$(\cC_i,w_i,\theta_i,\widehat{P}(w_i))_{i=1}^N$.
\end{algorithmic}
\end{algorithm}

\begin{restatable}[Initialization]{lemma}{initproperties}\label{lem:regular-init}
Consider \init running until it is finished, without stopping at $T$.
With probability at least $1-1/T$, it either
returns \emph{no trigger}, with $\opt=O(1/K)$, or returns a scale
$\thres>\eta$ satisfying $\thres\le\opt<6\thres$ and a list of
$N=O(\sqrt{\thres/\eta}\log(2T))$ fixed candidates
$(\cC_i,w_i,\theta_i,\widehat{P}(w_i))$.
Each $\cC_i\subset\operatorname{int}(\cT)$ is compact and convex,
contains its root $w_i$, and has a stored empirical estimate
$\widehat{P}(w_i)$ of $P(w_i)$. The candidates satisfy:
\begin{enumerate}[label=(\roman*),ref=(\roman*)]
\item \label{item:regular-init-auxiliary}\textbf{Auxiliary-profit guarantees.}
For every candidate $i$ and $|u|\le10^{-3}$,
$r_{\lambda_i e^u}\ge0.99\opt$ throughout $\cC_i$.
Some fixed candidate $i^\star$ satisfies
\[
 \max_{x\in\cC_{i^\star}}r_{\lambda_{i^\star}e^u}(x)
 \ge\opt-3\eta
 \qquad (|u|\le10^{-3}\sqrt{\eta/\thres}).
\]
\item \label{item:regular-init-crosses}\textbf{Crossed-query bounds.}
For any two points of $\cC_i$, both crosses belong to $\cT$.
Their probabilities of trade, and the probabilities at the original points,
belong to $[\widehat{P}(w_i)/10,10\widehat{P}(w_i)]$. Moreover,
\[
 \thres\le\widehat{P}(w_i)\le1,\qquad
 \frac{\thres}{q-p}\le10\widehat{P}(w_i)
 \quad((p,q)\in\cC_i).
\]
\end{enumerate}
On a valid triggered transcript, initialization incurs
$O(K\log^2(T))$ regret.
\end{restatable}
\subsection{Successive elimination}\label{sec:Successive elimination}

Consider a candidate from the list constructed by Algorithm~\ref{alg:regular-initialization-sketch} with parameter
\(
 \lambda_i=e^{\theta_i}\lambda_{w_i}.
\)
We know the offset $\theta_i$, but the root's balancing parameter
$\lambda_{w_i}$ depends on unknown marginal probabilities.
\Cref{eq:regular-observable} shows that, for any multiplier $\rho>0$,
a \textit{reference} pair of prices $z$ can be used to produce
observations of $1/r_{\rho\lambda_z}$.
To approximate the candidate's intended objective $1/r_{\lambda_i}$,
we thus choose $\rho$ by estimating
$\lambda_i/\lambda_z=e^{\theta_i}\lambda_{w_i}/\lambda_z$.
Initially $z=w_i$, so this multiplier is simply $e^{\theta_i}$.
The following \emph{successive elimination} algorithm repeatedly refines
active candidates and eliminates those with low estimated auxiliary profit, up until the final trading round $T$.

\begin{algorithm}[H]
\captionsetup{labelformat=default,font=normalsize}
\caption{\textsc{Successive Opt: Refinement and Successive Elimination}}
\label{alg:regular-refinement}
\small
\begin{algorithmic}
\Statex \textbf{Input:} Horizon $T$, target accuracy $\eta$, the output
of initialization, and access to one-bit feedback.
\Statex \textbf{Output:} Price pairs posted in the remaining trading rounds.
\Statex Stop at trading round $T$, even while producing a loss observation.

\If{\init returned ``no trigger''}
    \State Post $(0,0)$ in every remaining round; \textbf{stop}.
\EndIf
\State Mark all candidates active and initialize their references
$z_i\gets w_i$.
\For{$j=0,1,\ldots$ while trading rounds remain}
    \State Set the accuracy $\eps_j\gets\thres2^{-j}$ and run length
    $n\gets4^j\lceil C_0\log^4(2T)\rceil$, for a sufficiently large universal constant $C_0$.
    \For{each active candidate $i$}
        \State \textbf{Calibrate.} Using offset $\theta_i$, fresh trade
        bits at crossed prices formed by $w_i$ and $z_i$, estimate
        $\rho_i\approx\lambda_i/\lambda_{z_i}$.

        \State \textbf{Refine.} Keep $z_i$ and $\rho_i$ fixed throughout
        the following independent runs.
        \For{each of $\lceil C_0\log(2T)\rceil$ runs}
            \State Run $\COPT$ on $\cC_i$ to minimize
            $1/r_{\rho_i\lambda_{z_i}}$ for $n$ loss observations.
            Construct each observation from fresh trade bits at $z_i$
            and its crossed prices with the optimizer's query.
            \State Record the average query point and the average
            loss observation from this run.
        \EndFor

        \State \textbf{Select.} Choose the run with the smallest
        empirical average loss $\widehat L$. Let $y_i$ be its
        average query point and set
        $\widehat r_i\gets1/\max\{\widehat L,1/(10\thres)\}$.
    \EndFor

    \State \textbf{Eliminate.} Retain precisely the candidates satisfying
    $\widehat r_i\ge
    \max_{\ell\ {\rm active}}\widehat r_\ell-10(\eta+\eps_j)$.
    \State \textbf{Update references.} Set $z_i\gets y_i$ for every
    surviving candidate, ready for calibration in the next phase.
\EndFor
\end{algorithmic}
\end{algorithm}

\paragraph{Unbiased estimator}
After each elimination phase, surviving references are updated to match
the candidates' improved profits. \emph{Calibration} estimates
$\lambda_i/\lambda_z$ from the crosses between the root $w_i$ and the
updated reference $z$. The following lemma bounds the calibration error, the observations'
number of rounds and regret.

\begin{lemma}[Convex loss observations]\label{lem:regular-observation}
Assume the guarantees of the high probability event in Lemma~\ref{lem:regular-init} hold, with \init triggering before time $T$.
Fix a candidate $i$ and a reference $z\in\cC_i$.
\begin{enumerate}[label=(\roman*),ref=(\roman*)]
\item \label{item:regular-calibration}\textbf{Calibration.}
Calibration can be implemented so as to incur $O(\log(T)/\eta)$ regret, returning a multiplier
$\rho$ such that, except with probability at most $1/T$,
\begin{equation}\label{eq:regular-effective}
 \left|\log\frac{\rho\lambda_z}{\lambda_i}\right|
 \le10^{-3}\sqrt{\eta/\thres}.
\end{equation}
\item \label{item:regular-loss-observations}\textbf{Loss observations.}
For a constant $C$ and for $\rho$ satisfying \Cref{eq:regular-effective}, set
$\lambda=\rho\lambda_z$ and $L=1/r_\lambda$.
Each query $x\in\cC_i$ admits an observation $Y_x$ using at least one
trading round at prices in $\cT$, with
\begin{equation}\label{eq:regular-oracle}
 0\le Y_x\le C/\thres,\qquad
 \E{Y_x\mid\text{\normalfont past},x}=L(x),
\end{equation}
with the expected number of trading rounds to produce $Y_x$ bounded by $C/\thres$, again conditionally on the choice of $x$ and on the history before generating $Y_x$.
Every price pair $y$ posted to produce $Y_x$ moreover satisfies
\begin{equation}\label{eq:regular-round-regret}
 \opt-\mu(y)\le C\bigl(\opt-r_\lambda(z)\bigr)
 +C\thres^2\left(L(x)-\min_{\cC_i}L\right).
\end{equation}
\end{enumerate}
\end{lemma}

\begin{proof}
Throughout this proof, write $w=(a,b)$ for the fixed candidate's root and
$z=(u,v)$ for the reference.
For part~\ref{item:regular-calibration}, sample each crossed pair
$y\in\{(u,b),(a,v)\}$ for
$\lceil C\thres\log(2T)/(\eta\widehat P(w))\rceil$ fresh rounds.
Let $\overline P(y)$ be the fraction of these new rounds in which a
trade occurs, and set
\[
 \widehat d=\frac12\log\frac{\overline P(u,b)}{\overline P(a,v)},
\]
with $\widehat d=0$ if either frequency is zero.
By \Cref{eq:regular-balance-ratio}, this estimates
$\frac12\log(P(u,b)/P(a,v))=\log(\lambda_z/\lambda_w)$.
Both trade probabilities are at least $\widehat P(w)/10$ by
\Cref{lem:regular-init}\ref{item:regular-init-crosses}, so the expected
number of trades at each cross $y$ is at least $C\thres\log(2T)/(10\eta)$.
The multiplicative Chernoff bound (see \Cref{prop:chernoff}) and
a union bound therefore give, for a sufficiently large universal $C$,
\[
 \left|\frac{\overline P(y)}{P(y)}-1\right|
 \le\frac{10^{-3}}2\sqrt{\eta/\thres}
 \qquad(y\in\{(u,b),(a,v)\}),
\]
except with probability $1/T$.
Since $\eta<\thres$, both frequencies are then positive, and
$|\log(1+t)|\le2|t|$ for $|t|\le1/2$ gives
\[
 \left|\widehat d-\log\frac{\lambda_z}{\lambda_w}\right| = \left\lvert \frac{1}{2}\left(\log\left(\frac{\overline P(u, b)}{P(u, b)}\right)-\log\left(\frac{\overline P(a, v)}{P(a, v)}\right)\right) \right\rvert
 \le\frac12\sum_{y\in\{(u,b),(a,v)\}}
       \left|\log\frac{\overline P(y)}{P(y)}\right|
 \le10^{-3}\sqrt{\eta/\thres}.
\]
Since $\lambda_i=e^{\theta_i}\lambda_w$, choosing $\rho=e^{\theta_i-\widehat d}$
gives $\log(\rho\lambda_z/\lambda_i)=\log(\lambda_z/\lambda_w)-\widehat d$,
so the preceding bound proves \Cref{eq:regular-effective}.

By \Cref{lem:regular-init}\ref{item:regular-init-crosses},
$\widehat P(w)\ge\thres$, so the total number of rounds is at most
$2\lceil C\log(2T)/\eta\rceil=O(\log(T)/\eta)$.
The same point forces both crosses in $\cT$, so that their expected profit is non-negative, immediately implying an $O(\log(T)/\eta)$ regret bound.

For part~\ref{item:regular-loss-observations}, fix a calibration
satisfying \Cref{eq:regular-effective} and set $\lambda=\rho\lambda_z$.
Since $\eta<\thres$, its error satisfies
$|\log(\lambda/\lambda_i)|\le10^{-3}\sqrt{\eta/\thres}\le10^{-3}$,
so the perturbed-parameter guarantee in
\Cref{lem:regular-init}\ref{item:regular-init-auxiliary} applies.
Together with $r_\lambda(x)\le\mu(x)$ from \Cref{lem:regular-auxiliary}, it gives
\begin{equation}\label{eq:regular-calibrated-profit-range}
 0.99\thres\le0.99\opt\le r_\lambda(x)
 \le\mu(x)\le\opt\le6\thres
 \qquad(x\in\cC_i).
\end{equation}
The latter lemma also gives convexity of $L=1/r_\lambda$;
continuity follows from continuity of the CDFs and positivity of the
denominators on $\cC_i$.

We now construct an observation of $L(x)$ for a query $x=(p,q)\in \cK_i$.
\Cref{eq:regular-observable} expresses $L(x)$ as a known
multiple of $P(z)$ times the square of a weighted sum of reciprocal
crossed probabilities. We represent these factors by one bit $B_0$
and two independent copies $B,B'$: multiplying the copies will produce
the square in expectation.
All queried prices are $z$, $(p,v)$, or $(u,q)$: by
\Cref{lem:regular-init}\ref{item:regular-init-crosses}, they are in $\cT$ and their probabilities lie between $\widehat P(w)/10$ and
$10\widehat P(w)$.

First, represent the reference probability by a bit
\[
 B_0\sim\Ber{\frac{P(z)}{20\widehat P(w)}}.
\]
Its mean is at most $1/2$ by the probability bounds above.
If $20\widehat P(w)<1$, the Bernoulli factory from Theorem 1 in \citet{regular-huber} generates this bit from trade bits at
$z$ in $O(1/\widehat P(w))$ expected trading rounds.
Otherwise, simply retain a trade bit's success with independent
probability $1/(20\widehat P(w))$, with the same cost bound.

Next, represent the weighted reciprocal sum. To generate each of
$B,B'$, independently choose $y=(p,v)$ with probability
$\rho/(\rho+\rho^{-1})$ and $y=(u,q)$ otherwise.
An adaptation of the same procedure from \citet{regular-huber} we have used above can generate a bit of mean
$\widehat P(w)/(20P(y))\le1/2$ from trade bits at $y$ (see Propositions 1, 2, and 4 in \citet{regular-ldl}).
It uses $O(1/P(y))=O(1/\widehat P(w))$ expected rounds, since
$P(y)\ge\widehat P(w)/10$.
Averaging over the two crossed price pairs gives
\[
 \E{B}=\E{B'}=
 \frac{\widehat P(w)}{20(\rho+\rho^{-1})}
 \left(\frac\rho{P(p,v)}+\frac{\rho^{-1}}{P(u,q)}\right).
\]
Use fresh independent randomness and trade bits for all three bits.
If the factories request no trade bits, post $z$ once and discard its
bit, so each observation uses at least one trading round.
Their product now has the required factors in its mean; rescale it by
returning
\begin{equation}\label{eq:regular-observation}
 Y_x=\frac{2000(\rho+\rho^{-1})^2}
 {(q-p)\widehat P(w)}\,B_0BB'.
\end{equation}
To verify unbiasedness, independence of the bits gives
\[
 \begin{aligned}
 \E{Y_x}
 &=\frac{2000(\rho+\rho^{-1})^2}{(q-p)\widehat P(w)}
   \frac{P(z)}{20\widehat P(w)}\bigl(\E{B}\bigr)^2\\
 &=\frac{2000}{20^3}\frac{P(z)}{q-p}
   \left(\frac\rho{P(p,v)}+\frac{\rho^{-1}}{P(u,q)}\right)^2
 =L(x),
 \end{aligned}
\]
where $2000/20^3=1/4$ and the last equality is \Cref{eq:regular-observable}.

It remains to bound the observation size and expected duration. For the size, we first control the multiplier in
\Cref{eq:regular-observation}. Evaluating \Cref{eq:regular-balance}
at $z$ and using $\lambda=\rho\lambda_z$ gives
$r_\lambda(z)=\mu(z)\operatorname{sech}^2(\log\rho)
=4\mu(z)/(\rho+\rho^{-1})^2$.
Rearranging and using \Cref{eq:regular-calibrated-profit-range} yields
\begin{equation}\label{eq:regular-multiplier-bound}
 (\rho+\rho^{-1})^2=\frac{4\mu(z)}{r_\lambda(z)}\le\frac4{0.99}.
\end{equation}
Combining this with $B_0BB'\le1$ and
$\thres/(q-p)\le10\widehat P(w)$ from
\Cref{lem:regular-init}\ref{item:regular-init-crosses} gives, picking $C\ge 81000$:
\[
 0\le Y_x\le\frac{2000(\rho+\rho^{-1})^2}{(q-p)\widehat P(w)}
 \le2000\cdot\frac4{0.99}\cdot\frac{10}{\thres}
 \le\frac C\thres.
\]
For the duration, sum the three $O(1/\widehat P(w))$ factory costs, then use
$\widehat P(w)\ge\thres$ from the same initialization guarantee.
This proves all the observation bounds in \Cref{eq:regular-oracle}.

Finally, to bound the regret of each trading round, $L=1/r_\lambda$ gives
\[
 \begin{aligned}
 \max_{\cC_i}r_\lambda-r_\lambda(x)
 &=r_\lambda(x)\bigl(\max_{\cC_i}r_\lambda\bigr)
   \left(L(x)-\min_{\cC_i}L\right)\\
 &\le36\thres^2\left(L(x)-\min_{\cC_i}L\right),
 \end{aligned}
\]
using $r_\lambda\le6\thres$ from \Cref{eq:regular-calibrated-profit-range}.
Also, $\opt-\max_{\cC_i}r_\lambda\le\opt-r_\lambda(z)$.
Thus \Cref{lem:regular-cross} gives, for either cross $y$ and using again \Cref{eq:regular-calibrated-profit-range},
\[
 \begin{aligned}
 \opt-\mu(y)
 &\le4\bigl(\opt-r_\lambda(x)+\opt-r_\lambda(z)\bigr)\\
 &\le8\bigl(\opt-r_\lambda(z)\bigr)
      +144\thres^2\left(L(x)-\min_{\cC_i}L\right).
 \end{aligned}
\]
The reference $z$ has gap $\opt-\mu(z)\le\opt-r_\lambda(z)$, so the same
bound holds for every round used to produce $Y_x$, proving
\Cref{eq:regular-round-regret}.
\end{proof}

\paragraph{Refining and elimination}
In \refine, $\COPT$ is invoked several times in order to
successively refine our estimates while minimizing the regret.
Here, refinement means to find a point with nearly maximal auxiliary
profit within one candidate's region and to estimate the auxiliary
profit of that point.
More precisely, fix a candidate $i$, its current reference $z_i$, and
a calibrated multiplier $\rho_i$, and write $\lambda=\rho_i\lambda_{z_i}$.
We keep $z_i$ and $\rho_i$ fixed throughout and minimize the convex
loss $L=1/r_\lambda$ on $\cC_i$ using $\COPT$ from \Cref{sec:MHR}.

The loss is convex by \Cref{lem:regular-auxiliary}, continuous by
continuity of the CDFs, and defined on a compact convex region $\cC_i$
by \Cref{lem:regular-init}.
\Cref{lem:regular-observation}\ref{item:regular-loss-observations}
gives conditionally unbiased observations bounded by $C/\thres$.
Multiplying the loss $L(x)$ and the observation $Y_x$ by $\thres$ bounds the absolute value of the two by a constant, by \Cref{lem:regular-init,lem:regular-auxiliary} and \Cref{lem:regular-observation}\ref{item:regular-loss-observations}. This then trivially implies a constant upper bound on the subgaussian norm (see \Cref{def:subgaussian_norm}) of $\thres(Y_x-L(x))$, so that we can safely apply \Cref{prop:mhr-convex-bandit} in the affine hull of $\cC_i$, similarly to what we did in \Cref{sec:MHR}.

Dividing the associated regret bound by $\thres$ yields, for the non-scaled loss $L$ with $\tau$ indexing the ``virtual'' convex optimization time like in \Cref{sec:MHR}:
\begin{equation}\label{eq:regular-copt}
 \E{\sum_{\tau=1}^n\left(L(x_\tau)-\min_{\cC_i}L\right)}
 \le\frac C\thres\sqrt n\,\log^2(2n).
\end{equation}

We then use independent repetitions to turn this expected guarantee into a reliable point and profit estimate, via Markov's inequality:
in phase $j$, at accuracy $\eps_j=\thres2^{-j}$, run
$\lceil C_0\log(2T)\rceil$ independent copies of $\COPT$, each for
\begin{equation}\label{eq:regular-run-length}
 n=4^j\lceil C_0\log^4(2T)\rceil
\end{equation}
observations, where $C_0$ is a sufficiently large universal constant, chosen after the constants in the observation and convex-optimization bounds are fixed. For each copy, average its queries and its loss observations.
Select the copy with the smallest empirical average $\widehat L$, and
return its query average $y_i$ and the corresponding auxiliary-profit estimate
\(
 \widehat r_i=1/{\max\{\widehat L,1/(10\thres)\}}.
\)
$y_i$ becomes the new reference only if candidate $i$ survives the
subsequent elimination step.

We discard candidates whose estimates are more than $10(\eta+\eps_j)$
below the largest estimate, then set $z_i=y_i$ for each survivor.
Starting with $\eps_0=\thres$, we halve the accuracy after each phase
and continue refinement and elimination until trading round $T$.

The following lemma supplies guarantees on the refinement procedure
and elimination rule. Part~\ref{item:regular-refinement} bounds the
accuracy of one refinement, while part~\ref{item:regular-survival} ensures
that a nearly optimal candidate remains available, i.e. the $i^\star$ from \Cref{lem:regular-init} satisfying \[
 \max_{x\in\cC_{i^\star}}r_{\lambda_{i^\star}e^u}(x)
 \ge\opt-3\eta
 \qquad (|u|\le10^{-3}\sqrt{\eta/\thres}).
\]
When needed to disambiguate, we use the superscript to link each iteration of the loop in \Cref{alg:regular-refinement} with the value of the considered parameters at its beginning.

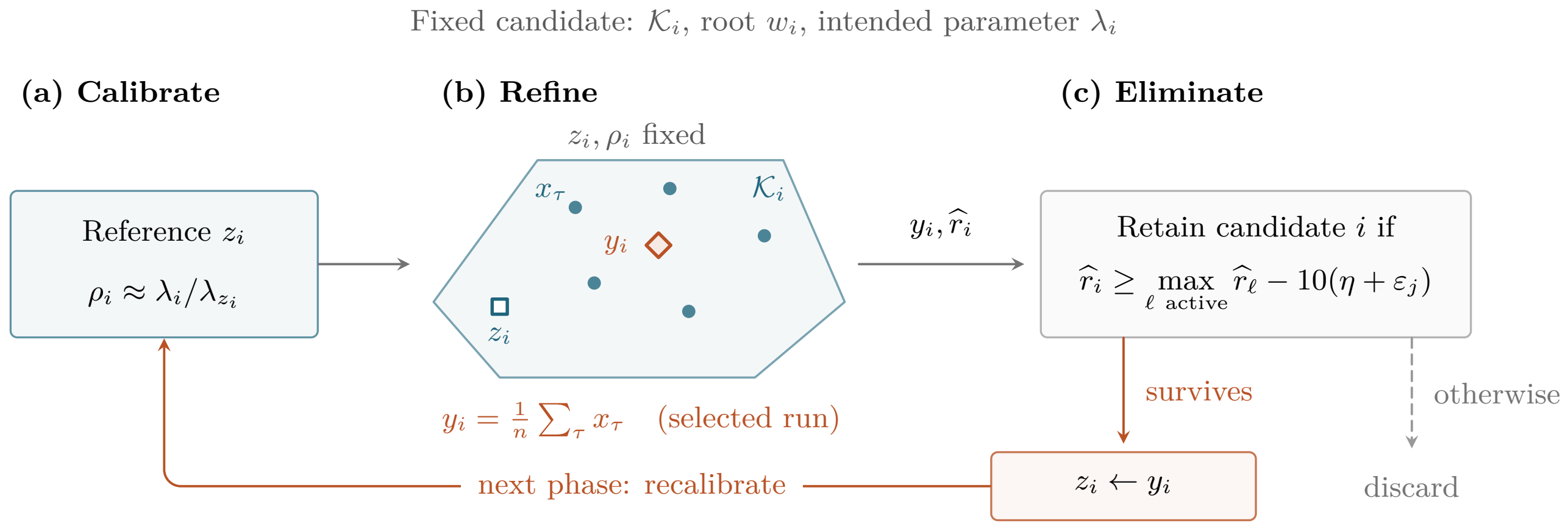
\begin{figure}[t]
\centering
\begingroup
\definecolor{regcycleblue}{RGB}{31,101,125}
\definecolor{regcycleorange}{RGB}{185,82,37}
\resizebox{\linewidth}{!}{%
\begin{tikzpicture}[font=\small, line cap=round, line join=round, >=stealth,
  cycle arrow/.style={->, draw=black!55, line width=.7pt},
  cycle box/.style={draw=black!30, fill=white, rounded corners=2pt,
    line width=.6pt, align=center, inner sep=7pt},
  cycle heading/.style={font=\bfseries\small, anchor=west}]
  \path[use as bounding box] (-.1,-.1) rectangle (16.0,5.65);
  \node[text=black!65] at (8.0,5.45)
    {Fixed candidate: $\cC_i$, root $w_i$, intended parameter $\lambda_i$};

  \node[cycle heading] at (0,4.7) {(a) Calibrate};
  \node[cycle heading] at (4.45,4.7) {(b) Refine};
  \node[cycle heading] at (11.0,4.7) {(c) Eliminate};

  \node[cycle box, draw=regcycleblue!70, fill=regcycleblue!5,
    minimum width=3.25cm, minimum height=1.55cm] (calibrate) at (1.65,2.9)
    {Reference $z_i$\\[7pt]$\rho_i\approx\lambda_i/\lambda_{z_i}$};
  \draw[cycle arrow] (calibrate.east) -- (4.25,2.9);

  \filldraw[draw=regcycleblue!60, fill=regcycleblue!5, line width=.65pt]
    (4.5,2.5) -- (5.2,1.7) -- (7.9,1.7) -- (8.85,2.5)
    -- (8.2,4.0) -- (5.6,4.0) -- cycle;
  \node[text=black!65] at (6.65,4.25) {$z_i,\rho_i$ fixed};
  \node[text=regcycleblue] at (8.05,3.7) {$\cC_i$};
  \draw[regcycleblue, fill=white, line width=1pt]
    (5.12,2.37) rectangle (5.28,2.53);
  \node[anchor=north, text=regcycleblue, inner sep=3pt] at (5.2,2.34) {$z_i$};
  \foreach \px/\qy in {6.0/3.5,6.2/2.7,7.0/3.7,7.2/2.4,8.0/3.2}
    \fill[regcycleblue!80] (\px,\qy) circle[radius=2pt];
  \node[anchor=south east, text=regcycleblue, inner sep=2pt]
    at (6.0,3.5) {$x_\tau$};
  \path[draw=regcycleorange, fill=regcycleorange!12, line width=1pt]
    (6.88,3.23) -- (7.01,3.10) -- (6.88,2.97) -- (6.75,3.10) -- cycle;
  \node[anchor=east, text=regcycleorange, inner sep=4pt]
    at (6.72,3.1) {$y_i$};
  \node[text=regcycleorange] at (6.7,1.25)
    {$y_i=\frac1n\sum_\tau x_\tau$\quad(selected run)};
  \draw[cycle arrow] (9.0,2.9) -- (10.75,2.9)
    node[midway, above=4pt] {$y_i,\widehat r_i$};

  \node[cycle box, fill=black!2, minimum width=4.55cm,
    minimum height=1.55cm] (compare) at (13.2,2.9)
    {Retain candidate $i$ if\\[5pt]
     $\widehat r_i\ge\max\limits_{\ell\ \mathrm{active}}\widehat r_\ell
       -10(\eta+\eps_j)$};
  \draw[cycle arrow, draw=regcycleorange] (11.8,2.12) -- (11.8,1.0)
    node[midway, right=3pt, text=regcycleorange] {survives};
  \node[cycle box, draw=regcycleorange!80, fill=regcycleorange!5,
    minimum width=2.8cm, minimum height=.72cm] (update) at (11.8,.55)
    {$z_i\gets y_i$};
  \draw[cycle arrow, densely dashed, draw=black!40]
    (14.85,2.12) -- (14.85,.95)
    node[midway, right=3pt, text=black!55] {otherwise};
  \node[text=black!55] at (14.85,.55) {discard};
  \draw[cycle arrow, draw=regcycleorange, rounded corners=5pt]
    (update.west) -- (1.65,.55) -- (calibrate.south);
  \node[fill=white, inner xsep=5pt, text=regcycleorange]
    at (6.6,.55) {next phase: recalibrate};
\end{tikzpicture}%
}
\endgroup
\caption{One cycle for candidate $i$. With $z_i,\rho_i$ fixed,
refinement optimizes $1/r_{\rho_i\lambda_{z_i}}$ and returns the selected
run's average query $y_i$ and estimate $\widehat r_i$.
If $i$ survives comparison with the other candidates, $y_i$ becomes its
reference for the next phase. The region, root, and intended parameter
remain fixed.}
\label{fig:regular-refinement-cycle}
\end{figure}

\begin{lemma}[Refinement and successive elimination]\label{lem:regular-refine}
Assume initialization triggers with the guarantees of
Lemma~\ref{lem:regular-init}.
\begin{enumerate}[label=(\roman*),ref=(\roman*)]
\item \label{item:regular-refinement}\textbf{Refinement.}
Fix a candidate $i$ and associated reference and valid calibrated parameter $\lambda$.
In any phase $j$ started before $T$, refinement returns
$y\in\cC_i$ and $\widehat r_i$ satisfying, except with probability $1/T$,
\begin{equation}\label{eq:regular-refinement}
 \max_{\cC_i}r_\lambda-\eps_j
 \le\widehat r_i\le r_\lambda(y)+\eps_j.
\end{equation}
This guarantee refers to refinement run to completion.

\item \label{item:regular-survival}
\label{item:regular-reference-improvement}\textbf{Elimination.}
If all calibrations and refinements through a completed phase $j$
satisfy \Cref{eq:regular-effective,eq:regular-refinement},
then $i^\star$ survives to phase $j+1$ and every retained and updated reference $z_i^{(j+1)}$ satisfies
\begin{equation}\label{eq:regular-reference-gap}
 \opt-r_{\rho^{(j+1)}_i\lambda_{z^{(j+1)}_i}}\left(z_i^{(j+1)}\right)\le65(\eta+\eps_{j+1}),
\end{equation}
where $\rho^{(j+1)}_i$ is the output of the next valid calibration
for candidate $i$. The same bound holds after a valid calibration at the start of phase $0$.
\end{enumerate}
\end{lemma}

\begin{proof}
For part~\ref{item:regular-refinement}, fix the starting history and
write $\eps=\eps_j$. In one optimizer copy, let $x_\tau\in\cC_i$ denote
its queries and $Y_\tau$ the corresponding loss observations constructed
as in \Cref{lem:regular-observation}\ref{item:regular-loss-observations}. Since this phase starts before round $T$, a run of length $n/4$ has already finished if $j\ge1$, so $n\le4T$. For $j=0$, $n=\lceil C_0\log^4(2T)\rceil$.
By \Cref{eq:regular-copt} and the choice of $n$,
\begin{equation}\label{eq:regular-average-loss}
 \E{\frac1n\sum_{\tau=1}^n L(x_\tau)-\min_{x\in\cC_i}L(x)}
 \le\frac C{\thres\sqrt n}\log^2(T)
 \le\frac{\eps}{800\thres^2},
\end{equation}
for sufficiently large $C_0$ in \Cref{eq:regular-run-length}.
Applying Markov's inequality to this nonnegative excess loss gives
\[
 \frac1n\sum_{\tau=1}^n L(x_\tau)
 \le\min_{x\in\cC_i}L(x)+\frac{\eps}{200\thres^2}
\]
with probability at least $3/4$.
Call a copy satisfying this inequality \emph{good}.
Since the $\lceil C_0\log(2T)\rceil$ copies are independent, the probability
that none is good is at most
\[
 (1/4)^{\lceil C_0\log(2T)\rceil}
 \le(2T)^{-C_0\log4}\le\frac1{2T},
\]
for sufficiently large universal $C_0$.

We cannot compute the average true loss of a copy directly. Instead,
we use its empirical average loss
$\widehat L=n^{-1}\sum_{\tau=1}^nY_\tau$, computed from the observations
collected during that run.
Although each query $x_\tau$ may depend on earlier observations,
\Cref{eq:regular-oracle} gives
$\E{Y_\tau-L(x_\tau)\mid\text{past},x_\tau}=0$ and
$|Y_\tau-L(x_\tau)|\le C/\thres$, so that $\{\sum_{\tau\leq n} (Y_\tau-L(x_\tau))\}_n$ is a martingale with bounded increments. Azuma-Hoeffding inequality (see \Cref{prop:azuma}) and a
union bound therefore give, simultaneously for all copies, for sufficiently large $C_0$:
\[
 \left|\widehat L-\frac1n\sum_{\tau=1}^nL(x_\tau)\right|
 \le\frac{\eps}{400\thres^2},
\]
except with probability $1/(2T)$.

Assume both events hold, which happens with probability at least $1-\nicefrac{1}{T}$. There is a good copy, and its empirical
average exceeds its true average by at most $\eps/(400\thres^2)$.
Consequently, its empirical average is at most
\[
 \min_{x\in\cC_i}L(x)
 +\frac{\eps}{200\thres^2}+\frac{\eps}{400\thres^2}
 =\min_{x\in\cC_i}L(x)+\frac{3\eps}{400\thres^2}.
\]
As specified in the refinement procedure, select the copy with the
smallest empirical average. From now on, $x_\tau$ and $\widehat L$
refer to this selected copy.
Hence
\(
 \widehat L\le\min_{x\in\cC_i}L(x)+3\eps/({400\thres^2}).
\)

Next we relate the selected copy's empirical average to the loss at
the returned point. Since the concentration bound holds for every
copy, it also holds for the selected one, giving
\(
 \frac1n\sum_{\tau=1}^nL(x_\tau)
 \le\widehat L+{\eps}/({400\thres^2}).
\)
Recall that the returned point in \refine is $y=n^{-1}\sum_{\tau=1}^nx_\tau$ and belongs to $\cC_i$ by convexity of the region. Convexity of $L$ then gives
\[
 L(y)\le\frac1n\sum_{\tau=1}^nL(x_\tau)
 \le\widehat L+\frac{\eps}{400\thres^2}.
\]
Combining the two bounds on the selected empirical average yields
\[
 L(y)-\frac{\eps}{400\thres^2}
 \le\widehat L
 \le\min_{x\in\cC_i}L(x)+\frac{3\eps}{400\thres^2}.
\]
The map $a\mapsto1/\max\{a,1/(10\thres)\}$ is decreasing and
$100\thres^2$-Lipschitz. Since true losses are at least
$1/(6\thres)$ by \Cref{eq:regular-calibrated-profit-range} (specifically, \Cref{lem:regular-auxiliary,lem:regular-init}), applying
this map on $\hat L$ to define $\hat r$ yields
\begin{equation}
\label{eq:refinement_guarantees}
 \max_{\cC_i}r_\lambda-3\eps/4
 \le\widehat r\le r_\lambda(y)+\eps/4.
\end{equation}
This proves \Cref{eq:regular-refinement}, with total failure
probability at most $1/T$.

For part~\ref{item:regular-survival}, assume inductively that
$i^\star$ is active at the start of phase $j$.
For each active candidate $i$, let $\widehat r_i$ denote the estimate
returned by refinement in this phase (i.e. the procedure described above), using the calibrated parameter
$\rho_i^{(j)}\lambda_{z_i^{(j)}}$; similarly let $r_i$ be
$r_{\rho_i^{(j)}\lambda_{z_i^{(j)}}}$.
Here $z_i^{(j)}$ denotes the reference at the start of phase $j$.
Then \Cref{lem:regular-init}\ref{item:regular-init-auxiliary},
the logarithmic calibration bound in
\Cref{lem:regular-observation}\ref{item:regular-calibration},
and the lower bound in \Cref{eq:regular-refinement} give
\begin{equation}
\label{eq:istar_subopt}
 \widehat r_{i^\star}\ge\opt-3\eta-\eps_j.
\end{equation}
Every candidate has $\widehat r_i\le\opt+\eps_j$ by
\Cref{eq:regular-refinement,lem:regular-auxiliary}.
Thus the largest estimate exceeds
$\widehat r_{i^\star}$ by at most $3\eta+2\eps_j$, so the threshold
$10(\eta+\eps_j)$ preserves $i^\star$.

Now assume that $i$ is retained. Since $i$ is retained, then
$\hat r_i\geq\hat r_{i^\star}-10(\eta+\eps_j)
\geq\opt-3\eta-\eps_j-10(\eta+\eps_j)$,
where the second inequality follows from \Cref{eq:istar_subopt}.
Together with the refinement upper bound from
\Cref{eq:regular-refinement}, this yields, with the updated
reference $z_i^{(j+1)}=y_i$:
\begin{equation}
\label{eq:crossed_difference}
 \opt-r_i(z_i^{(j+1)})
 \le\opt-\widehat r_i+\eps_j
 \le13\eta+12\eps_j\le16(\eta+\eps_j).
\end{equation}

Finally, notice that this bound evaluates the updated reference --- associated to iteration $j+1$ --- using the
auxiliary-profit function from phase $j$.
We must transfer it to the parameter obtained by the next calibration.
Let $\rho_i^{(j+1)}$ be the output of that calibration, using
$z_i^{(j+1)}$.
Both valid calibrations approximate the same fixed intended
parameter $\lambda_i$. Thus, by \Cref{eq:regular-effective} in \Cref{lem:regular-observation}, we get:
\[
 \left|
 \log\frac{\rho_i^{(j+1)}\lambda_{z_i^{(j+1)}}}
              {\rho_i^{(j)}\lambda_{z_i^{(j)}}}
 \right|
 \le
 \left|
 \log\frac{\rho_i^{(j+1)}\lambda_{z_i^{(j+1)}}}{\lambda_i}
 \right|
 +
 \left|
 \log\frac{\rho_i^{(j)}\lambda_{z_i^{(j)}}}{\lambda_i}
 \right|\le2\cdot10^{-3}\sqrt{\eta/\thres}.
\]
The stability bound from \Cref{eq:regular-stability} in
\Cref{lem:regular-auxiliary}, applied at the updated reference,
therefore gives
\begin{align*}
 \opt-r_{\rho_i^{(j+1)}\lambda_{z_i^{(j+1)}}}
          (z_i^{(j+1)})
 &\le2\bigl(\opt-r_i(z_i^{(j+1)})\bigr)
      +2\opt\left(2\cdot10^{-3}\sqrt{\eta/\thres}\right)^2 \\
 &\le32(\eta+\eps_j)
      +8\cdot10^{-6}\opt\frac{\eta}{\thres}\tag{By \Cref{eq:crossed_difference}}\\
 &\le(32+48\cdot10^{-6})\eta+64\eps_{j+1}\\
 &\le65(\eta+\eps_{j+1}),
\end{align*}
using $\opt\le6\thres$ and $\eps_j=2\eps_{j+1}$.

Initially, all candidates are active, $z_i^{(0)}=w_i$,
and $\eps_0=\thres$. Using $\mu^\star\leq 6\thres$ from \Cref{lem:regular-init}, we get:
\[
 \opt-r_{\rho_i^{(0)}\lambda_{w_i}}(w_i)
 \le\opt\le6\thres\le65(\eta+\eps_0),
\]
so this same bound holds at the start of phase $0$.
Induction then concludes the proof of \ref{item:regular-survival}.
\end{proof}

\subsection{Regret bound}

We now combine the preceding lemmata to bound the regret from initialization, calibration, and refinement, accounting also for failed guarantees. Choosing $\eta=T^{-2/5}$ then yields the bound in \Cref{thm:regular}. In the following, let $C$ be a sufficiently large universal constant which upper bounds all the other constants we have considered in the lemmata of this section.  

\begin{proof}[Proof of \Cref{thm:regular}]
Fix $\eta\in (0, 1)$ with $K=\lceil1/\eta\rceil\le T$.
Recall that we stop at round $T$ and post only prices in $\cT$, using $(0,0)$ for the remaining rounds if an expression or price is invalid or no candidates remain. Thus every round has regret gap in $[0,1]$.

\paragraph{Initialization.}
Analyze \init as if it were allowed to run all iterations. Its actual
execution may be a prefix of this full execution, as it stops at trading round $T$.
By Lemma~\ref{lem:regular-init}, initialization fails with probability
at most $1/T$. Charging the entire horizon on this event contributes
at most $T\cdot(1/T)=1$ to expected regret.
If its eventual valid output is \emph{no trigger}, then
$\opt\le C/K\le C\eta$, so the entire horizon costs at most $CT\eta$,
even if initialization is interrupted before returning that output.
If instead it returns a scale and candidates satisfying the guarantees
of Lemma~\ref{lem:regular-init}, its full regret is at most $O(K\log^2(T))$.
Non-negativity of the instantaneous regret gives the same bound for any prefix; we do not need initialization to actually finish before $T$.

It remains to consider the case of \init triggering when initialization completes before $T$. We thus now proceed to the analysis of \refine. Fix the completed transcript of \init, so $\thres$, the candidates,
and their number $N$ are fixed, with
\begin{equation}\label{eq:regular-theorem-candidates}
 N\le O\left(\sqrt{\thres/\eta}\log(T)\right).
\end{equation}
All subsequent expectations and probabilities are conditional on
this transcript. In the following, we call a calibration or refinement output valid when it satisfies \Cref{eq:regular-effective,eq:regular-refinement}, respectively.

\paragraph{Regret during one optimizer run.}
We first count the costs of single optimizer calls that start with all preceding outputs valid, including the call producing the first failed output. We will bound the cost of all rounds after that first failure separately.

At the start of phase $j$, a valid calibration gives, via \Cref{lem:regular-refine}\ref{item:regular-reference-improvement}:
\[
 \opt-r_\lambda(z_i)\le65(\eta+\eps_j).
\]
Therefore, for a query $x$ and every price pair $y$ posted to produce its
observation, substituting this bound into \Cref{eq:regular-round-regret} from \Cref{lem:regular-observation}\ref{item:regular-loss-observations} gives:
\begin{equation}\label{eq:fullthm_oneround_ineq}
 \opt-\mu(y)\le65C\eta+65C\eps_j
       +C\thres^2\left(L(x)-\min_{a\in\cC_i}L(a)\right).
\end{equation}
Charge the $65C\eta$ term only over actual trading rounds, giving at
most $65CT\eta$ in total. To bound the last two terms of \Cref{eq:fullthm_oneround_ineq}, allow each started optimizer run to finish all its $n$ observations, including any observation interrupted at time $T$. Their contribution on the actual prefix is at most their contribution on this full run.

Conditional on the past history and query $x$, the other two terms in \Cref{eq:fullthm_oneround_ineq} are fixed,
and producing the observation takes at most $C/\thres$ expected
trading rounds by \Cref{eq:regular-oracle} in \Cref{lem:regular-observation}\ref{item:regular-loss-observations}. Their expected contribution for this query is therefore at most
\[
 65C^2\frac{\eps_j}{\thres}
 +C^2\thres\left(L(x)-\min_{a\in\cC_i}L(a)\right).
\]
Summing over the $n$ queries and applying \Cref{eq:regular-copt} from the stochastic convex bandits regret bound in \Cref{prop:mhr-convex-bandit}
bounds the expected contribution of these terms, conditional on the
run's starting history, by
\[
 \begin{aligned}
 65C^2 n\frac{\eps_j}{\thres}
 +C^2\thres\E{\sum_{\tau=1}^n
       \left(L(x_\tau)-\min_{x\in\cC_i}L(x)\right)}
 &\le65C^2 n\frac{\eps_j}{\thres}+O\left(\sqrt n\log^2(2n)\right),
 \end{aligned}
\] 
The choice of $n$ from \Cref{eq:regular-run-length} then gives
\(
 n/4^j\le\lceil C\log^4(2T)\rceil
 \le2C\log^4(2T).
\)
Since $\eps_j=\thres2^{-j}$, this implies
\(
 n{\eps_j}/{\thres}
 =\sqrt n\sqrt{n/4^j}
 \le\sqrt{2C}\sqrt n\log^2(2T).
\)
Thus the expected contribution of the last two terms of \Cref{eq:fullthm_oneround_ineq} over one run is at most
\begin{equation}\label{eq:regular-optimizer-run-cost}
 65C^2\sqrt{2C}\sqrt n\log^2(2T)
 +O\left(\sqrt n\log^2(2n)\right).
\end{equation}

\paragraph{Summing the optimizer runs.}
Next, we bound the sum of the numbers of prescribed observations over all optimizer runs started before trading round $T$.
Notice that, after fixing a valid initialization transcript, the following counting argument holds for every realization of the subsequent execution,
regardless of whether calibrations or refinements fail.

Every completed run uses at least $n$ trading rounds, since each
observation uses at least one round. Hence the prescribed lengths of
completed runs sum to at most $T$. There is at most one unfinished
run. If it is in phase $j\ge1$, using \Cref{eq:regular-run-length}, a run of length $n/4$ must have
completed in the preceding phase, so $n\le4T$. In phase $0$, its
length is at most $2C\log^4(2T)$.
Using $\log^4(2T)\le10T$, we then get, using $n_k$ as the prescribed number of loss observations in optimizer run $k$:
\begin{equation}\label{eq:regular-total-run-length}
 \sum_{k\text{ is a started run}} n_k
 \le T+4T+2C\log^4(2T)
 \le(5+20C)T\le25CT.
\end{equation}
In particular, $n_k\le25CT$ for every started run, so
$\log(n_k)=O(\log(T))$, which can be used to bound the logarithm in \Cref{eq:regular-optimizer-run-cost}.
Choosing $C$ sufficiently large, the expected contribution of the last two
terms of \Cref{eq:fullthm_oneround_ineq} over run $k$ is at most
\[
 \bigl(65C^2\sqrt{2C}+C^3\bigr)\sqrt{n_k}\log^2(2T)
 \le100C^3\sqrt{n_k}\log^2(2T).
\]
Starting phase $j\ge1$ requires completing a run of at least
$4^{j-1}$ observations in the preceding phase. Hence $4^{j-1}\le T$,
so
\begin{equation}\label{eq:regular-started-phases}
 \text{number of started phases}\le2+\log_4T\le3\log(T).
\end{equation}
Each refinement uses at most $\lceil C\log(2T)\rceil\le2C\log(T)$
optimizer copies. With at most $N$ active candidates per phase, there
are therefore at most $6CN\log^2(T)$ started runs.
Summing this single-run bound, Cauchy--Schwarz and the
run-length bound \Cref{eq:regular-total-run-length} give
\begin{equation}\label{eq:regular-optimizer-total-cost}
 \begin{aligned}
 \sum_{k\text{ is a started run}}100C^3\sqrt{n_k}\log^2(2T)
 &\le100C^3\log^2(2T)
       \sqrt{\sum_{\text{started runs}}1}
       \sqrt{\sum_{\text{started runs}}n_k}\\
 &\le100C^3\log^2(2T)
       \sqrt{6CN\log^2(T)}\sqrt{25CT}\\
 &\le100\sqrt{150}\,C^4\sqrt{NT}\log^3(2T)\\
 &\le1300C^4\sqrt{NT}\log^3(2T).
 \end{aligned}
\end{equation}
With \Cref{eq:regular-optimizer-run-cost} we have bounded the expected regret contribution of a single optimizer run, conditional on the run’s starting history, whenever preceding outputs are valid. We now want to drop the conditioning, while preserving the one on the \init transcript and while still restricting the expectation to runs started with valid preceding outputs. We can do this by using standard properties\footnote{We use the following fact: for a nonnegative random variable
$R$ and an event $G$ measurable with respect to a sigma-algebra $\cH$, we have $\E{R\mathbf{1}_G\mid\cH}=\mathbf{1}_G\E{R\mid\cH}$. If $\E{R\mid\cH}\le B$ on $G$, for a constant $B$, total expectation then gives
$\E{R\mathbf{1}_G}
=\E{\mathbf{1}_G\E{R\mid\cH}}
\le \E{B\mathbf{1}_G}$.
Here, $\cH$ can be seen as the history at the run's start and $G$ as the event that all preceding outputs are valid.} of conditional expectations (see Chapter 9 in \citet{Williams_1991}); we then get the following bound, with $\clean_k$ being the clean event that all preceding outputs are valid for run $k$ and with $R_k$ being the actual regret contribution of run $k$ before time $T$: 
\begin{align*}
 \E{\sum_{k\text{ is a started run}}R_k\mathbf{1}_{\clean_k}}&\leq  \E{\sum_{k\text{ is a started run}} 100C^3\log^2(2T)\sqrt{n_k}\mathbf{1}_{\clean_k}}+65CT\eta\\
& \le 1300C^4\sqrt{NT}\log^3(2T)+65CT\eta,
\end{align*}
where we considered the extra term from \Cref{eq:fullthm_oneround_ineq} and where we used that the counting argument holds for any realization.

Now, every trading round used by refinement before the first failed output or the end of the time horizon belongs to an optimizer run that started with all preceding outputs valid. The fact that regret is non-negative then immediately gives:
\begin{equation}\label{eq:regular-refinement-total-regret}
 \begin{aligned}
 \E{\text{refinement regret up to first failure or time }T}&\leq \E{\sum_{k\text{ is a started run}}R_k\mathbf{1}_{\clean_k}}\\
 &\le1300C^4\sqrt{NT}\log^3(2T)+65CT\eta.
 \end{aligned}
\end{equation}

\paragraph{Calibration and failures.}
By \Cref{eq:regular-started-phases}, there are at most
$3N\log(T)$ calibration calls, since each phase calibrates at most $N$ candidates.
\Cref{lem:regular-observation}\ref{item:regular-calibration}
bounds each call's regret by $C\log(T)/\eta$, including
any interrupted prefix. Thus their total regret up to the first
failed output or time $T$ is at most
\begin{equation}\label{eq:regular-calibration-total-cost}
 3N\log(T)\cdot\frac{C\log(T)}\eta
 =3CN\frac{\log^2(T)}{\eta}.
\end{equation}

There are also at most $3N\log(2T)$ refinement calls, hence at most
$6N\log(2T)$ calibration and refinement calls combined.
For each calibration or refinement started with valid preceding
outputs, the conditional probability that its output fails is at
most $1/T$, by \Cref{lem:regular-observation,lem:regular-refine}. Taking the union over which call produces the \emph{first} failed output therefore gives:
\begin{equation}\label{eq:regular-failure-probability}
 \P{\text{a completed calibration or refinement fails}}
 \le\frac{6N\log(2T)}T.
\end{equation}

After the first failed output, charge every remaining trading round
at the worst-case gap of one. Thus by
\Cref{eq:regular-failure-probability}, the additional expected regret is at most
\begin{equation}
\label{eq:failure-regret}
 T\cdot\frac{6N\log(2T)}T=6N\log(2T).
\end{equation}

\paragraph{Combining the bounds.}
Combining the initialization regret, \Cref{eq:regular-refinement-total-regret}, \Cref{eq:regular-calibration-total-cost} and \Cref{eq:failure-regret} on any completed valid triggered initialization transcript,
the total expected regret is at most
\begin{equation}\label{eq:regular-total-regret-bound}
 \begin{aligned}
O(K\log^2(T))+O\left(N\frac{\log^2(T)}{\eta}\right)+O(\sqrt{NT}\log^3(T))+O(T\eta)+O(N\log(T)).
 \end{aligned}
\end{equation}
Choose $\eta=T^{-2/5}$, implying $K=\lceil T^{2/5}\rceil$. \Cref{eq:regular-theorem-candidates} then gives
$N= O(T^{1/5}\log(T))$ uniformly over all valid triggered transcripts.
The terms in \eqref{eq:regular-total-regret-bound} therefore satisfy:
$$
O(T^{2/5}\log^2(T)) + O(T^{3/5}\log^3(T)) + O(T^{3/5}\log^{7/2}(T)) + O(T^{3/5}) + O(T^{1/5}\log^2T) \leq O(T^{3/5}\log^{7/2}(T)), 
$$
which proves the regret bound.
\end{proof}
\endgroup

\section{Separation Between MHR and Regular Distributions}
\label{sec:separation}

    In this section, we prove a separation between MHR and regular distributions in bilateral trade. More precisely, we prove the following result. 

\begin{theorem}\label{thm:regular_lower}
    There exists a universal constant $c>0$ such that, for every
    sufficiently large horizon $T$ and every learning algorithm $\cA$,
    there exists a product distribution $\cD$ with regular seller
    and buyer marginals satisfying
    \[ R_T(\cA,\cD)\ge cT^{\nicefrac35}. \]
\end{theorem}
    \begin{proof}
    We prove the result for an even stronger class of algorithms, i.e., algorithms that receive two bits of information at each step, indicating
    separately whether the seller and the buyer accepted the corresponding prices. Since we are proving a lower bound for the stochastic setting, where the agents' valuations are drawn i.i.d. from the same fixed distribution, it is enough to construct a mixture of i.i.d. instances which is hard for any deterministic algorithm. Indeed, the hardness then extends to any (possibly randomized) algorithm by conditioning on any realization of the algorithm and averaging, so that the algorithm has to fail against at least one of the instances in the mixture.

         We first define a baseline distribution $\cD_0$ (for simplicity, we keep the same name for the random variables, but change the underlying measure, so that with $\mathbb{E}^0$ we mean the expectation when $(S,B)$ is sampled according to $\cD_0$). Under $\cD_0$, the seller distribution is uniform in $[0, 1]$ and is thus regular and MHR. The buyer distribution is instead defined by the following (differentiable) survival function $S_0(b) = 1-F_0(b)$, which induces the virtual value function $\phi_0(b)$ as follows:
         \begin{equation}
         \label{eq:base_cdf_def}
         S_0(b) = \begin{cases}
             1-\dfrac{8}{3}b, \ & b\in [0, \nicefrac{1}{4}]\\
             \dfrac{1}{48b^2}, & b\in [\nicefrac{1}{4}, \nicefrac{1}{2}]
             \vspace{0.1cm}\\
             \dfrac{1-b}{12b}, & b\in [\nicefrac{1}{2}, 1].
         \end{cases},
         \quad \phi_0(b) = b-\frac{S_0(b)}{-S_0'(b)} = 
         \begin{cases}
             2b-\dfrac{3}{8}, \ & b\in [0, \nicefrac{1}{4}]\\
             \dfrac{b}{2}, & b\in [\nicefrac{1}{4}, \nicefrac{1}{2}]
             \vspace{0.1cm}\\
             b^2, & b\in [\nicefrac{1}{2}, 1].
         \end{cases}
         \end{equation}

        The virtual valuation is increasing over the domain, and the distribution is thus regular (but not MHR). We now want to construct a family of distributions $\cD_j$ which coincide with $\cD_0$ on all the support, with the exception of the values of $b$ in certain intervals. To this end, we introduce the following function 
        \(
        \rho(u)=
        16u^2(1-u)^2 \cdot \ind{u \in [0,1]}
        \). {Its properties are summarized in the following, immediate, claim}.
        \begin{claim}
        \label{cl:prop_rho}
            The function $\rho$ is twice continuously differentiable in $(0, 1)$, has unique maximum $1$ at $u=\nicefrac 12$ and its first derivative vanishes at $0$ and $1$. Moreover, \(\|\rho'\|_\infty\le4\), \( \|\rho''\|_\infty\le32\).
        \end{claim}
        
        Fix an integer parameter $K$ we set later, and partition uniformly the interval $[\nicefrac 14, \nicefrac 12]$ into $K$ sub-intervals $I_1, I_2, \dots, I_K$. We would like to add small local ``bumps'' to the baseline buyer distribution, one for each interval in $\{I_j\}_{j=1}^K$. Specifically, let $Y_0(b) = \nicefrac{1}{S_0(b)}$ for any $b \in [0, 1)$ and $w = \nicefrac{1}{4K}$; for any $j$, define $Y_j$ as follows: 
        \begin{equation}
        \label{eq:y_perturbation}
            Y_j(b) = Y_0(b)-\dfrac{1}{2}w^2\rho\left(\dfrac{b-a_j}{w}\right),
        \end{equation}
        where $a_j$ is the left-endpoint of $I_j$, and $b \in [0, 1)$. 

        We have all the ingredients to define the distributions $\cD_j$ for $j\in [K]$. The seller distribution is uniform in $[0,1]$, while the buyer distribution is characterized by the survival function $S_j(b)$, defined as $S_j(b) = S_0(b)$ for $b\notin I_j$ and $S_j(b) = \nicefrac{1}{Y_j(b)}$ for $b\in I_j$. 

        Before continuing, we show that the buyer marginal for any perturbation is regular. 
        \begin{claim}
        \label{cl:buyer_distro_reg}
        The buyer distribution under $\cD_j$ is well-defined and regular. 
        \end{claim}
        \begin{proof}
        To prove that the distribution is well-defined, it is enough to prove that $S_j$ is non-negative, continuous, and decreasing in $I_j$. For the latter, it suffices proving that $Y_j'$ is strictly positive. Since $I_j\subseteq [\nicefrac{1}{4}, \nicefrac{1}{2}]$, it follows by \Cref{eq:base_cdf_def} that $Y_0'(b) \geq 24$ for any $b \in I_j$: together with $\tfrac{d\rho(\nicefrac{b-a_j}{w})}{db} \leq \nicefrac{1}{w}\cdot \lVert\rho'\rVert_\infty$, this suffices to conclude that $S_j$ is indeed decreasing, recalling that $\lVert\rho'\rVert_\infty\leq 12$ by Claim \ref{cl:prop_rho} and noticing that $w = \nicefrac{1}{4K}\leq 1$. Non-negativity then follows simply by continuity of $S_j$.
         
        For regularity, we only need to show that $\phi'_j \geq 0$ in the interior of $I_j$ and that $\phi_j$ converges to $\phi_0$ at the boundaries of $I_j$. To prove the first sufficient condition, notice that differentiating yields: 
        $$
        \phi'_j(b) = \dfrac{Y_j(b)Y_j''(b)}{(Y_j'(b))^2},
        $$
        for any $b$ in the interior of $I_j$, so that proving non-negativity of $Y_j''$ suffices. This can be shown analogously to the previous lower bound on $Y_j'$, using that\footnote{This induces the main constraint of our lower bound. To guarantee convexity of $Y$ and thus regularity, we need to cancel the $\nicefrac{1}{w^2}$ factor stemming from the differentiation of $\rho$. In turn, this means that $\rho$ needs to be scaled by a factor $O(w^2)$ in \Cref{eq:y_perturbation}. In other words, considering \Cref{eq:diff_lower_bound}, the ``bump'' in expected profit can only be $O(\nicefrac{1}{K^2})$.} $\tfrac{d^2\rho(\nicefrac{b-a_j}{w})}{db^2} \leq \nicefrac{1}{w^2}\cdot \lVert\rho''\rVert_\infty$ and that $\lVert \rho''\rVert_\infty\leq 32$. To prove continuity, notice that we may write $\phi_j(b) = b-\nicefrac{Y_j(b)}{Y'_j(b)}$: we can thus just focus on the difference $\lvert \nicefrac{Y_0(b)}{Y'_0(b)}-\nicefrac{Y_j(b)}{Y'_j(b)}\rvert$, which vanishes since $\rho$ and $\rho'$ vanish approaching $0$ or $1$, by Claim \ref{cl:prop_rho}. 
        \end{proof}

        Now that we have introduced the baseline distribution $\cD_0$ and the family of distributions $\cD_j$ parameterized for $j=1, \dots, K$, we can study the corresponding expected profit: 
        \[
            \Ej{\prof(p,q)} = (q-p)pS_j(q).
        \]
        From this formula, it is easy to see that for any buyer price $q$, the best seller price comes from setting $p=\nicefrac{q}{2}$, for an expected profit of $h^j(q) = \nicefrac{q^2}{4}\cdot S_j(q)$. This observation implies that the underlying optimization problem of finding the best posted prices boils down to finding the best $q$ for $h^j$, for the considered instances. Moreover, the generic $h^j$ functions only differ from $h^0$ in the corresponding interval $I_j$. If we introduce the suboptimality gap $\Delta = \nicefrac{w^2}{384 \cdot 12}$, we can get the following result. 
        
    \plateaux*
    \begin{proof}
        Point (i) follows directly from \Cref{eq:base_cdf_def}. Indeed, we have: 
        $$
        h^0(q) =
        \begin{cases}
             \dfrac{q^2}{4}\left(1-\dfrac{8}{3}q\right), \ & q\in [0, \nicefrac{1}{4}]\\
             \dfrac{1}{192}, & q\in [\nicefrac{1}{4}, \nicefrac{1}{2}]
             \vspace{0.1cm}\\
             \dfrac{q(1-q)}{48}, & q\in [\nicefrac{1}{2}, 1]
        \end{cases}
        $$
        Differentiating, one easily checks that the first branch is indeed increasing and the last is indeed decreasing.

        To prove point (ii), notice that by Claim \ref{cl:prop_rho}, we have $\rho(\nicefrac{(q^\star_j-a_j)}{w}) = 1$. Therefore, $Y_j(q^\star_j) = 48(q^\star_j)^2-\nicefrac{w^2}{2}$. Using the definition of $S_j$ for $j\in [K]$, we immediately get:
        \begin{equation}
        \label{eq:diff_lower_bound}
            h^{j}(q^\star_{j}) - \max_{q\notin I_j} h^j(q) \geq h^j(q_j^\star)-\dfrac{1}{192} = \dfrac{(q_j^\star)^2}{4\cdot \left(48(q_j^\star)^2-\nicefrac{w^2}{2}\right)}-\dfrac{1}{192} = \dfrac{w^2}{384\cdot\left(48(q^\star_j)^2-\nicefrac{w^2}{2}\right)}\geq \dfrac{w^2}{384\cdot 12},
        \end{equation}
        using that $q^\star_j\leq \nicefrac{1}{2}$ since $I_j\subseteq [\nicefrac{1}{4}, \nicefrac{1}{2}]$.
        \end{proof}

        Fix any deterministic algorithm $\cA$, we prove that it fails against the uniform mixture of $\{\cD_j\}_{j=1}^K$. To this end, denote with $N_j$ the random variable counting the number of times that $\cA$ posts a buyer price in $I_j$. The crucial ingredient to analyze the regret of $\cA$ onto the mixture resides in arguing that the algorithm manages to differentiate between the generic $\cD_j$ and the baseline $\cD_0$ only by posting buyer prices in the ``right'' interval $I_j$. Formally, we have: 
        
        \pinskerkl*
        \begin{proof}[Proof of Claim~\ref{cl:KL}]
        We denote with $z_t$ the feedback received by the algorithm at time $t$, and with $\mathbb{P}^j_{z_1,\dots, z_{t-1}}$ the distribution on the feedback random variables up to time $t-1$, induced by the deterministic algorithm and the valuation random variables up to time $t-1$, under the distribution $\cD_j$. To be clear, $\mathbb{P}^j_{z_1,\dots, z_{t-1}}$ is a push-forward measure on $(\{0, 1\}^2)^{t-1}$. Moreover, notice that, with $q_t$ being the posted buyer price at time $t$, $\ind{q_t\in I_j}$ is a measurable function of $\{z_t\}_{t=1}^{t-1}$, since the algorithm is deterministic. We then have the following chain of inequalities:
            \begin{align*}
                \Ej{N_j} - \Eo{N_j}&= \sum_{t=2}^T \mathbb{P}^j_{z_1, \dots, z_{t-1}}(q_t \in I_j) - \mathbb{P}^0_{z_1, \dots, z_{t-1}}(q_t \in I_j) \tag{By linearity}\\
                &\le \sum_{t=2}^T \TV{\mathbb{P}^j_{z_1, \dots, z_{t-1}} -\mathbb{P}^0_{z_1, \dots, z_{t-1}}} \tag{By definition of Total Variation}\\
                &\le \sum_{t=2}^T\sqrt{\frac{1}2\KL{\mathbb{P}^0_{z_1, \dots, z_{t-1}}}{{\mathbb{P}^j_{z_1, \dots, z_{t-1}}}}} \tag{By Pinsker Inequality}\\
                &=\sum_{t=2}^T \frac{1}{\sqrt{2}}\sqrt{\KL{\mathbb{P}^0_{z_1}}{{\mathbb{P}^j_{z_1}}}\!+\!\sum_{\tau=2}^{t-1} \Eo{\KL{\mathbb{P}^0_{z_\tau|z_1, \dots, z_{\tau-1}}}{{\mathbb{P}^j_{z_{\tau}|z_1, \dots, z_{\tau-1}}}}}} \tag{Chain Rule of KL Divergence}
            \end{align*}
            We focus on the single divergence terms, starting from the first one. The algorithm observes some difference in the distribution of the buyer only if it posts $q_1$ in the interval $I_j$. The distribution of the seller feedback is instead invariant, since the seller distribution is uniform under both $\cD_0$ and $\cD_j$. We thus have:
            \begin{align}
                \KL{\mathbb{P}^0_{z_1}}{{\mathbb{P}^j_{z_1}}} &= \ind{q_1 \in I_j} \cdot \KL{\bern(S_0(q_1))}{\bern(S_j(q_1))}\notag\\
                &\le \ind{q_1 \in I_j}\frac{(S_0(q_1) - S_j(q_1))^2}{S_j(q_1)(1-S_j(q_1))}, \label{eq:kl_first_bound}
            \end{align}
            where the last inequality follows by the fact that $\log x \leq x-1$, see \Cref{prop:chi_squared} in \Cref{app:inequalities} for a formal proof. 
            
            We now proceed to bound the above expression by upper and lower bounding $Y_0$ and $Y_j$. For the former, since $I_j\subseteq[\nicefrac{1}{4}, \nicefrac{1}{2}]$, \Cref{eq:base_cdf_def} implies $3\leq Y_0(q) \leq 12$ for any $q\in I_j$. For the latter, $\lVert \rho\rVert_\infty\leq 1$ by Claim \ref{cl:prop_rho} and $w^2\leq1$ 
            imply that $2\leq Y_j(q)\leq 12$. Taking reciprocals, we get $S_j(q_1)\in [\nicefrac{1}{12}, \nicefrac{1}{2}]$, so that $S_j(q_1)(1-S_j(q_1))\geq \nicefrac{1}{24}$. Moreover, using the same bounds, we also get: 
            \begin{align*}
            \lvert S_0(q_1) - S_j(q_1)\rvert = \left\lvert \frac{1}{Y_0(q_1)}-\frac{1}{Y_j(q_1)}\right\rvert = \frac{\lvert Y_j(q_1)-Y_0(q_1)\rvert}{Y_0(q_1)Y_j(q_1)} \leq \frac{w^2}{6}.
            \end{align*}
            Plugging the above derivation into \Cref{eq:kl_first_bound}, we get:
            $$
            \KL{\mathbb{P}^0_{z_1}}{{\mathbb{P}^j_{z_1}}} \leq w^4\ind{q_1 \in I_j} = \left(\frac{1}{4K}\right)^4\ind{q_1 \in I_j}
            $$

            An identical argument holds also for the generic term, so that we have: 
            \begin{align*}
                \Ej{N_j} - \Eo{N_j} &\le \sum_{t=2}^T \frac{1}{\sqrt 2}\sqrt{\KL{\mathbb{P}^0_{z_1}}{{\mathbb{P}^j_{z_1}}}\!+\!\sum_{\tau=2}^t \Eo{\KL{\mathbb{P}^0_{z_\tau|z_1, \dots, z_{\tau-1}}}{{\mathbb{P}^j_{z_{\tau}|z_1, \dots, z_{\tau-1}}}}}} \\
                &\le \frac{T}{22K^2}\sqrt{\Eo{N_j}}
            \end{align*}
        \end{proof}

        By Claim \ref{cl:prop_exp_prof} and recalling that $\Delta = \nicefrac{w^2}{(384\cdot 12)}$ is the ``suboptimality gap'' of playing outside the right interval, the regret of the deterministic algorithm $\cA$ against the generic distribution $\cD_j$ is
        \[
            R_T(\cA,\cD_j) \ge (T-\Ej{N_j})\Delta,
        \]
        so that if we take the expectation with respect to the uniform choice of $j\in [K]$, we get:
        \begin{align*}
            \frac{1}{K} \sum_{j=1}^K R_T(\cA,\cD_j)&\geq \frac{1}{K}\sum_{j=1}^K(T-\Ej{N_j})\Delta\\
            &\ge \frac{1}{K}\sum_{j=1}^K\left(T-\Eo{N_j}- \frac{T}{22K^2}\sqrt{\Eo{N_j}}\right)\Delta \tag{By Claim~\ref{cl:KL}}\\
            &\ge \frac{K-1}{K} T \Delta - \frac{1}{K}\sum_{j=1}^K \frac{T\Delta}{22K^2} \sqrt{\Eo{N_j}} \tag{As $\sum_j \Eo{N_j}\le T$}\\
            &\ge \frac{K-1}{K} T \Delta -  \frac{T\Delta}{22K^2}\sqrt{\frac{1}{K}\sum_{j=1}^K\Eo{N_j}} \tag{By Jensen's Inequality}\\
            &\ge \frac{K-1}{K} T \Delta -  \frac{T\Delta}{22K^2}\sqrt{\frac TK} \tag{As $\sum_j \Eo{N_j}\le T$}\\
            &= T\Delta\left(1 - \frac{1}{K} - \frac{1}{22}\sqrt{\frac {T}{K^5}}\right)\ge \frac{T\Delta}{2} \tag{By choosing $K = \left\lceil T^{\nicefrac{1}{5}}\right\rceil$}, 
        \end{align*}
        where the last inequality follows for all $T$ such that $\left\lceil T^{\nicefrac{1}{5}}\right\rceil \geq 3$. To conclude the proof, recall that $\Delta \propto w^2 \propto \tfrac{1}{K^2} \propto \tfrac{1}{T^{\nicefrac{2}{5}}}$.
    \end{proof}

\section*{Acknowledgment}

    The authors acknowledge OpenAI for providing complimentary access through the ``ChatGPT for Academic Researchers'' program. The work is partly supported by AI-PHOQUS PN RIC 2021-2027, co-financed by EU Italian Ministry MIMIT and MUR. S.D.G. was also supported in part by the Institute for Complex Systems (Italian National Research Council).
    
\section*{AI Disclosure and Methodology}
The authors used several versions of ChatGPT through the ``Plus'', ``Pro'', and ``Academic Researchers'' account tiers. In particular, the MHR regret upper bound (\Cref{sec:MHR}) and the lower bound for regular distributions (\Cref{sec:separation}) were discovered through repeated interactions with ChatGPT 5.6 Sol. The authors then considerably simplified, shortened, and polished the original proofs.

After \citet{CastiglioniLunghi2026Breaking} broke the $T^{\nicefrac 34}$ barrier, the authors tried interacting with ChatGPT 6 Astra Pro through the chat interface to understand whether a $\widetilde O(T^{\nicefrac 23})$ regret upper bound was possible. These conversations did not lead to any nontrivial results. One of the authors then used the agentic ``Work'' interface of ChatGPT 6 Astra to coordinate multiple Astra agents, providing the existing proofs of the MHR upper bound and the regular lower bound (with a proof of $\tilde O(T^{\nicefrac 2 3})$ regret assuming independence), together with a long Markdown document produced during previous ``Work'' sessions summarizing the unsuccessful approaches considered so far. At that point, the model proposed the first version of what is now the Localized Sampling Lemma (\Cref{lem:local_intro,lem:efficient_estimation}), together with a Bayesian minimax argument for obtaining the $\widetilde O(T^{\nicefrac 23})$ regret bound in the stochastic i.i.d.\ setting. The observation that successive elimination could be used directly to obtain the same rate came later from the authors (\Cref{subsec:stochastic_sketch}).

From the form of the minimax argument, the authors realized that it could also yield an adversarial regret bound. They subsequently simplified the proof of the Localized Sampling Lemma with LLM assistance, approximately halving its length. They also asked whether the adversarial result could be made constructive, and further interactions with the AI assistant led to the algorithm presented in \Cref{sec:adversarial}.

Finally, the authors obtained the regret upper bound for regular distributions (\Cref{sec:Regular}) by trying to generalize the techniques used for the MHR upper bound, again with help from ChatGPT 6 Astra through the ``Work'' interface.

The authors also used ChatGPT 6 Astra to help polish and write the text, and to enforce correctness. Astra also provided first drafts of the technical overview for the stochastic i.i.d.\ and adversarial settings (\Cref{subsec:stochastic_sketch,subsec:adversarial_sketch}), the formal description of the successive-elimination algorithm (see \Cref{sec:stochastic-warmup-proof}), and the comparison with dimension-based bandit bounds in \Cref{sec:dimension-comparison}.

The authors verified the correctness of all mathematical arguments, checked the references and the originality of the results, and take full responsibility for the content of the paper.

\bibliography{references}

\appendix 
\crefalias{section}{appendix}
\crefalias{subsection}{appendix}
\section{Appendix}

\subsection{Useful Inequalities}
\label{app:inequalities}

    \begin{proposition}
        \label{prop:chi_squared}
        For any choice of parameters $p,q \in (0,1)$, we have the following bound on the KL divergence of two independent Bernoulli random variables of parameters $p$ and $q$:
        \[
            \KL{\textnormal{Bern}(p)}{ \textnormal{Bern}(q)} \le \frac{(p-q)^2}{q(1-q)}.
        \]
    \end{proposition}
    \begin{proof}
        The proof follows immediately from the definition of divergence and the fact that $\log(x) \le x-1$ for $x > 0$:
        \begin{align*}
            \KL{\textnormal{Bern}(p)}{ \textnormal{Bern}(q)} &= p \log \frac{p}{q} + (1-p) \log \frac{1-p}{1-q}\\
            &\le p\left(\frac{p}{q}-1\right) + (1-p)\left(\frac{1-p}{1-q}-1\right)\\
            &= \frac{(p-q)^2}{q(1-q)}. \qedhere
        \end{align*}
    \end{proof}

\subsection{Concentration Inequalities}
\label{app:concentrations}
To prove \Cref{prop:mhr-concentration,cl:monotone_estimation,lem:regular-observation,lem:regular-refine}, we use Bernstein's inequality, Hoeffding's inequality and the stronger Azuma-Hoeffding's inequality, see e.g., Theorem 1.2, Problem 1.9 and Theorem 5.1 of \citet{Dubhashi_Panconesi_2009}.

\begin{proposition}[Bernstein's inequality]
    \label{prop:bernstein}
    Let the random variables $X_1, X_2, \dots, X_n$ be independent with $X_i - \E{X_i}<b$ for each $i=1, 2,\dots, n$. Let $X=\sum_i X_i$, and let $\sigma^2=\sum_i \sigma_i^2$ be the variance of $X$. Then, for any $t>0$, 
    \[
        \P{X > \E{X} + t} \le \exp{-\frac{t^2}{2\sigma^2+\tfrac{2}{3}bt}}.
    \]
\end{proposition}

\begin{proposition}[Hoeffding's Inequality]
\label{prop:hoeffding}
    Let the random variables $X_1, X_2, \dots, X_n$ be independent with $X_i\in [a, b]$ for $a, b\in \mathbb R$ and for each $i=1, 2,\dots, n$. Letting $X=\sum_i X_i$, for any $t>0$, 
    \[
    \P{\lvert X-\E{X}\rvert>t}
    \le 2\cdot \exp{-\frac{2t^2}{n(b-a)^2}}.
    \]
\end{proposition}
\begin{proposition}[Azuma-Hoeffding inequality]
\label{prop:azuma}
Let $\{X_0, X_1, \ldots\}$ be a martingale satisfying, for any $i\geq 1$, $X_i-X_{i-1}\in [a, b]$ almost surely, for two fixed values $a, b\in \mathbb R$. Then we have the following concentration inequality, for any $t>0$:
$$
\P{X_n > X_0 + t }, \P{X_n < X_0 -t }\leq \exp{-\frac{2t^2}{n(b-a)^2}}
$$   
\end{proposition}

We also use the following version of the multiplicative Chernoff–Hoeffding bound (restatement of Theorem 1.1. of \citet{Dubhashi_Panconesi_2009}).

\begin{proposition}[Multiplicative Chernoff bound]
\label{prop:chernoff}
    Let $X$ be defined as the sum of $n$ random variables $X_i$, independently distributed in $[0,1]$. Then the following inequalities hold, for $\eps \in (0, 1)$:
    \begin{align*}
    \P{X < (1-\eps) \E{X}} &\le \exp{-\tfrac 12 \eps^2\E{X}}\\
    \P{X > (1+\eps) \E{X}} &\le \exp{-\tfrac 13 \eps^2\E{X}}
    \end{align*}
\end{proposition}

\subsection{The Discretization Lemma}
\label{app:discretization}

In this lemma, we formalize the fact that the action space $\AA = G_K \cap \cT$ yields a discretization error of order $\nicefrac{T}{K}.$
\begin{lemma}[Discretization Error]
\label{lem:discretization_err}
For every integer $K\ge1$ and every $(p,q)\in\cT$, there is a pair $(p',q')\in\AA$ such that, for every valuation pair $(s_t,b_t)$,
\[
    \prof_t(p,q)\le\prof_t(p',q')+2/K.
\]
Consequently, restricting the comparator to $\AA$ adds at most $2T/K$ to regret.
\end{lemma}\begin{proof}
Fix $(p,q)\in\cT$ and put $\bar p=\lceil Kp\rceil/K$ and $\bar q=\lfloor Kq\rfloor/K$. If $\bar p\le\bar q$, use $(p',q')=(\bar p,\bar q)$. Every trade accepted at $(p,q)$ is still accepted, and its profit decreases by at most $2/K$. Any additional trade has nonnegative profit, so the inequality holds for every valuation pair.

If $\bar p>\bar q$, then $q-p<2/K$. In this case use $(p',q')=(0,0)$, whose zero profit is within $2/K$ of the profit at $(p,q)$. The chosen grid pair depends only on $(p,q)$, so summing over rounds and taking expectations proves the regret claim.
\end{proof}

    \subsection{Missing Result from \texorpdfstring{\Cref{sec:adversarial}}{Section \getrefnumber{sec:adversarial}}}
    \label{app:adversarial}
    \monotoneestimation*
    \begin{proof}
        Let $g\in \cG$ with domain $\cY_g = \{y_1, \ldots, y_N\}\subseteq \cY$. We want to construct an estimator $\hat g$ for $g$. Let the integer $n$ be a sample complexity parameter to be set later. We first start the estimation from the boundary of the domain, assuming that $N\geq 2$, and let 
        \begin{equation}
        \label{eq:avg_monotone_claim}
            \hat g(y_1) = \frac{1}{n}\sum_{i=1}^n Z_{g, y_1}^{(i)} \qquad \hat g(y_N) = \frac{1}{n}\sum_{i=1}^n Z_{g, y_N}^{(i)}.
        \end{equation}

        Let $\clean$ be the clean event in which $\lvert\nicefrac{1}{n}\sum_{i=1}^n Z_{g, y}^{(i)}-g(y)\rvert\leq \nicefrac{\alpha}{8}$, for any $g\in \cG$ and $y\in \cY_g$. For now, assume $\clean$. Then, if $\lvert \hat{g}(y_1)-\hat{g}(y_N)\rvert\leq \nicefrac{\alpha}{2}$, the triangle inequality implies $\lvert g(y_1)-g(y_N)\rvert\leq \nicefrac{\alpha}{2}+\nicefrac{\alpha}{4} = \nicefrac{3\alpha}{4}$, and we can set $\hat g(y) = \hat g(y_1)\, \forall\, y\in \cY_g\setminus\{y_N\}$, guaranteeing the following, by monotonicity:
       \begin{equation}
       \label{eq:triangle_monotone}
           \lvert \hat g(y)-g(y)\rvert = \lvert \hat g(y_1)-g(y)\rvert \leq \lvert \hat g(y_1)-g(y_1)\rvert + \lvert g(y_1)-g(y)\rvert\leq \frac{\alpha}{8}+\frac{3\alpha}{4}<\alpha
       \end{equation}
       In other words, assuming that the estimated variation is small enough, this simple procedure is enough to guarantee $\lVert \hat g-g\rVert_\infty\leq \alpha$, under $\clean$.

       If $\lvert \hat g(y_1)-\hat g(y_N)\rvert > \nicefrac{\alpha}{2}$ and $N>2$, split the domain in two sets $\{y_1, y_2, \ldots, y_{\lceil\nicefrac{N}{2}\rceil}\}$, $\{y_{\lceil\nicefrac{N}{2}\rceil},\ldots, y_N\}$ and recurse over them with the same procedure explained above and the same halving rule, retaining the estimates on already estimated boundary points. The recursion stops when $N\leq 2$ or when the estimated variation is under $\nicefrac{\alpha}{2}$. Under $\clean$, both termination conditions enforce $\lVert g-\hat g\rVert_\infty\leq \alpha$. Indeed, if the variation is small enough, then an argument identical to \Cref{eq:triangle_monotone} yields the guarantee; if $N\leq 2$, we either have one point or two, so that each can be directly estimated to precision $\nicefrac{\alpha}{8}$.

       We now need to count the number of points in $\cY$ whose value of $g$ we directly estimate using this procedure, assuming $\clean$. First, notice that the recursion tree is a binary tree with maximum depth bounded by $1+\log_2 (\lvert \cY\rvert)$, due to the fact that each pair of children halves the parent node set. Let us focus on a specific depth. If an interval at that depth is split, then the estimated variation must be at least $\nicefrac{\alpha}{2}$ and the reverse triangle inequality together with $\clean$ implies that the real variation must be at least $\nicefrac{\alpha}{4}$. Moreover, the considered intervals have disjoint interiors, so that any split at that level of the tree consumes at least $\nicefrac{\alpha}{4}$ of $\text{TV}(g)$. This means that, at any given depth, there can only be $\nicefrac{4\text{TV}(g)}{\alpha}$ splits.

       Any split adds at most one more mean to estimate, and the start of the recursion at the root requires the sample means of at most two points. Therefore, summing over the number of functions in $\cG$, we need to estimate the average of a number of random variables bounded by: 
        \begin{equation}
        \label{eq:upper_bound_estimation}
            2\lvert \cG\rvert+\frac{4\sum_{g\in \cG}\text{TV}(g)}{\alpha}\cdot (1+\lceil\log_2|\cY|\rceil) \leq 2\lvert \cG\rvert+\frac{4W}{\alpha}\cdot (1+\lceil\log_2|\cY|\rceil)
        \end{equation}
        Clearly, this bound holds only under $\clean$, but we can force the procedure to deterministically halt after having estimated enough averages since $W$ is known, so the cap in \Cref{eq:upper_bound_estimation} holds independently of the occurrence of the failure event $\clean^c$. 

        Since each observation lies in an interval of length at most $4$, \Cref{prop:hoeffding} gives \[\P{\left|\frac1n\sum_{i=1}^n Z_{g,y}^{(i)}-g(y)\right|>\alpha/8} \le 2\exp{-\frac{n\alpha^2}{512}}.\] Choose \(n=\left\lceil\frac{512}{\alpha^2} \log\frac{2|\cG||\cY|}{\delta'}\right\rceil.\) A union bound over all function--point pairs gives $\P{\clean}\ge1-\delta'$. Multiplying this sample size by the bound on the number of estimated points proves the claim.     
    \end{proof}

    \subsection{Missing Results from \texorpdfstring{\Cref{sec:MHR}}{Section \getrefnumber{sec:MHR}}}
    \label{app:mhr}
    \begingroup
    \let \cC \cK
 \convexmhrroutine*
        \begin{proof}
The result is a corollary of Theorem 1 from \citet{Lattimore021}, which requires the domain to be a convex body containing the unit radius Euclidean ball. To enforce this, let $d$ be the affine dimension of $\cC$. If $d=0$, then $\cC$ is a singleton and the regret is zero. Otherwise, John's ellipsoid theorem~\citep{john1948extremum}, applied in the affine hull of $\cC$, allows an affine and invertible change of coordinates such that
\[
    \mathbb B_2^d\subseteq\cC\subseteq d\mathbb B_2^d,
\]
where $\mathbb B_2^d$ is the Euclidean unit ball. This domain change preserves convexity and regret, due to the map being affine and bijective. In these coordinates, $\cC$ contains the unit ball and has diameter at most $2d$.

Write the known interval containing the image of $L$ as $[b,b+B]$, and set
\[
    S=\max\{1,B,\gamma\},
    \qquad
    \widetilde L(x)=\frac{L(x)-b}{S}.
\]
The normalized loss is continuous, convex, and takes values in $[0,1]$.

Run the algorithm of \citet[Theorem~1]{Lattimore021} for $n$ rounds on this normalized problem. In its round $\tau$, when it selects $x_\tau$, perform one query and obtain the observation $L(x_\tau)+\xi_\tau$. We compute
\[
\frac{L(x_\tau) + \xi_\tau -b}{S} =  \widetilde L(x_\tau) + \frac{\xi_\tau}{S},
\]
and return it to the algorithm. 

Conditional on the previous queries and observations and on $x_\tau$, the returned noise $\nicefrac{\xi_\tau}{S}$ has mean zero and $\psi_2$ norm at most one. These are precisely the noise assumptions of \citep[Theorem~1]{Lattimore021}. Since $d$ is a constant and $\cC$ after the change of coordinates has diameter at most $2d$, the theorem gives
\[
    \mathbb E\left[\sum_{\tau=1}^n
        \left(\widetilde L(x_\tau)
        -\min_{x\in\cC}\widetilde L(x)\right)\right]
    =O\!\left(\sqrt n\,\log^2(n)\right),
\]
with a constant depending only on $d$. Every original loss difference is exactly $S$ times its normalized counterpart. Multiplying by $S\le\max\{1,B\}(1+\gamma)$ proves the claim.
\end{proof}

\mhrconcentration*
\begin{proof}
Fix a scale $\delta$ and a sampled pair $(i,j)\in\mathcal S$.
Write $P_{ij}=P(i/K,j/K)$, and let $\overline P_{ij}$ be the
average of $N$ fresh trade indicators at that pair.
Since each indicator is Bernoulli with mean $P_{ij}$, its second
moment equals $P_{ij}$. This second-order bound enables
Bernstein's inequality (see \Cref{prop:bernstein}): for any $\varepsilon>0$, we have
\[
    \mathbb P\!\left(
        |\overline P_{ij}-P_{ij}|>\varepsilon
    \right)
    \le
    2\exp{
        \frac{-N\varepsilon^2}
             {2P_{ij}+\tfrac23\varepsilon}}.
\]
In order for the claimed statement to hold, we bound
the right-hand side by
$2(2T)^{-4}=2e^{-4\log(2T)}$.
Therefore, a sufficient condition is:
\[
    \frac{N\varepsilon^2}
         {2P_{ij}+\tfrac23\varepsilon}
    \ge 4\log(2T)
    \quad\Longleftrightarrow\quad
    \varepsilon
    \ge
    \frac{4\log(2T)}{3N}
    +
    \sqrt{
        \frac{8P_{ij}\log(2T)}{N}
        +\frac{16\log^2(2T)}{9N^2}
    }.
\]
We set $\varepsilon$ equal to the expression on the right.
Thus, with probability at least $1-2(2T)^{-4}$,
\begin{align*}
    |\overline P_{ij}-P_{ij}|
    &\le
    \frac{4\log(2T)}{3N}
    +
    \sqrt{
        \frac{8P_{ij}\log(2T)}{N}
        +\frac{16\log^2(2T)}{9N^2}
    }\\
    &\le
    \sqrt{\frac{8P_{ij}\log(2T)}{N}}
    +\frac{8\log(2T)}{3N}\\
    &\le
    \sqrt{\frac{P_{ij}\delta}{2^{13}}}
    +\frac{\delta}{2^8}
    &&\tag{by our choice of $N$}\\
    &\le
    \frac{P_{ij}}{128}
    +\frac{2\delta}{2^8}\tag{$2\sqrt{ab}\le a+b$ with
$a=\nicefrac{P_{ij}}{2^7}$ and $b=\nicefrac{\delta}{2^8}$}\\
    &=
    \frac{P_{ij}+\delta}{128}.
\end{align*}

There are at most $|\mathcal S|\le(K+1)^2\le4T^2$ sampled pairs
and $1+\lfloor\log_2(K)\rfloor\le T$ possible scales.
A union bound over all pairs
and scales gives total failure probability at most
$2|\mathcal S|T(2T)^{-4}\le \nicefrac{1}{T}$.
Thus, with probability at least $1-\sedit{\nicefrac{1}{T}}$, at every reached scale,
\begin{equation}\label{eq:mhr-frequency-error}
    |\overline P_{ij}-P_{ij}|
    \le\frac{P_{ij}+\thres}{128}
    \qquad\text{for all }(i,j)\in\mathcal S.
\end{equation}
We prove the claimed profit bound deterministically on this high probability event.

We first show that for each interior grid pair, the reconstruction formula estimates its trade probability $P_{ij}$ with error at most $(P_{ij}+\thres)/16$. Fix any reached scale $\thres$ and indices $1\le i<j\le K-1$,
and let $k$ be the midpoint of the first split separating
$i$ and $j$, as explained in \Cref{alg:mhr-bootstrap}. The construction includes $(i,k)$, $(k,j)$, and
$(k,k)$ in $\mathcal S$. Independence and monotonicity give
\begin{equation}
\label{eq:independence_monotonicity}
    P_{ik}P_{kj}=P_{ij}P_{kk}
     \qquad 0\le P_{ik},P_{kj},P_{ij}\le P_{kk}
\end{equation}
Denote the estimator's denominator by
\(
    D=\max\{\overline P_{ik},\overline P_{kj},\overline P_{kk}\},
\)
and notice that $0\le\overline P_{ik}\overline P_{kj}/D
\le\min\{\overline P_{ik},\overline P_{kj}\}$,
with $0/0=0$. Therefore, if $P_{kk}<\thres/32$, then
\[
    \left|\frac{\overline P_{ik}\overline P_{kj}}{D}-P_{ij}\right|
    \le P_{kk}+\frac{P_{kk}+\thres}{128}
    <\frac{\thres}{16},
\]
where the first inequality follows from $P_{ik}\leq P_{kk}$ and \Cref{eq:mhr-frequency-error}, so that both $P_{ij}$ and $\overline P_{ik}\overline P_{kj}/D$ belong to $[0, P_{kk}+\nicefrac{(P_{kk}+\thres)}{128}]$.

To handle $P_{kk}\ge\thres/32$, \Cref{eq:mhr-frequency-error} implies, via \Cref{eq:independence_monotonicity}:
\begin{equation}\label{eq:mhr-denominator-error}
    |D-P_{kk}|\le\frac{P_{kk}+\thres}{128}.
\end{equation}
Adding and subtracting and applying the error bounds from \Cref{eq:mhr-frequency-error} yields:

\begin{align*}
    \lvert\overline P_{ik}\overline P_{kj}-P_{ij}D\rvert
    & = \lvert P_{ik}P_{kj} + P_{kj}(\overline P_{ik}-P_{ik}) + P_{ik}(\overline P_{kj}-P_{kj}) + (\overline P_{kj}-P_{kj})(\overline P_{ik}-P_{ik})  -P_{ij}P_{kk}-P_{ij}(D-P_{kk})\rvert\\
    & = \lvert P_{kj}(\overline P_{ik}-P_{ik}) + P_{ik}(\overline P_{kj}-P_{kj}) + (\overline P_{kj}-P_{kj})(\overline P_{ik}-P_{ik}) -P_{ij}(D-P_{kk})\rvert \tag{By \Cref{eq:independence_monotonicity}}\\
    &\le\frac{P_{kj}(P_{ik}+\thres)+P_{ik}(P_{kj}+\thres)
          +P_{ij}(P_{kk}+\thres)}{128} +\frac{(P_{ik}+\thres)(P_{kj}+\thres)}{128^2}\tag{Triangle Inequality and \Cref{eq:mhr-frequency-error,eq:mhr-denominator-error}}\\
    &\le\left(\frac3{128}+\frac{34}{128^2}\right)
        P_{kk}(P_{ij}+\thres)
    \le\frac{P_{kk}(P_{ij}+\thres)}{32} \tag{By \Cref{eq:independence_monotonicity} and $\thres^2\leq \delta\leq 32P_{kk}$}.
\end{align*}

\noindent To conclude, \Cref{eq:mhr-denominator-error} and $\delta \leq 32P_{kk}$ give
\(
    D\ge P_{kk}-\nicefrac{(P_{kk}+\thres)}{128}
    \ge\nicefrac{95P_{kk}}{128}\ge\nicefrac{P_{kk}}2>0,
\)
so that dividing the chain of inequalities above by $D$ bounds the trade-probability error by $(P_{ij}+\thres)/16$.

In both cases, multiplying by $(j-i)/K\le1$ gives
\[
    |\widehat\mu(i/K,j/K)-\mu(i/K,j/K)|
    \le\frac{j-i}{K}\,\frac{P_{ij}+\thres}{16}
    \le\frac{\mu(i/K,j/K)+\thres}{16}.
\]
For $i=0$, $j=K$, or $i=j$, both the true profit and its estimate
are zero. Thus the bound holds for every $e\in G_K$ at every
reached scale on the same event, proving the proposition.
\end{proof}

\logestimation*

\begin{proof}
The case $Q=1$ is immediate, so assume $Q<1$.
First, taking the expectation yields:
\[
    \E{\mathsf H_{N-1}}
    =\sum_{k\ge1}\frac{\P{N>k}}{k}
    =\sum_{k\ge1}\frac{(1-Q)^k}{k}
    =-\log Q.
\]

It remains to bound the centered noise $\xi=\mathsf H_{N-1}+\log Q$. The integral bounds for harmonic sums give
\[
    \log N\le\mathsf H_{N-1}\le1+\log N.
\]
We use these inequalities to control the two tails separately.

For the upper tail, $\mathsf H_{N-1}\le1+\log N$ gives, for every
$u\ge0$,
\[
\begin{aligned}
    \P{\xi\ge u}
    &\le \P{N\ge e^{u-1}/Q}\\
    &=(1-Q)^{\lceil e^{u-1}/Q\rceil-1}\\
    &\le e^{-Q(\lceil e^{u-1}/Q\rceil-1)}\\
    &\le e^{Q-e^{u-1}}
    \le e^{1-e^{u-1}}.
\end{aligned}
\]
Here we used $1-Q\le e^{-Q}$,
$\lceil \nicefrac{e^{u-1}}{Q}\rceil\ge \nicefrac{e^{u-1}}{Q}$, and $Q\le1$.
For $u\ge2$, the inequality $e^{u-1}-1\ge u^2/4$ bounds the last
expression by $e^{-u^2/4}$. For $0\le u<2$, we use the trivial
probability bound $1$. Thus
\begin{equation}\label{eq:mhr-harmonic-upper-tail}
    \P{\xi\ge u}
    \le
    \begin{cases}
        1, & 0\le u<2,\\
        e^{-u^2/4}, & u\ge2.
    \end{cases}
\end{equation}

For the lower tail, $\mathsf H_{N-1}\ge\log N$ gives, for
$0\le u\le-\log Q$:
\[
    \P{\xi\le-u}
    \le \P{N\le e^{-u}/Q}
    \le Q\frac{e^{-u}}Q
    = e^{-u}.
\]
The second inequality is a union bound over the first
$\lfloor e^{-u}/Q\rfloor$ trials, each of which succeeds with
probability $Q$. In this range, $u\le1-\log Q$, so
$e^{-u}\le e^{-u^2/(1-\log Q)}$.
For $u>-\log Q$, we have
$\xi=\mathsf H_{N-1}+\log Q\ge\log Q>-u$, so
$\P{\xi\le-u}=0$. Consequently,
\begin{equation}\label{eq:mhr-harmonic-lower-tail}
    \P{\xi\le-u}
    \le e^{-u^2/(1-\log Q)}
    \qquad(u\ge0).
\end{equation}

We now combine \Cref{eq:mhr-harmonic-upper-tail,eq:mhr-harmonic-lower-tail}. Since $1-\log Q\ge1$, for
$u\ge2$, summing over disjoint events gives
\[
    \P{|\xi|\ge u}
    \le e^{-u^2/4}+e^{-u^2/(1-\log Q)}
    \le2e^{-u^2/[4(1-\log Q)]}.
\]
For $0\le u<2$, the above probability is trivially at most $1$, whereas
$3e^{-u^2/[4(1-\log Q)]}\ge3/e>1$. Hence, for every $u\ge0$ we can claim:
\[
    \P{|\xi|\ge u}
    \le3e^{-u^2/[4(1-\log Q)]}.
\]

Applying this tail bound with
$u=4\sqrt{(1-\log Q)\log t}$ gives
\[
\begin{aligned}
    \E{e^{\xi^2/[16(1-\log Q)]}}
    &=1+\int_1^\infty
        \P{|\xi|>4\sqrt{(1-\log Q)\log t}}\,dt\\
    &\le1+3\int_1^\infty
        \operatorname{exp}\left(
            -\frac{16(1-\log Q)\log t}{4(1-\log Q)}
        \right)\,dt\\
    &=1+3\int_1^\infty t^{-4}\,dt
    =2.
\end{aligned}
\]
\noindent By the definition of the $\psi_2$ norm, this proves
$\|\xi\|_{\psi_2}\le4\sqrt{1-\log Q}$.
\end{proof}

\subsection{Missing Results from \texorpdfstring{\Cref{sec:Regular}}{Section \getrefnumber{sec:Regular}}}
\label{app:missing_regular}
\initproperties*
\begin{proof}
\init calls \range unchanged to find the profit scale, then
performs a separate estimation step at that scale before constructing
the candidates. We analyze the steps of
Algorithm~\ref{alg:regular-initialization-sketch} in order, allowing
initialization to run to completion.

\paragraph{Step~\ref{step:regular-init-scale}: the profit scale.}
We give a guarantee on the scale $\thres$ in relation to the maximal expected profit.
For the call to \range, the concentration event of
Proposition~\ref{prop:mhr-concentration} has failure probability at most
$8T^3(2T)^{-4}=1/(2T)$, by the union bound in its proof.
On this event, the scale argument in the proof of
Lemma~\ref{lem:mhr-safe-body} gives
$\opt<6/K$ if there is no trigger, and $\thres\le\opt<6\thres$ otherwise.

For the later retention and hull bounds, we also compare the grid
optimum $\opt_K=\max_{e\in\grid\cap\cT}\mu(e)$ with $\opt$ and $\thres$.
Lemma~\ref{lem:discretization_err} (Discretization Error) gives
\begin{equation}\label{eq:regular-grid-approximation}
 \opt_K\ge\opt-2/K.
\end{equation}
At a trigger, the $3\thres/2$ threshold and the error bound in
Proposition~\ref{prop:mhr-concentration} give
\begin{equation}\label{eq:regular-grid-scale}
 \frac{3\thres}{2}\le\frac{17\opt_K+\thres}{16}
 \quad\Longrightarrow\quad \opt_K\ge\frac{23}{17}\thres.
\end{equation}
The additional \emph{no trigger} test checks whether $\thres<2^{16}/K$;
then $\opt<6\cdot2^{16}/K=O(1/K)$.
Henceforth assume $\thres\ge2^{16}/K\ge2^{15}\eta>\eta$.

\paragraph{Step~\ref{step:regular-init-estimates}: more accurate empirical estimates.}
We use the estimates of \range with a smaller error tolerance. Keep its sampled Sierpinski index set $\cS$, with $|\cS|=O(K\log(K))$, and fix the returned scale $\thres$. Post each sampled pair for $n=\lceil2^{38}\log(2T)/\thres\rceil$ fresh rounds. Write $P_{ij}=P(i/K,j/K)$ and let $\overline P_{ij}$ be its empirical trade frequency.

For this concentration calculation, condition on the completed transcript of \range, so that $\thres$ and $\cS$ are fixed; all additional samples are fresh. The Bernstein and union-bound calculation in the proof of Proposition~\ref{prop:mhr-concentration}, with this sample size, gives
\[
 |\overline P_{ij}-P_{ij}|
 \le\frac{P_{ij}+\thres}{2^{18}}
 \qquad((i,j)\in\cS)
\]
with conditional failure probability at most $1/(2T)$. Let $Q$ be the event that both the concentration guarantee for the initial call and the displayed bound for the additional samples hold. The first guarantee fails with probability at most $1/(2T)$. Averaging the conditional failure bound for the additional samples over the initial transcript gives the same bound $1/(2T)$ for the second guarantee. A union bound therefore yields
\[
 \P{Q^c}\le\frac1{2T}+\frac1{2T}=\frac1T.
\]
Assume $Q$ for the remainder of the proof.

For each grid pair $(i/K,j/K)\in\cT$ with $0<i<j<K$, independence and monotonicity give
\begin{equation}\label{eq:regular-reconstruction}
 P_{ij}=\frac{P_{ik}P_{kj}}{P_{kk}},\qquad
 P_{ij}\le P_{ik},P_{kj}\le P_{kk},
\end{equation}
where all three pairs on the right are sampled. We use exactly the empirical reconstruction
\eqref{eq:regular-empirical-estimate} from \range, which always belongs to $[0,1]$.
The reconstruction calculation in the proof of Proposition~\ref{prop:mhr-concentration} applies with frequency-error coefficient $2^{-18}$ in place of $1/128$: its split at $P_{kk}=\thres/32$ becomes $P_{kk}=4\cdot2^{-18}\thres$, and both cases give error at most $8\cdot2^{-18}(P_{ij}+\thres)$.
Since $8\cdot2^{-18}=2^{-15}$, multiplying by $0\le q-p\le1$ gives the bound on $|\widehat\mu(e)-\mu(e)|$ in \eqref{eq:regular-estimation}: for $e=(p,q)$, use $(q-p)P(e)=\mu(e)$ and $(q-p)\thres\le\thres$.
With the prescribed diagonal and boundary estimates, these bounds hold for every $e\in\grid\cap\cT$:
\begin{equation}\label{eq:regular-estimation}
 \begin{aligned}
 0\le\widehat P(e)&\le1,&
 |\widehat P(e)-P(e)|&\le2^{-15}\bigl(P(e)+\thres\bigr),\\
 &&|\widehat\mu(e)-\mu(e)|&\le2^{-15}\bigl(\mu(e)+\thres\bigr),
 \end{aligned}
\end{equation}
where $\widehat\mu(p,q)=(q-p)\widehat P(p,q)$.
In particular,
\begin{equation}\label{eq:regular-relative-accuracy}
 (1-2^{-14})P(e)\le\widehat P(e)\le(1+2^{-14})P(e)
 \qquad\text{if }P(e)\ge\thres.
\end{equation}

\paragraph{Step~\ref{step:regular-init-retain}: retained prices.}
We show that every grid maximizer is retained and every retained pair has nearly optimal profit.
By \eqref{eq:regular-grid-scale}, $\thres\le\opt_K$.
Thus \eqref{eq:regular-estimation} implies
\begin{equation}\label{eq:regular-profit-estimation}
 |\widehat\mu(e)-\mu(e)|\le2^{-14}\opt_K
 \qquad(e\in\grid\cap\cT),
\end{equation}
and consequently
\begin{equation}\label{eq:regular-estimated-maximum}
 (1-2^{-14})\opt_K\le\widehat\mu_K^\star\le(1+2^{-14})\opt_K,
 \qquad \widehat\mu_K^\star=\max_{e\in\grid\cap\cT}\widehat\mu(e).
\end{equation}
The retention threshold $1-2^{-13}$ leaves room for errors in both directions. Indeed, at a grid maximizer $e^\star$,
\[
 \widehat\mu(e^\star)\ge(1-2^{-14})\opt_K
 \ge(1-2^{-13})(1+2^{-14})\opt_K
 \ge(1-2^{-13})\widehat\mu_K^\star.
\]
Conversely, every retained pair $e$ satisfies
\[
 \begin{aligned}
 \mu(e)&\ge\widehat\mu(e)-2^{-14}\opt_K
 &&\quad\text{by \eqref{eq:regular-profit-estimation}}\\
 &\ge\bigl((1-2^{-13})(1-2^{-14})-2^{-14}\bigr)\opt_K
 &&\quad\text{by retention and \eqref{eq:regular-estimated-maximum}}\\
 &\ge(1-2^{-12})\opt_K\ge(1-0.0003)\opt.
 \end{aligned}
\]
The last inequality uses \eqref{eq:regular-grid-approximation}: its
relative loss is at most $2/(K\opt)\le2^{-15}$, since $\opt\ge\thres\ge2^{16}/K$, and $2^{-12}+2^{-15}<0.0003$. We have therefore proved
\begin{equation}\label{eq:regular-retained}
 \begin{aligned}
 \operatorname*{arg\,max}_{e\in\grid\cap\cT}\mu(e)
 &\subseteq
 \{e\in\grid\cap\cT:\widehat\mu(e)\ge(1-2^{-13})\widehat\mu_K^\star\}\\
 &\subseteq
 \{e\in\grid\cap\cT:\mu(e)\ge(1-0.0003)\opt\}.
 \end{aligned}
\end{equation}
Moreover, this implies that every retained pair belongs to $\operatorname{int}(\cT)$.

\paragraph{Step~\ref{step:regular-init-group}: grouping by balance.}
We first check that each assigned pair has balance close to its root's. Process the retained pairs in a fixed order. Assign $x=(p,q)$ to an existing root $w=(u,v)$ if both crosses belong to $\cT$, both have estimated profit at least $0.99\widehat\mu_K^\star$, and
\begin{equation}\label{eq:regular-group}
 \left|\frac12\log\frac{\widehat P(p,v)}{\widehat P(u,q)}\right|
 \le0.01.
\end{equation}
If no root passes these tests, start a new group with root $x$. For a cross $y$ passing the profit test,
\eqref{eq:regular-profit-estimation} and
\eqref{eq:regular-estimated-maximum} give
\[
 P(y)\ge\mu(y)\ge0.99\widehat\mu_K^\star-2^{-14}\opt_K
 \ge\bigl(0.99(1-2^{-14})-2^{-14}\bigr)\opt_K>\thres,
\]
using \eqref{eq:regular-grid-scale} in the last step. The relative-error bound \eqref{eq:regular-relative-accuracy} therefore applies to both crosses. Combining it with the balance test
\eqref{eq:regular-group} and identity \eqref{eq:regular-balance-ratio} gives
\begin{equation}\label{eq:regular-assigned-balance}
 \left|\log\frac{\lambda_x}{\lambda_w}\right|
 \le0.01+\frac12\log\frac{1+2^{-14}}{1-2^{-14}}<0.011
\end{equation}
for every pair in the group rooted at $w$. We use this bound in Step~\ref{step:regular-init-hulls} to show that these pairs have high auxiliary profit at the common parameter $\lambda_w$.

To bound the number of groups, we first show that any existing root $w$ with $|\log(\lambda_x/\lambda_w)|\le0.005$ satisfies the feasibility, estimated-profit, and estimated-balance conditions in Step~\ref{step:regular-init-group} for assigning $x$ to its group.
Both $x$ and $w$ are retained, so their actual profit gaps are at most $0.0003\opt$ by \eqref{eq:regular-retained}.
At the root, $\lambda_w$ is exactly balanced, so $r_{\lambda_w}(w)=\mu(w)$ and $\opt-r_{\lambda_w}(w)\le0.0003\opt$. At $x$, using $\lambda_w$ instead of $\lambda_x$ introduces an additional loss bounded by Lemma~\ref{lem:regular-auxiliary}:
\[
 \begin{aligned}
 \opt-r_{\lambda_w}(x)
 &=\bigl(\opt-\mu(x)\bigr)+\bigl(\mu(x)-r_{\lambda_w}(x)\bigr)\\
 &\le0.0003\opt+\mu(x)\left(\log\frac{\lambda_w}{\lambda_x}\right)^2\\
 &\le(0.0003+0.005^2)\opt=0.000325\opt.
 \end{aligned}
\]
Both auxiliary profits exceed $0.99\opt$, so Lemma~\ref{lem:regular-cross} makes both crosses feasible with gaps at most $4(0.0003+0.000325)\opt=0.0025\opt$.
We now verify the two estimated-profit tests. For either cross $y$, its price difference is at most one, and \eqref{eq:regular-grid-scale} gives
\[
 P(y)\ge\mu(y)\ge(1-0.0025)\opt
 \ge(1-0.0025)\frac{23}{17}\thres>\thres.
\]
Thus \eqref{eq:regular-relative-accuracy} applies to both crosses and yields
\[
 \begin{aligned}
 \widehat\mu(y)
 &\ge(1-2^{-14})\mu(y)\ge(1-2^{-14})(1-0.0025)\opt\\
 &>0.99(1+2^{-14})\opt\ge0.99\widehat\mu_K^\star,
 \end{aligned}
\]
where the last inequality uses \eqref{eq:regular-estimated-maximum}. The balance test also passes: the same probability-estimate comparison used above, together with \eqref{eq:regular-balance-ratio}, gives
\[
 \begin{aligned}
 \left|\frac12\log\frac{\widehat P(p,v)}{\widehat P(u,q)}\right|
 &\le\left|\log\frac{\lambda_x}{\lambda_w}\right|
       +\frac12\log\frac{1+2^{-14}}{1-2^{-14}}\\
 &\le0.005+\frac12\log\frac{1+2^{-14}}{1-2^{-14}}<0.01.
 \end{aligned}
\]
Consequently, a retained pair within log distance $0.005$ of an existing root joins an existing group. Since a new root is created only when all existing roots fail the tests, distinct roots have log balances more than $0.005$ apart.

To bound the interval containing these log balances, \eqref{eq:regular-retained} and $\opt\ge2^{16}/K$ from Step~\ref{step:regular-init-scale} give $\mu(x)\ge(1-0.0003)\opt\ge1/K$ at every retained pair.
Each marginal acceptance probability is at least this profit and at most one. Hence \eqref{eq:regular-auxiliary} gives
\[
 |\log\lambda_x|
 =\frac12\left|\log\frac{\Fs(p)}{1-\Fb(q)}\right|
 \le\frac12\log K.
\]
Thus all root log balances lie in an interval of length $\log K$. Since they are more than $0.005$ apart, there are at most $1+\log K/0.005=O(\log(T))$ roots, using $K\le T$.

\paragraph{Step~\ref{step:regular-init-hulls}: convex hulls.}
We now extend the profit and balance guarantees to every point of each convex hull. Each group's convex hull is compact, contains its root, and lies in $\operatorname{int}(\cT)$. Every vertex $x$ is a retained pair in the group rooted at $w$. Its profit is at least $(1-0.0003)\opt$ by \eqref{eq:regular-retained}, and its log balance is within $0.011$ of the root's by \eqref{eq:regular-assigned-balance}.
The balancing identity \eqref{eq:regular-balance} therefore gives
\begin{equation}\label{eq:regular-hull-profit}
 \begin{aligned}
 r_{\lambda_w}(x)
 &=\mu(x)\operatorname{sech}^2\!\left(\log\frac{\lambda_w}{\lambda_x}\right)\\
 &\ge(1-0.0003)\opt\operatorname{sech}^2(0.011)
 \ge(1-1/2000)\opt.
 \end{aligned}
\end{equation}
By \eqref{eq:regular-hull-profit}, the reciprocal $1/r_{\lambda_w}$ is at most $1/((1-1/2000)\opt)$ at every vertex. Its convexity, from Lemma~\ref{lem:regular-auxiliary}, extends this upper bound throughout the hull, and hence extends the corresponding lower bound on $r_{\lambda_w}$. Since $\mu(x)\le\opt$, \eqref{eq:regular-balance} then gives
\(
 \tanh^2\!\left(\log\frac{\lambda_x}{\lambda_w}\right)
 =1-\frac{r_{\lambda_w}(x)}{\mu(x)}
 \le 1/{2000}.
\)
Thus $|\log(\lambda_x/\lambda_w)|\le D$ throughout the hull. For any offset $|\theta_i|\le D$ selected in Step~\ref{step:regular-init-offsets} and $|u|\le10^{-3}$,
the triangle inequality gives
\[
 \left|\log\frac{e^{\theta_i+u}\lambda_w}{\lambda_x}\right|
 \le|\theta_i|+|u|+\left|\log\frac{\lambda_w}{\lambda_x}\right|
 \le2D+10^{-3}.
\]
Using $\mu(x)\ge r_{\lambda_w}(x)\ge(1-1/2000)\opt$ in
\eqref{eq:regular-balance}, it follows that
\[
 r_{e^{\theta_i+u}\lambda_w}(x)
 \ge(1-1/2000)\opt\operatorname{sech}^2(2D+10^{-3})>0.99\opt.
\]
This proves the first assertion of part~\ref{item:regular-init-auxiliary}.

For part~\ref{item:regular-init-crosses}, we first compare probabilities with the true root probability $P(w)$. By \eqref{eq:regular-hull-profit}, every hull point has auxiliary profit above $0.99\opt$ at the common parameter $\lambda_w$, so Lemma~\ref{lem:regular-cross} applies to any two hull points. Applying it to $x$ and $w$ gives
\[
 \frac{P(w)}3\le P(x)\le3P(w).
\]
Applying it to $x$ and another hull point $z$ makes both crosses feasible and gives $P(x)/3\le P(y)\le3P(x)$ for either cross $y$. Combining the two comparisons yields
\[
 \frac{P(w)}9\le\frac{P(x)}3\le P(y)\le3P(x)\le9P(w).
\]
Thus the original points and their crosses all have probabilities between $P(w)/9$ and $9P(w)$.

Next we verify $\thres\le\widehat P(w)\le1$ and replace $P(w)$ by its stored empirical estimate. The root's price difference is at most one, so the retention rule, \eqref{eq:regular-estimated-maximum}, and
\eqref{eq:regular-grid-scale} give
\[
 \widehat P(w)\ge\widehat\mu(w)
 \ge(1-2^{-13})\widehat\mu_K^\star
 \ge(1-2^{-13})(1-2^{-14})\opt_K>\thres.
\]
Also $\widehat P(w)\le1$ by construction. The retained-profit bound \eqref{eq:regular-retained} and \eqref{eq:regular-grid-scale} give $P(w)\ge\mu(w)>\thres$, so \eqref{eq:regular-relative-accuracy} implies
\[
 \frac{\widehat P(w)}{1+2^{-14}}\le P(w)
 \le\frac{\widehat P(w)}{1-2^{-14}}.
\]
Combining this with the factor-$9$ comparison above yields, for every original or crossed pair $y$,
\[
 \frac{\widehat P(w)}{10}
 \le\frac{\widehat P(w)}{9(1+2^{-14})}
 \le P(y)\le\frac{9\widehat P(w)}{1-2^{-14}}
 <10\widehat P(w).
\]

Finally, we prove the bound on $\thres/(q-p)$. By \eqref{eq:regular-hull-profit}, actual profit throughout the hull satisfies $\mu(p,q)\ge r_{\lambda_w}(p,q)\ge(1-1/2000)\opt>\thres$; the last inequality uses $\opt\ge(23/17)\thres$ from
\eqref{eq:regular-grid-scale}. Consequently,
\[
 \frac{\thres}{q-p}
 =\frac{\thres}{\mu(p,q)}P(p,q)
 \le P(p,q)\le10\widehat P(w).
\]
This proves part~\ref{item:regular-init-crosses}.

\paragraph{Step~\ref{step:regular-init-offsets}: offsets and a near-optimal candidate.}
We show that the offset mesh gives few candidates while preserving a nearly optimal one. Divide $[-D,D]$ into equal intervals of length at most $\tfrac12\sqrt{\eta/\thres}$, using the smallest possible number, and use their endpoints as offsets. Each root gives $O(\sqrt{\thres/\eta})$ candidates, hence $N=O(\sqrt{\thres/\eta}\log(2T))$ in total.
To identify the near-optimal candidate, choose a grid maximizer $e^\star$, retained by \eqref{eq:regular-retained}. By the balance bound in Step~\ref{step:regular-init-hulls}, its log balancing parameter relative to its assigned root lies in $[-D,D]$. Choose the nearest mesh endpoint as the offset. Its distance from this log balancing parameter is at most half the interval length, so this fixes a candidate $i^\star$ with
\[
 \left|\log\frac{\lambda_{i^\star}}{\lambda_{e^\star}}\right|
 \le\frac14\sqrt{\eta/\thres}.
\]
For every $|u|\le10^{-3}\sqrt{\eta/\thres}$, the triangle inequality bounds the perturbed mismatch by $(\tfrac14+10^{-3})\sqrt{\eta/\thres}$. Thus \eqref{eq:regular-balance}, \eqref{eq:regular-grid-approximation}, and $\opt<6\thres$ give
\[
 r_{\lambda_{i^\star}e^u}(e^\star)
 \ge\opt-\frac2K
       -6\thres\left(\tfrac14+10^{-3}\right)^2\frac{\eta}{\thres}
 \ge\opt-3\eta.
\]
The last inequality uses $1/K\le\eta$ and $6(\tfrac14+10^{-3})^2<1$.
This proves the remaining assertion of part~\ref{item:regular-init-auxiliary}.

\paragraph{Regret.}
Only Steps~\ref{step:regular-init-scale} and~\ref{step:regular-init-estimates} use trading rounds. The proof of Lemma~\ref{lem:mhr-safe-body} shows that, when \range returns scale $\thres$, it has used $O(K\log^2(T)/\thres)$ trading rounds in total. The additional sampling in Step~\ref{step:regular-init-estimates} uses $n|\cS|=O(K\log^2(T)/\thres)$ further trading rounds. All remaining steps use the stored estimates. Every posted pair is feasible and has regret at most $\opt<6\thres$, giving $O(K\log^2(T))$ regret. Together with $\P{Q}\ge1-1/T$, this proves the lemma.
\end{proof}

\endgroup

\section{The Zooming Dimension of Bilateral Trade}
\label{sec:dimension-comparison}

    In this appendix, we show that the tighter notion of \emph{zooming dimension} used in \citet{KleinbergSU19} does not improve over the ``trivial'' $T^{\nicefrac 34}$ regret bound for the stochastic setting. This implies that the generalized notion for the adversarial setting \citep{PodimataS21} does not help as well. 

    We then focus on the stochastic i.i.d. setting, and write $\Delta(x)=\mu^\star-\mu(x)$ for the suboptimality gap, and $N_r(S)$ for the minimum number of sets of diameter less than $r$ covering $S$. The stochastic zooming dimension $z$ in \citet[Definition~4.1]{KleinbergSU19} controls
    \[
        N_{\nicefrac{r}8}\bigl(\{x:\nicefrac{r}2<\Delta(x)\le r\}\bigr)\le C_0 r^{-z}.
    \]
    Their Theorem~4.2 gives regret $\widetilde O(T^{\nicefrac{z+1}{z+2}})$. Their metric-only refinement, the max-min-covering dimension, is still $2$ on $\cT$: every nonempty relatively open subset of this triangle contains a two-dimensional ball, and the triangle itself has covering dimension $2$.

\citet{PodimataS21} replace stochastic gaps by realized average gaps over prefixes of the horizon. Their adversarial zooming dimension covers arms whose gap is $\widetilde O(r)$ on some prefix of length $\Omega(r^{-2})$, using a finite representative set. It yields the same exponent $\nicefrac{z+1}{z+2}$. In the stochastic case their comparison lemma identifies this exponent with that obtained from the sublevel sets $\{\Delta\le O(r)\}$, allowing polylogarithmic changes in the covering multiplier. We distinguish these sublevel sets from the shells used above.

\paragraph{One-sided discretization is not a symmetric Lipschitz assumption.} Before addressing the zooming dimension bound, we remark that profit is not even Lipschitz, but only one sided. For deterministic valuations $(s,b)=(\nicefrac{1}4,\nicefrac{3}4)$, the optimal pair $(\nicefrac{1}4,\nicefrac{3}4)$ earns $\nicefrac 12$, whereas $(\nicefrac{1}4-\eta,\nicefrac{3}4)$ earns zero for every $0<\eta\le1/4$. Thus even the optimal-arm condition $\Delta(x)\le L\|x-x^\star\|$, allowed by \citet[Theorem~4.3]{KleinbergSU19}, fails for every fixed $L$. The directed rounding argument in Section~\ref{sec:challenges} avoids this problem by increasing the seller price and decreasing the buyer price. This observation alone does not dispose of \citet{PodimataS21}: their full version also treats one-sided Lipschitzness for one-sided pricing. The next example isolates the dimension obstruction even when the mean profit is Lipschitz.

\begin{proposition}\label{prop:dimension-two-profit}
There is a joint distribution of valuations in $[0,1]^2$ whose mean profit is Lipschitz on $\cT$, has a unique maximizer, and has stochastic zooming dimension $2$ in the price metric. The sublevel-set covering exponent is also $2$.
\end{proposition}
\begin{proof}
    Define the rectangle $R=[0,\nicefrac 14]\times[\nicefrac 34,1]$, the reference price $x_0=(\nicefrac18, \nicefrac78)$, and the function 
\[
 h(x)=
 \begin{cases}e^{-1/\|x-x_0\|_2^2},&x\ne x_0,\\0,&x=x_0,\end{cases}
\]
Consider the joint distribution with trade probability $H$ defined as follows on $R$:
\[
 H(p,q)=c\cdot\frac{(1-h(p,q))}{q-p},
\]
where $c$ is a normalization constant equal to $1/[2(1-e^{-32})]$. It can be verified analytically that there exists an explicit distribution supported on $R$ whose trade probability is $H$.

For arbitrary prices, define $p'=\min\{p,1/4\}$ and $q'=\max\{q,3/4\}$. The resulting trade probability is $H(p',q')$, so it is Lipschitz on $[0,1]^2$. For generic prices $(p,q)\in\cT$, we have the following bound on the expected profit
\[
 \mu(p,q)=(q-p)H(p',q')
 \le(q'-p')H(p',q')=c(1-h(p',q'))\le c.
\]
Equality holds only at $x_0$, so that the expected gap in $R$ has the following clean form $\Delta(x)=c\cdot h(x)$; outside $R$ it is bounded away from zero because the projected point lies on the boundary of $R$.
Consequently, for all sufficiently small $r$, the shell $\{\nicefrac r2<\Delta\le r\}$ is an annulus centered at $x_0$, of area
\[
 \pi\left(\frac1{\log(c/r)}-\frac1{\log(2c/r)}\right)
 =\frac{\pi\log2}{\log(c/r)\log(2c/r)}
 =\Theta\!\left(\frac1{\log^2(1/r)}\right).
\]
A set of diameter $r/8$ has area $O(r^2)$, hence the shell's covering number is at least $\Omega(r^{-2}/\log^2(1/r))$. This exceeds $C_0r^{-z}$ for every fixed $C_0$ and every $z<2$. The ambient two-dimensional covering bound supplies the reverse exponent. Likewise, $\{\Delta\le r\}$ is a disk of area $\pi/\log(c/r)$, so its covering exponent is $2$ as well.
\end{proof}

Thus the stochastic zooming bound retains the exponent $3/4$ on an admissible instance with correlated valuations. The stochastic specialization of the adversarial zooming bound has the same obstruction, including its finite-horizon multiplier. To see this quantitatively, take $r=T^{-\nicefrac 14}$ for sufficiently large $T$. The sublevel disk requires
\[
 N_r(\{\Delta\le r\})
 =\Omega\!\left(\frac{\sqrt{T}}{\log T}\right).
\]
Consequently, an exponent $z<2$ in a bound $N_r\le\gamma(T)r^{-z}$ requires
\[
 \gamma(T)=\Omega\!\left(\frac{T^{(2-z)/4}}{\log T}\right).
\]
This is an admissible scale for the stochastic comparison in \citet{PodimataS21}: $r^{-2}=T^{1/2}\le T$, the representative mesh of order $1/T$ resolves the disk, and the uniform stochastic fluctuation on their polynomial-size representative set at time $T$ is $O(\sqrt{\log T/T})=o(r)$. Their comparison transfers the covering obstruction to the adversarial notion up to polylogarithmic factors. The multiplier in their regret bound enters as $\gamma^{1/(z+2)}$, so the corresponding bound retains a factor of order at least
\[
 \widetilde\Omega\!\left(
 T^{(z+1)/(z+2)}T^{(2-z)/(4(z+2))}
 \right)=\widetilde\Omega(T^{3/4}).
\]
This concerns the size of the dimension-based guarantee, not the actual regret. In particular, a dimension bound of at most $1$ with a polylogarithmic multiplier, which would yield our worst-case exponent $\nicefrac 23$, cannot follow from the bilateral-trade structure alone. 

\section{Complete Analysis of the Stochastic Warm-Up}
\label{sec:stochastic-warmup-proof}

We give the full successive-elimination proof of
the warm-up result in the stochastic setting. Recall that in this setting the seller and buyer valuations may be arbitrarily
correlated within a round, but i.i.d. across rounds.

Fix an integer $K$ and recall the (discretized) action space \[
 \AA=G_K\cap\cT=\{(\nicefrac{i}K,\nicefrac{j}K):0\leq i\leq j\leq K\}.
\]
For a valuation distribution $\cQ$, write $\mu_\cQ(a)$ for the expected profit of $a$,
$\mu_\cQ^\star=\max_{a\in\AA}\mu_\cQ(a)$, and
$\Delta_\cQ(a)=\mu_\cQ^\star-\mu_\cQ(a)$ for its suboptimality gap when the agents' valuations are drawn according to $\cQ$. Finally, recall that the discretization lemma (\Cref{lem:discretization_err}) tells us that if we fix upfront the grid step-size $\nicefrac 1K$ and pay the additive  $\nicefrac{2T}{K}$ in the regret, then we can assume that the underlying distribution $\cD$ is supported on the uniform grid $G_K$, without loss of generality.

\paragraph{The local estimation primitive.}
    We repeatedly use the Localized Sampling Lemma (\Cref{lem:efficient_estimation}, \Cref{rem:sampling_lem_better_guarantee}) for various reference distributions $
\cQ$, scales
$\varepsilon$, and failure probabilities $\delta'$.  In each one of these calls, it estimates $\mu_{\cD}(a)$, for the real underlying distribution $\cD$, to error $\nicefrac{\varepsilon}{64}$, simultaneously on
\[
 S(\cQ,\varepsilon)=\{a\in\AA:\Delta_\cQ(a)\leq\varepsilon\},
\]
while querying only actions satisfying $\Delta_\cQ(a)\leq2\varepsilon$.
In each one of these calls, the number of queries is bounded as follows (for some universal constant $C$):
\begin{equation}\label{eq:stoch-local-cost}
 N(\varepsilon,\delta')
 \leq C \log^2(K)\log(K/\delta')
 \left(\frac K{\varepsilon^2}+\frac{\log^3(K)}{\varepsilon^3}\right).
\end{equation}

\paragraph{Confidence sets and phases.}
Let $H=\lfloor\log_4K\rfloor$, set $\varepsilon_h=4^{-h}$ for
$h=0,\ldots,H$, and let $\delta=T^{-2}$.
Initially, $\mathcal C_0$ is the simplex of all distributions on the finite
grid.  Phase $h$ consists of the following steps.
\begin{enumerate}[label=\arabic*.,leftmargin=*]
 \item Choose any $\cQ_h\in\mathcal C_h$, and set
 $S_h=S(\cQ_h,\varepsilon_h)$.
 \item Run the local estimator from \Cref{lem:efficient_estimation} with reference $\cQ_h$, scale
 $\varepsilon_h$, and failure probability $\delta/(H+1)$.
 Write $\widehat\mu_h(a)$ for its estimates on $S_h$.
 \item Intersect the current confidence set with the new constraints:
 \begin{equation}\label{eq:stoch-confidence-set}
 \mathcal C_{h+1}=
 \left\{\cQ\in\mathcal C_h:
  |\mu_\cQ(a)-\widehat\mu_h(a)|\leq\nicefrac{\varepsilon_h}{64}
  \text{ for every }a\in S_h\right\}.
 \end{equation}
\end{enumerate}
If the horizon ends during a phase, stop immediately.  If every phase
finishes, use an empirical maximizer of $\widehat\mu_H$ on $S_H$ for the
remaining rounds.  If a confidence set is empty, use any fixed action in
$\AA$ thereafter.

Let $\mathcal E$ be the event that every completed estimator satisfies its
simultaneous error guarantee.  Conditional on the history at the start of
a phase, $\cQ_h$ is fixed and all subsequent observations are generated by
fresh samples from $\cD$.  Thus the estimator guarantee and a union bound give
$\P{\mathcal E}\geq1-\delta$.  Conditioning on $\mathcal E$, the true distribution $\cD$ always belongs to
the maintained confidence sets $\cC_h$, so the fallback for an empty set is never used.
The enlarged query region makes it necessary to retain \emph{all} confidence
constraints, including those from earlier phases.  Indeed, \Cref{eq:stoch-confidence-set} implicitly considers the previous filtering of distributions, and thus the old constraints. Once an action is
excluded, these constraints prevent a later reference distribution from making it
look promising again.

\begin{lemma}\label{lem:stoch-nesting}
For every completed phase $h$ and every $\cQ\in\mathcal C_{h+1}$,
\begin{equation}\label{eq:stoch-excluded-gap}
 a\notin S_h \quad\Longrightarrow\quad
 \Delta_\cQ(a)>\tfrac34\varepsilon_h.
\end{equation}
Consequently, for $h\geq1$, every query in phase $h$ belongs to $S_{h-1}$.
Conditioning on $\mathcal E$, each such query $a$ has true gap
\begin{equation}\label{eq:stoch-query-regret}
 \Delta_{\cD}(a)\leq\tfrac94\varepsilon_h.
\end{equation}
The same bound holds in phase zero.
\end{lemma}
\begin{proof}
    We start from the first statement, proving it by induction on the level. For $h=0$, there is nothing to prove as $\varepsilon_0=1$ and all profits in $\AA$ lie in $[0,1]$, $S_0=\AA$ and \Cref{eq:stoch-excluded-gap} holds vacuously. Suppose now that  \Cref{eq:stoch-excluded-gap} holds at the generic level $h-1$.  Since $\cQ_h\in\mathcal C_h$, all actions
outside $S_{h-1}$ have $\cQ_h$-gap greater than
$3\varepsilon_{h-1}/4=3\varepsilon_h$.  Thus $S_h\subseteq S_{h-1}$,
every maximizer of $\mu_{\cQ_h}$ belongs to $S_{h-1}$, and every query in
phase $h$, whose $\cQ_h$-gap is at most $2\varepsilon_h$, belongs to
$S_{h-1}$ as well.
Fix $\cQ\in\mathcal C_{h+1}$.  If $a\notin S_{h-1}$, the induction
hypothesis and $\mathcal C_{h+1}\subseteq\mathcal C_h$ already give
$\Delta_\cQ(a)>3\varepsilon_{h-1}/4>3\varepsilon_h/4$.
Otherwise, let $a\in S_{h-1}\setminus S_h$ and choose a maximizer $a_h^\star$ of
$\mu_{\cQ_h}$.  Both $a$ and $a_h^\star$ belong to $S_{h-1}$, and both $\cQ$ and
$\cQ_h$ satisfy the constraints from phase $h-1$.  Hence
\[
 |\mu_\cQ(x)-\mu_{\cQ_h}(x)|\leq\frac{\varepsilon_{h-1}}{32}
 \qquad(x\in S_{h-1}).
\]
It follows that
\[
 \Delta_\cQ(a)\geq\mu_\cQ(a_h^\star)-\mu_\cQ(a)
 \geq\Delta_{\cQ_h}(a)-\frac{\varepsilon_{h-1}}{16}
 >\varepsilon_h-\frac{\varepsilon_h}4
 =\frac34 \varepsilon_h,
\]
completing the induction.

For the regret bound, work on $\mathcal E$ and fix a phase $h\geq1$.
The true distribution $\cD$ belongs to $\mathcal C_h$, so its maximizer $a^\star$
also belongs to $S_{h-1}$, by the induction hypothesis.  For any queried
action $a$, both $a$ and $a^\star$ are in $S_{h-1}$.  The same confidence
constraints therefore imply
\begin{align*}
 \Delta_{\cD}(a)=\mu_{\cD}(a^\star)-\mu_{\cD}(a)\leq\mu_{\cQ_h}(a^\star)-\mu_{\cQ_h}(a)
       +\frac{\varepsilon_{h-1}}{16}\leq2\varepsilon_h+\frac{\varepsilon_h}4
 =\frac94\varepsilon_h.
\end{align*}
In phase zero the gap is at most $1\leq9\varepsilon_0/4$.
\end{proof}

\begin{warmupresult}[Stochastic Setting]
Consider the problem of learning the profit-maximizing posted prices in bilateral trade in the stochastic setting. There exists a learning algorithm which enjoys $\widetilde O(T^{\nicefrac 23})$ regret.
\end{warmupresult}

\begin{proof}
Choose $K=\lceil T^{1/3}\rceil$.
The bounds above apply also to a phase interrupted by the horizon: the
quality of its queries depends only on the confidence constraints from
completed phases.  By \Cref{lem:stoch-nesting} and
\eqref{eq:stoch-local-cost}, on $\mathcal E$ the total conditional mean
regret during all estimation phases is at most
\begin{align*}
 \sum_{h=0}^H\frac94\varepsilon_h
       N(\varepsilon_h,\nicefrac{\delta}{H})
 &\leq C' \log^2(K)\log(KH/\delta)
    \sum_{h=0}^H
    \left(\frac K{\varepsilon_h}
                  +\frac{\log^3(K)}{\varepsilon_h^2}\right)\\
 &\leq C'' \log^2(K)\log(KH/\delta)
    \left(\frac K{\varepsilon_H}
                  +\frac{\log^3(K)}{\varepsilon_H^2}\right)\\
 &=\widetilde O(K^2).
\end{align*}
The final equality uses $\nicefrac 1K\leq\varepsilon_H<\nicefrac{4}K$.
If all phases complete, \Cref{eq:stoch-excluded-gap} applied to $\cD$ shows
that a true grid optimum lies in $S_H$.  The empirical maximizer therefore
has gap at most $\nicefrac{\varepsilon_H}{32}$, so that the remaining rounds contribute
at most $O(\nicefrac{T}K)$.

On $\mathcal E^c$, regret is at most $T$, so taking
expectations adds at most $T\delta\leq1$.  
Finally, adding the discretization error $\nicefrac{2T}K$ gives
\[
 R_T\in\widetilde O(K^2+\nicefrac{T}K)
 =\widetilde O(T^{2/3}).
\]
\end{proof}

\end{document}